\documentclass[11pt,a4paper]{article}
\usepackage[margin=1.0in]{geometry}
\usepackage{aas_macros,amsmath,amssymb,cite,hyperref,mathrsfs,microtype,wasysym}
\usepackage{xcolor}
\usepackage{comment}
\usepackage{graphicx}
\numberwithin{equation}{section}

\let\originalleft\left
\let\originalright\right
\renewcommand{\left}{\mathopen{}\mathclose\bgroup\originalleft}
\renewcommand{\right}{\aftergroup\egroup\originalright}

\newcommand{\ab}[1]{\left|#1\right|}
\newcommand{\av}[1]{\left\langle#1\right\rangle}
\newcommand{\br}[1]{\left[#1\right]}
\newcommand{\cu}[1]{\left\{#1\right\}}
\newcommand{\pa}[1]{\left(#1\right)}
\newcommand{\pd}{\mathop{}\!\partial}
\newcommand{\ed}{\mathop{}\!\mathrm{d}}

\DeclareMathOperator{\HeunC}{HeunC}
\DeclareMathOperator{\HeunG}{HeunG}
\DeclareMathOperator{\im}{Im}
\DeclareMathOperator{\re}{Re}
\newcommand{\mc}[1]{\mathcal{#1}}
\newcommand{\msc}[1]{\mathscr{#1}}
\newcommand{\ol}{\overline}

\newcommand{\wt}{\widetilde}

\usepackage[normalem]{ulem}

\begin{document}

\title{\textbf{Gravitational Waves on Kerr Black Holes II:\\
Metric Reconstruction with Cosmological Constant}}
\author{Roman Berens\footnote{\href{mailto:roman.berens@vanderbilt.edu}{roman.berens@vanderbilt.edu}}, Trevor \textbf{}Gravely\footnote{\href{mailto:trevor.gravely@vanderbilt.edu}{trevor.gravely@vanderbilt.edu}}, and Alexandru Lupsasca\footnote{\href{mailto:alexandru.v.lupsasca@vanderbilt.edu}{alexandru.v.lupsasca@vanderbilt.edu}}}
\date{\textit{Department of Physics \& Astronomy, Vanderbilt University, Nashville TN 37212, USA}}

\maketitle

\begin{abstract}
    In this second paper of our series started with \cite{Berens2024}, we investigate linearized gravitational perturbations of a rotating Kerr black hole in a non-asymptotically flat spacetime with (anti-)de Sitter boundary conditions.
    Here, we explicitly write down the metric components (in both ingoing and outgoing radiation gauge) of the perturbations that correspond to a given mode of either Weyl scalar.
    We provide formulas involving Hertz potentials (intermediate quantities with a holographic interpretation) as well as some that involve only the separated radial and angular modes.
    We expect these analytic results to prove useful in numerical studies of black hole perturbation theory in the context of the holographic correspondence.
\end{abstract}

\newpage
\tableofcontents

\parskip=1em
\newpage

\section{Introduction and summary}
\label{sec:Introduction}

This paper is the second in a series that started with Paper I \cite{Berens2024}.
Paper I explicitly reconstructed the metric components of the linearized gravitational perturbations of a rotating (Kerr) black hole with asymptotically flat boundary conditions.
In this paper, we extend this perturbative analysis to the background of a Kerr black hole embedded in a spacetime with a nonzero cosmological constant $\Lambda$, which we allow to be either negative or positive, resulting in the either the Kerr--anti-de Sitter ($\Lambda<0$) or Kerr--de Sitter ($\Lambda>0$) solutions, respectively.

This paper closely follows the structure of Paper I.
However, while the new results resemble those of Paper I at first glance, the addition of a nonzero cosmological constant fundamentally changes the nature of the spacetime, and one cannot merely repeat the prior analysis with an extra parameter $\Lambda\neq0$: many aspects of the computations---such as the solution to the radial and angular equations of motion---are qualitatively altered.

The Kerr--(anti-)de Sitter metric describes the spacetime geometry around a rotating black hole in an (anti)-de Sitter background.
In Boyer-Lindquist coordinates $(t,r,\theta,\phi)$, its line element $ds^2=g_{\mu\nu}\ed x^\mu\ed x^\nu$ takes the form
\begin{subequations}
\label{eq:KerrdS}
\begin{gather}
    ds^2=\epsilon_g\pa{-\frac{\Delta}{\Xi^2\Sigma}\pa{\ed t-a\sin^2{\theta}\ed\phi}^2+\frac{\Sigma}{\Delta}\ed r^2+\frac{\Sigma}{\Upsilon}\ed\theta^2+\frac{\Upsilon\sin^2{\theta}}{\Xi^2\Sigma}\br{\pa{r^2+a^2}\ed\phi-a\ed t}^2},\notag\\
    \Delta=r^2-2Mr+a^2-\frac{\Lambda}{3}r^2\pa{r^2+a^2},\qquad
    \Sigma=r^2+a^2\cos^2{\theta}
    =\zeta\ol{\zeta},\\
    \zeta=r-ia\cos{\theta},\qquad
    \Upsilon=1+\frac{\Lambda a^2}{3}\cos^2{\theta},\qquad
    \Xi=1+\frac{\Lambda a^2}{3},
\end{gather}
\end{subequations}
where $\epsilon_g=+1$ if the metric has signature $(-,+,+,+)$, or $\epsilon_g=-1$ if its signature is $(+,-,-,-)$.
Throughout this paper, we keep $\epsilon_g$ arbitrary to accommodate either choice of convention.

To study the gravitational perturbations of this spacetime, we insert the ansatz $g_{\mu\nu}+h_{\mu\nu}$ into Einstein's field equations and expand to linear order in the perturbation $h_{\mu\nu}$ around the Kerr--(anti-)de Sitter background $g_{\mu\nu}$, obtaining the linearized vacuum Einstein equations (App.~\ref{app:LinearizedGravity})
\begin{align}
    \label{eq:LinearizedEE}
    -\nabla^2h_{\mu\nu}+2\nabla^\rho\nabla_{(\mu}h_{\nu)\rho}-\nabla_\mu\nabla_\nu h+g_{\mu\nu}\pa{-\nabla^\rho\nabla^\sigma h_{\rho\sigma}+\nabla^2h}+\epsilon_g\Lambda\pa{g_{\mu\nu}h-2h_{\mu\nu}}=0.
\end{align}
As discussed in Paper I, this is a coupled system of ten second-order partial differential equations for the ten independent components of the metric perturbation $h_{\mu\nu}$.
Despite this complexity, however, the metric perturbation really contains only two physical (propagating) degrees of freedom, so this redundancy must be eliminated by fixing a gauge.

In a major breakthrough (reviewed in Paper I), Teukolsky \cite{Teukolsky1973,Teukolsky1974} realized that black hole perturbation theory could be dramatically simplified by changing the fundamental variables in the problem.
Rather than directly solving for the metric components $h_{\mu\nu}$ of a perturbation, he instead considered its associated Weyl curvature tensor, encoded in two (complex) Weyl scalars:%
\begin{subequations}
\label{eq:WeylScalars}
\begin{align}
    \psi_0&=\epsilon_gC_{\mu\nu\rho\sigma}^{(1)}l^\mu m^\nu l^\rho m^\sigma, \\
    \psi_4&=\epsilon_gC_{\mu\nu\rho\sigma}^{(1)}n^\mu\ol{m}^\nu n^\rho\ol{m}^\sigma.
\end{align}
\end{subequations}
Here, $C_{\mu\nu\rho\sigma}^{(1)}$ is the linearized Weyl tensor, expressible via derivatives of $h_{\mu\nu}$ as in Eq.~\eqref{eq:LinearizedWeyl} below, while $\cu{l,n,m,\ol{m}}$ is a Newman--Penrose tetrad \cite{Newman1962}: a complex null tetrad that consists of a pair of real null vectors $\cu{l,n}$ and a complex null vector $m$ obeying the conditions
\begin{subequations}
\label{eq:Tetrad}
\begin{gather}
    l^\mu l_\mu=n^\mu n_\mu
    =m^\mu m_\mu
    =\ol{m}^\mu\ol{m}_\mu 
    =l^\mu m_\mu 
    =l^\mu\ol{m}_\mu 
    =n^\mu m_\mu 
    =n^\mu\ol{m}_\mu 
    =0,\\
    -l^\mu n_\mu=m^\mu\ol{m}_\mu
    =\epsilon_g.
\end{gather}
\end{subequations}
Together, these conditions imply that the metric can be decomposed as
\begin{align}
    g_{\mu\nu}=\epsilon_g\pa{-2l_{(\mu}n_{\nu)}+2m_{(\mu}\ol{m}_{\nu)}}.
\end{align}
Teukolsky's seminal approach continues to work for perturbations of the Kerr--(anti-)de Sitter black hole with line element \eqref{eq:KerrdS}.
For this spacetime, our Newman--Penrose tetrad will be an extension of the standard Kinnersley tetrad introduced by Suzuki, Takasugi, and Umetsu \cite{Suzuki1998}:
\begin{subequations}
\label{eq:KinnersleyTetrad}
\begin{align}
    l&=l^\mu\pd_\mu
    =\frac{\Xi\pa{r^2+a^2}}{\Delta}\pd_t+\pd_r+\frac{a\Xi}{\Delta}\pd_\phi,\\
    n&=n^\mu\pd_\mu
    =\frac{1}{2\Sigma}\br{\Xi\pa{r^2+a^2}\pd_t-\Delta\pd_r+a\Xi\pd_\phi},\\
    m&=m^\mu\pd_\mu
    =\frac{1}{\sqrt{2\Upsilon}\,\ol{\zeta}}\pa{ia\Xi\sin{\theta}\pd_t+\Upsilon\pd_\theta+\frac{i\Xi}{\sin{\theta}}\pd_\phi}.
\end{align}
\end{subequations}
Just as in the original Kerr spacetime, when the linearized Einstein equations \eqref{eq:LinearizedEE} are recast as equations for the Weyl scalars \eqref{eq:WeylScalars}, they miraculously decouple \textit{and} separate.
In fact, both $\Psi^{(+2)}=\psi_0$ and $\Psi^{(-2)}=\zeta^4\psi_4$ obey a \textit{single} decoupled equation: the Teukolsky master equation
\begin{align}
    \label{eq:TME}
    E^{(s)}\Psi^{(s)}=4\pi\Sigma T^{(s)},
\end{align}
where the spin-$s$ equations of motion are packaged in the differential operator (here, a prime denotes $\pd_r$ or $\pd_\theta$ according to whether it acts on a function of $r$ only or $\theta$ only, respectively)
\begin{align}
    E^{(s)}&=\Xi^2\br{\frac{\pa{r^2+a^2}^2}{\Delta}-\frac{a^2\sin^2{\theta}}{\Upsilon}}\pd_t^2+2a\Xi^2\pa{\frac{r^2+a^2}{\Delta}-\frac{1}{\Upsilon}}\pd_t\pd_\phi\notag\\
    &\phantom{=}+\Xi^2\pa{\frac{a^2}{\Delta}-\frac{1}{\sin^2{\theta}}+\frac{\Lambda a^2}{3}\frac{\cot^2{\theta}}{\Upsilon}}\pd^2_\phi-\Delta^{-s}\pd_r\pa{\Delta^{s+1}\pd_r}-\frac{1}{\sin{\theta}}\pd_\theta\pa{\Upsilon\sin{\theta}\pd_\theta}\notag\\ 
    &\phantom{=}-2s\Xi\br{\frac{\pa{r^2+a^2}}{2}\frac{\Delta'}{\Delta}-2r+\frac{ia\Xi\cos{\theta}}{\Upsilon}}\pd_t-2s\Xi\br{\frac{a}{2}\frac{\Delta'}{\Delta}+\frac{i\cos{\theta}}{\sin^2{\theta}}\pa{2-\frac{\Xi}{\Upsilon}}}\pd_\phi\notag\\
    &\phantom{=}+s^2\Upsilon\pa{\frac{1}{2}\frac{\Upsilon'}{\Upsilon}+\cot{\theta}}^2-\frac{s}{2}\Delta''+\frac{2\Lambda}{3}\pa{2s^2+1}\Sigma,
\end{align}
while $T^{(s)}$ encodes the source for the spin-$s$ perturbation.\footnote{Besides describing the spin-2 gravitational perturbations when $s=\pm2$, this master equation also governs the perturbations of the Kerr--(anti-)de Sitter spacetime with arbitrary (half-)integer spin $s$: for instance, when $s=0$, it reduces to the  wave equation with conformal coupling $\pa{\nabla^2 -\frac{1}{6}R}\Psi^{(0)}=-4\pi\epsilon_gT^{(0)}$ for a scalar field $\Psi^{(0)}$, and when $s=\pm1$, it describes the two physical modes $\Psi^{(+1)}=F_{\mu\nu}l^\mu m^\nu$ and $\Psi^{(-1)}=\pa{r-ia\cos{\theta}}^2F_{\mu\nu}\ol{m}^\mu n^\nu$ within a spin-1 electromagnetic perturbation $F_{\mu\nu}$.}
In the vacuum (sourceless) case, $T^{(s)}=0$.
In this paper, we will restrict our attention to the linearized Einstein equations \eqref{eq:LinearizedEE} in vacuum, which correspond to the master equation \eqref{eq:TME} with $s=\pm2$ and $T^{(\pm2)}=0$.

Teukolsky's second ``miracle'' still occurs in the Kerr--(anti-)de Sitter spacetime, where the master equation \eqref{eq:TME} is once again separable.
After using the stationarity and axisymmetry of the background to decompose the Weyl scalars \eqref{eq:WeylScalars} into modes that behave as $e^{-i\omega t+im\phi}$, each $\Psi^{(\pm2)}$ has a radial and polar dependence that further separates as follows:
\begin{subequations}
\label{eq:ModeDecomposition}
\begin{align}
    \Psi^{(s)}(t,r,\theta,\phi)&=\int\ed\omega\sum_{\ell=\ab{s}}^\infty\sum_{m=-\ell}^{+\ell}c_{\ell m}^{(s)}(\omega)\Psi_{\omega\ell m}^{(s)}(t,r,\theta,\phi),\\
    \label{eq:SingleMode}
    \Psi_{\omega\ell m}^{(s)}(t,r,\theta,\phi)&=e^{-i\omega t+im\phi}R_{\omega\ell m}^{(s)}(r)S_{\omega\ell m}^{(s)}(\theta).
\end{align}
\end{subequations}
Indeed, plugging this ansatz into the master equation \eqref{eq:TME} separates it into two, second-order ordinary differential equations (ODEs) for the radial and angular modes $R_{\omega\ell m}^{(s)}(r)$ and $S_{\omega\ell m}^{(s)}(\theta)$:
\begin{align}
    \label{eq:RadialODE}
    \br{\Delta^{-s}\frac{d}{dr}\pa{\Delta^{s+1}\frac{d}{dr}}+V_r(r)-\lambda_{\omega\ell m}^{(s)}}R_{\omega\ell m}^{(s)}(r)&=0,\\
    \label{eq:AngularODE}
    \br{\frac{1}{\sin{\theta}}\frac{d}{d\theta}\pa{\Upsilon\sin{\theta}\frac{d}{d\theta}}-\frac{V_\theta(\theta)}{\Upsilon}-\frac{2\Lambda a^2}{3}\pa{2s^2+1}\cos^2{\theta}+s+\lambda_{\omega\ell m}^{(s)}}S_{\omega\ell m}^{(s)}(\theta)&=0,
\end{align}
where $\lambda_{\omega\ell m}^{(s)}$ denotes a separation constant, which is only known numerically, and we introduced radial and angular potentials
\begin{subequations}
\begin{align}
    \label{eq:Potentials}
    K&=\pa{r^2+a^2}\omega-am,\\
    V_r(r)&=\frac{\Xi^2K^2-is\Xi\Delta'K}{\Delta}+4is\omega\Xi r-\frac{2\Lambda}{3}(2s-1)(s-1)r^2+s\pa{\Delta''-2+\frac{\Lambda a^2}{3}},\\
    V_\theta(\theta)&=\pa{a\omega\Xi\sin{\theta}}^2-2a\omega\Xi^2\pa{m-s\cos{\theta}}+\frac{\br{m\Xi+s(2\Upsilon-\Xi)\cos{\theta}}^2}{\sin^2{\theta}}.
\end{align}
\end{subequations}

The radial and angular ODEs \eqref{eq:RadialODE} and \eqref{eq:AngularODE} reveal why the gravitational perturbations of the Kerr--(anti-)de Sitter spacetime are more complicated than their Kerr analogues.
In the Kerr spacetime, these ODEs each have two regular singular points at finite coordinate values, together with an irregular singular point at infinity.
As such, the Kerr radial and angular modes can be expressed in terms of the special function $\HeunC$, which is defined as a solution of the confluent Heun equation and was only recently implemented in \textsc{Mathematica}.
By contrast, in the Kerr--(anti-)de Sitter spacetime, these ODEs have five regular singular points (four at finite coordinate values and another at infinity).
A priori, this would suggest that they do \textit{not} belong to the Heun class of second-order linear ODEs with at most four singular points, and hence that their solutions may not be expressible using known special functions of Heun type.\footnote{The hypergeometric equation is the general second-order linear ODE with three regular singular points, and its solutions are very well understood.
The Heun equation is the generalization to four regular singular points, and its solutions are still the subject of active study.
Little is known when there are five regular singular points.}

Remarkably, this naive expectation is false due to a surprising property shared by the radial and angular ODEs \eqref{eq:RadialODE} and \eqref{eq:AngularODE}: for each of them, the characteristic exponents (roots of the indicial equation) of the regular singular point at infinity differ by an integer.
Thanks to this serendipitous property, this singular point may be entirely removed via a field redefinition discovered by Suzuki, Takasugi, and Umetsu \cite{Suzuki1998}.
As a result, both ODEs can be mapped to the general Heun equation with four regular singular points, whose solutions can be expressed in terms of the special function $\HeunG$ recently implemented in \textsc{Mathematica}.
We describe the requisite transformations for the angular and radial ODEs in Apps.~\ref{app:AngularHeun} and \ref{app:RadialHeun}, respectively.

Paper I reviewed in detail the motivation for solving the metric reconstruction problem (i.e., for finding the specific metric perturbation $h_{\mu\nu}$ associated with a given Weyl scalar).
Here, we give a brief summary of how to carry out this procedure in the Kerr--(anti-)de Sitter spacetime.
The relevant expressions are quite similar to those obtained in Paper I for the Kerr black hole; this similarity is largely explained by our continued use of the Geroch, Held, and Penrose (GHP) formalism \cite{Geroch1973}, which is largely unchanged by the presence of a nonzero cosmological constant.

The key result of metric reconstruction is the existence of two complex, symmetric, differential operators ${\mc{S}_0^\dag}_{\mu\nu}$ and ${\mc{S}_4^\dag}_{\mu\nu}$---given in Eqs.~\eqref{eq:AdjointOperator0}--\eqref{eq:AdjointOperator4} below---with the following properties.
Given a solution $\Psi^{(-2)}$ to the Teukolsky equation \eqref{eq:TME}, one forms the real, symmetric 2-tensor
\begin{align}
    \label{eq:MetricReconstructionIRG}
    h_{\mu\nu}^{\rm IRG}=2\epsilon_g\re\pa{{\mc{S}_0}_{\mu\nu}^\dag\Psi_{\rm H}}
    =2\epsilon_g\re\pa{{\mc{S}_0}_{\mu\nu}^\dag\Psi^{(-2)}}.
\end{align}
This field solves the linearized Einstein equations \eqref{eq:LinearizedEE} in the ``ingoing radiation gauge'' where
\begin{align}
    \label{eq:IRG}
    \text{IRG}:\qquad
    l^\mu h_{\mu\nu}^{\rm IRG}=0,\qquad
    g^{\mu\nu}h_{\mu\nu}^{\rm IRG}=0.
\end{align}
Likewise, given a solution $\Psi^{(+2)}$ to the Teukolsky equation \eqref{eq:TME}, the real, symmetric 2-tensor
\begin{align}
    \label{eq:MetricReconstructionORG}
    h_{\mu\nu}^{\rm ORG}=2\epsilon_g\re\pa{{\mc{S}_4}_{\mu\nu}^\dag\Psi_{\rm H}'}
    =2\epsilon_g\re\pa{{\mc{S}_4}_{\mu\nu}^\dag\zeta^4\Psi^{(+2)}}
\end{align}
solves the linearized Einstein equations \eqref{eq:LinearizedEE} in the ``outgoing radiation gauge'' where
\begin{align}
    \label{eq:ORG}
    \text{ORG}:\qquad
    n^\mu h_{\mu\nu}^{\rm ORG}=0,\qquad
    g^{\mu\nu}h_{\mu\nu}^{\rm ORG}=0.
\end{align}

As explained in Paper I, the ``reconstructed'' metrics \eqref{eq:MetricReconstructionIRG} and \eqref{eq:MetricReconstructionORG} do not have as their Weyl scalars the $\Psi^{(\pm2)}$ from which they are built via the application of ${\mc{S}_0^\dag}_{\mu\nu}$ or ${\mc{S}_4^\dag}_{\mu\nu}$.
To properly reconstruct the metric perturbation $h_{\mu\nu}$ associated with a given Weyl scalar $\psi_4$ or $\psi_0$, one must plug a \textit{different} solution $\Psi^{(-2)}$ or $\Psi^{(+2)}$ (known as a Hertz potential) into Eqs.~\eqref{eq:MetricReconstructionIRG} or \eqref{eq:MetricReconstructionORG}: one that is specifically engineered for the projections \eqref{eq:WeylScalars} to recover the desired Weyl scalars.

We continue to use $\Psi_{\rm H}$ to denote the solution $\Psi^{(-2)}$ to the master equation \eqref{eq:TME} with $s=-2$ that provides a Hertz potential for $h_{\mu\nu}^{\rm IRG}$, and will use $\Psi_{\rm H}'=\zeta^4\Psi^{(+2)}$ to denote the (rescaled) solution $\Psi^{(+2)}$ to the master equation \eqref{eq:TME} with $s=+2$ that provides a Hertz potential for $h_{\mu\nu}^{\rm ORG}$. 
The Weyl scalars $\psi_0$ and $\psi_4$ are related to the IRG Hertz potential $\Psi_{\rm H}$ by%
\begin{subequations}
\label{eq:IngoingPotential}
\begin{align}
    \label{eq:IngoingPotential0}
    \psi_0&=\frac{1}{4}l^4\ol{\Psi}_{\rm H},\\
    \label{eq:IngoingPotential4}
    \zeta^4\psi_4&=\frac{1}{4}\pa{\frac{1}{4}\mc{L}_{-1}\mc{L}_0\mc{L}_1\mc{L}_2\ol{\Psi}_{\rm H}-3M\Xi\pd_t\Psi_{\rm H}},
\end{align}
\end{subequations}
and similarly, they are also related to the ORG Hertz potential $\Psi_{\rm H}'$ by
\begin{subequations}
\label{eq:OutgoingPotential}
\begin{align}
    \label{eq:OutgoingPotential0}
    \psi_0&=\frac{1}{4}\pa{\frac{1}{4}\ol{\mc{L}}_{-1}\ol{\mc{L}}_0\ol{\mc{L}}_1\ol{\mc{L}}_2\frac{\ol{\Psi}_{\rm H}'}{\ol{\zeta}^4}+3M\Xi\pd_t\frac{\Psi_{\rm H}'}{\zeta^4}},\\
    \label{eq:OutgoingPotential4}
    \zeta^4\psi_4&=\frac{1}{4}\Delta^2\pa{\frac{\Sigma}{\Delta}n}^4\Delta^2\frac{\ol{\Psi}_{\rm H}'}{\ol{\zeta}^4},
\end{align}
\end{subequations}
where the differential operators $l$, $n$, and $m$ were defined in Eq.~\eqref{eq:KinnersleyTetrad}, and we also introduced
\begin{subequations}
\label{eq:CurlyLn}
\begin{align}
    \mc{L}_n&=\sqrt{2}\zeta\ol{m}+n\pa{\sqrt{\Upsilon}\cot{\theta}+\frac{\Upsilon'}{2\sqrt{\Upsilon}}}\\
    &=\sqrt{\Upsilon}\pd_\theta-\frac{ia\Xi\sin{\theta}}{\sqrt{\Upsilon}}\pd_t-\frac{i\Xi}{\sqrt{\Upsilon}\sin{\theta}}\pd_\phi+n\pa{\sqrt{\Upsilon}\cot{\theta}+\frac{\Upsilon'}{2\sqrt{\Upsilon}}}.
\end{align}
\end{subequations}
To the best of our knowledge, this is the first time that these relations appear in the literature.
We derive them in Sec.~\ref{subsec:KerrdS} below.
By manipulating them, one may completely eliminate the Hertz potentials $\Psi_{\rm H}$ and $\Psi_{\rm H}'$ to obtain two coupled, fourth-order, differential relations between the Weyl scalars $\psi_0$ and $\psi_4$ only: the Teukolsky--Starobinsky identities
\begin{subequations}
\label{eq:TeukolskyStarobinskyIdentities}
\begin{align}
    l^4\zeta^4\psi_4&=\frac{1}{4}\mc{L}_{-1}\mc{L}_0\mc{L}_1\mc{L}_2\psi_0-3M\Xi\pd_t\ol{\psi}_0,\\
    \Delta^2\pa{\frac{\Sigma}{\Delta}n}^4\Delta^2\psi_0&=\frac{1}{4}\ol{\mc{L}}_{-1}\ol{\mc{L}}_0\ol{\mc{L}}_1\ol{\mc{L}}_2\zeta^4\psi_4 +3M\Xi\pd_t\ol{\zeta}^4\ol{\psi}_4.
\end{align}
\end{subequations}
Here, they appear in their ``first form'' \cite{PriceThesis}; we give their ``second form'' in Eqs.~\eqref{eq:TSI2} below.

In summary, as in Paper I, given a $\psi_0$ (or $\psi_4$), one can solve Eq.~\eqref{eq:IngoingPotential0} (or \eqref{eq:IngoingPotential4}, resp.) for its IRG Hertz potential $\Psi_{\rm H}$, and then recover the corresponding $\psi_4$ (or $\psi_0$) via Eq.~\eqref{eq:IngoingPotential4} (or \eqref{eq:IngoingPotential0}, resp.), or reconstruct the physical metric perturbation $h_{\mu\nu}^{\rm IRG}$ in IRG via Eq.~\eqref{eq:MetricReconstructionIRG}.
Alternatively, given the same $\psi_0$ (or $\psi_4$), one can instead solve Eq.~\eqref{eq:OutgoingPotential0} (or \eqref{eq:OutgoingPotential4}, resp.) for its ORG Hertz potential $\Psi_{\rm H}'$, and then recover the corresponding $\psi_4$ (or $\psi_0$) via Eq.~\eqref{eq:OutgoingPotential4} (or \eqref{eq:OutgoingPotential0}, resp.), or reconstruct the physical metric perturbation $h_{\mu\nu}^{\rm ORG}$ in the ORG via Eq.~\eqref{eq:MetricReconstructionORG}.

As part of the metric reconstruction process, one must invert the relations \eqref{eq:IngoingPotential} and \eqref{eq:OutgoingPotential} to obtain the Hertz potentials $\Psi_{\rm H}$ and $\Psi_{\rm H}'$ in terms of the Weyl scalars $\psi_0$ and $\psi_4$.
As in Paper I, this is once again done at the mode level, that is, assuming that $\Psi^{(+2)}=\psi_0$ or $\Psi^{(-2)}=\zeta^4\psi_4$ consists of a single mode \eqref{eq:SingleMode}.
From a purely gravitational perspective, the Hertz potentials are intermediate quantities that can be eliminated and hence appear to play no fundamental physical role.
Nevertheless, these potentials do admit an interpretation in the context of anti-de Sitter holography.
Specifically, since the sources of the dual conformal field theory are related to the falloff coefficients of the metric perturbation near the anti-de Sitter boundary, they can be directly extracted from the Hertz potentials using \eqref{eq:MetricReconstructionIRG} or \eqref{eq:MetricReconstructionORG} \cite{Porfyriadis2014}.

Finally, we point to some related references that we found useful in carrying out this work.
Dias, Santos, and Stein \cite{Dias2012} also investigated metric perturbations of Kerr--anti-de Sitter, though with a specific focus on the near-horizon geometry of the extreme black hole and its conjectured holographic dual.
The follow-up paper \cite{Dias2013} examined the metric perturbations of Kerr--anti-de Sitter, providing expressions for Eqs.~\eqref{eq:MetricReconstructionIRG} and \eqref{eq:MetricReconstructionORG} in the Newman--Penrose formalism---see their Eqs.~(2.27) and (2.28)---but without providing the mode analysis in terms of Heun functions that we develop herein.
Dias and Santos \cite{Dias2013} also considered the quasinormal mode spectrum of the Schwarzschild--anti-de Sitter black hole, for which Kovtun \cite{Kovtun2005} had previously provided a holographic interpretation: perturbing the background spacetime is holographically equivalent to giving the boundary stress-tensor an expectation value $\av{T_{\mu\nu}}\propto h_{\mu\nu}$, which may be extracted from the limiting form of the metric perturbation at the boundary.
Earlier works in spherical symmetry by Cardoso and Lemos \cite{Cardoso2003}, and by Berti and Kokkotas \cite{Berti2003}, studied gravitational perturbations of the Schwarzschild--de Sitter and Reissner--Nordstr\"om--anti-de Sitter black holes, respectively.
Lastly, Stein \cite{Stein2024} recently showed that IRG and ORG are both horizon-locking gauges, implying that the location of the horizon is unchanged by perturbations in those gauges.

In the remainder of this paper, we present explicit expressions for the metric perturbation $h_{\mu\nu}$ of a Kerr--\mbox{(anti-)}de Sitter background associated with a single mode of a Weyl scalar $\psi_0$ or $\psi_4$. 
We give $h_{\mu\nu}$ in both IRG \eqref{eq:IRG} and ORG \eqref{eq:ORG}, and in three different coordinate systems: Boyer-Lindquist coordinates (Sec.~\ref{sec:Results}), as well as ingoing and outgoing coordinates (App.~\ref{app:IngoingOutgoing}). 
The structure of the main body of this paper is identical to that of Paper I and is laid out in Section 1.2 therein.
However, the appendices of this paper differ from those of Paper I.
In App.~\ref{app:LinearizedGravity}, we review the linearization of the Einstein equations with a nonzero cosmological constant. 
In App.~\ref{app:IngoingOutgoing}, we transform the components of $h_{\mu\nu}^{\rm IRG/ORG}$ to ingoing and outgoing coordinates, which have the distinct advantage of remaining regular across the horizon (unlike their Boyer-Lindquist counterparts). 
In App.~\ref{app:Roots}, we derive expressions for the roots of a generic quartic polynomial and use them to constrain the parameters of the spacetime. 
In App.~\ref{app:Fuchsian}, we review the theory of Fuchsian differential equations. 
In App.~\ref{app:Heun}, we discuss the general Heun equation and quote relevant results from Becker \cite{Becker1997}. 
We then apply these results to the angular ODE \eqref{eq:AngularODE} in App.~\ref{app:AngularHeun} and to the radial ODE \eqref{eq:RadialODE} in App.~\ref{app:RadialHeun}.
Finally, we connect our solutions to the angular and radial ODEs to those of Paper I in App.~\ref{app:KerrLimit}.
Beyond this paper, we also provide \textsc{Mathematica} notebooks with analytic expressions for the metric perturbation and checks of the linearized Einstein equations in \href{https://github.com/Metric-Reconstruction/metric-reconstruction-de-Sitter}{this Github}.

\section{Statement of results}
\label{sec:Results}

This section follows the same structure as Sec.~2 of Paper I.
We give explicit metric components $h_{\mu\nu}$ for a perturbation of the Kerr--(anti-)de Sitter spacetime \eqref{eq:KerrdS} with a particular associated Weyl scalar $\psi_0$ or $\psi_4$.
As in Paper I, we defer to Sec.~\ref{sec:Derivation} the derivations of these formulas, which use the full machinery of the Newman--Penrose and Geroch--Held--Penrose formalisms.

After some preliminary definitions in Sec.~\ref{subsec:Definitions}, we describe the angular modes $S_{\omega\ell m}^{(s)}(\theta)$ of the perturbations of the Kerr--(anti-)de Sitter geometry in Sec.~\ref{subsec:AngularModes} and then their radial modes $R_{\omega\ell m}^{(s)}(r)$ in Sec.~\ref{subsec:RadialModes}.
Then in Sec.~\ref{subsec:TeukolskyStarobinskyConstants}, we introduce the corresponding angular constants $D_{\omega\ell m}$, $D_{\omega\ell m}'$, and $\mc{D}_{\omega\ell m}$, together with radial constants $\msc{C}_{\omega\ell m}$, $\msc{C}_{\omega\ell m}'$, and $\mc{C}_{\omega\ell m}$, which appear in the angular and radial Teukolsky--Starobinsky identities that are presented in Sec.~\ref{subsec:TeukolskyStarobinsky} along with their unseparated version. 
Relations between the Weyl scalars $\psi_0$ and $\psi_4$ are given in Sec.~\ref{subsec:WeylRelations}. 
At last, we provide explicit metric reconstruction formulas from the modes of either Weyl scalar in Secs.~\ref{subsec:MetricReconstructionIRG} and \ref{subsec:MetricReconstructionORG} for ingoing and outgoing radiation gauge, respectively. 
Finally, we provide some consistency checks in Sec.~\ref{subsec:ConsistencyChecks} before concluding in Sec.~\ref{subsec:TeukolskyStarobinskyForms} with a discussion of the Teukolsky--Starobinsky identities in their various forms.

\subsection{Preliminary definitions}
\label{subsec:Definitions}

The horizons of the Kerr--(anti-)de Sitter metric \eqref{eq:KerrdS} are located at roots of $\Delta(r)$, and are complicated functions of the parameters $M$, $a$, and $\Lambda$.
Denoting these roots by $r_i$, we can write
\begin{align}
    \Delta(r) = -\frac{\Lambda}{3}\pa{r-r_1}\pa{r-r_2}\pa{r-r_3}\pa{r-r_4}.
\end{align}
Explicit expressions for the roots $r_i$ in terms of $M$, $a$, and $\Lambda$ are given in App.~\ref{subsec:ConstrainingDelta}.
For the Kerr--de Sitter metric (with $\Lambda>0$), we require that all four roots be real, whereas for the Kerr--anti-de Sitter metric (with $\Lambda<0$), we require that only two roots (which we label $r_1$ and $r_2$) be real.
These conditions place constraints on the spacetime parameters that we derive in App.~\ref{subsec:ConstrainingDelta}.
We also introduce the (anti)-de Sitter radius $L$ such that $\Lambda=3\sigma^2/L^2$, with
\begin{align}
    \label{eq:Sigma}
    \sigma=
    \begin{cases}
        1&\Lambda>0,\\
        i&\Lambda<0.
    \end{cases}
\end{align}
In the $\Lambda\to0$ (or $L\to\infty$) limit, the roots behave as\footnote{In the notation of Suzuki, Takasugi, and Umetsu \cite{Suzuki1998}, these roots are labeled $r_\pm$ and $r'_\pm$, respectively.}
\begin{align}
    \label{eq:RootExpansion}
    r_{1,2}=M\pm\sqrt{M^2-a^2}+\mc{O}\pa{\frac{1}{L^2}},\qquad
    r_{3,4}=\pm\sigma L-M+\mc{O}\pa{\frac{1}{L}}.
\end{align}
We see that the Kerr--de Sitter metric has three physical horizons at $r_1$, $r_2$, and $r_3$, while the Kerr--anti-de Sitter metric has two physical horizons at $r_1$ and $r_2$.
We can then define
\begin{align}
    \Omega_i=-\frac{g_{t\phi}}{g_{\phi\phi}}\bigg\vert_{\theta=0,\,r=r_i}
    =\frac{a^2}{r_i^2+a^2},
\end{align}
which may be interpreted as the angular velocity of the horizon located at $r=r_i$, provided that this radius is real and positive. 

The metric \eqref{eq:KerrdS} exhibits coordinate singularities at the horizons. 
Regular coordinates that ensure the metric remains smooth across the horizons are introduced in App.~\ref{app:IngoingOutgoing} using a tortoise coordinate
\begin{align}
    \label{eq:TortoiseCoordinate}
    r_*=\sum_{i=1}^2C_i\ln\pa{\frac{r-r_i}{2M}}+\sum_{i=3}^4C_i\ln\pa{\frac{r_i-r}{r_i}},\qquad
    C_i=\frac{\Xi\pa{r_i^2+a^2}}{\Delta'(r_i)},
\end{align}
as well as another coordinate
\begin{align}
    \label{eq:SharpCoordinate}
    r_\sharp=a\Xi\br{\sum_{i=1}^2\frac{1}{\Delta'(r_i)}\ln\pa{\frac{r-r_i}{2M}}+\sum_{i=3}^4\frac{1}{\Delta'(r_i)}\ln\pa{\frac{r_i-r}{r_i}}},
\end{align}
which are defined such that
\begin{align}
    \frac{dr_*}{dr}=\frac{\Xi\pa{r^2+a^2}}{\Delta},\qquad
    \frac{dr_\sharp}{dr}=\frac{a\Xi}{\Delta}.
\end{align}
For any choice of tetrad vectors $a,b\in\cu{l,n,m,\ol{m}}$, we will write $h_{ab}$ to denote the projection
\begin{align}
    \label{eq:TetradProjection}
    h_{ab}\equiv a^\mu b^\nu h_{\mu\nu}.
\end{align}
Since we only consider real metric perturbations $h_{\mu\nu}$, these projections are such that $\ol{h}_{ab}=h_{\ol{a}\ol{b}}$.

\subsection{Angular modes}
\label{subsec:AngularModes}

As in Kerr, for any fixed frequency $\omega\in\mathbb{C}$ and azimuthal angular momentum $m\in\mathbb{Z}$, there exists an infinite but discrete set of separation constants $\lambda^{(s)}(\omega, m)$ for which the angular ODE \eqref{eq:AngularODE} admits a solution that is regular at both poles ($\theta=0$ and $\pi$).
As always, each of these solutions has some number $n\geq0$ of zeros in that range (excluding the endpoints), allowing us to define an index $\ell\equiv n+\max\pa{|s|,|m|}$ such that $\ell\geq|s|$, and $-\ell\leq m\leq\ell$.
Together with $\omega$ and $m$, this index $\ell$ labels the regular solutions (or ``angular eigenmodes'') $S_{\omega\ell m}^{(s)}(\theta)$ and their associated separation constants (or ``eigenvalues'') $\lambda_{\omega\ell m}^{(s)}$.
These solutions take the form
\begin{align}
    \label{eq:AngularModes}
    \hat{S}_{\omega\ell m}^{(s)}(\theta)=\pa{1-u}^{\mu_1}\pa{1+u}^{\mu_2}\pa{\frac{u+\frac{i}{\sqrt{\alpha}}}{-1+\frac{i}{\sqrt{\alpha}}}}^{\mu_3}\pa{\frac{u-\frac{i}{\sqrt{\alpha}}}{-1-\frac{i}{\sqrt{\alpha}}}}^{\mu_4} H\pa{\frac{1}{2}\pa{1-\frac{i}{\sqrt{\alpha}}}\frac{u+1}{u-\frac{i}{\sqrt{\alpha}}}},
\end{align}
where the hat indicates that we are referring to the modes with this specific normalization, while $u=\cos{\theta}$, $\alpha=\frac{\Lambda a^2}{3}$, and
\begin{align}
\label{eq:AngularHeun}
    H(z)=\HeunG\pa{z_0,q,\alpha_{\rm H},\beta,2\mu_2+1,2\mu_1+1;z}
\end{align}
denotes the general Heun function as implemented in \textsc{Mathematica} (see App.~\ref{app:Heun} for details), which is normalized such that $H(0)=1$.
Here, we also introduced the parameters
\begin{subequations}
\label{eq:AngularModeParameters}
\begin{gather}
    \mu_1=\frac{|s+m|}{2},\qquad
    \mu_2=\frac{|s-m|}{2},\\
    \mu_3=\frac{i}{2}\pa{\frac{1+\alpha}{\sqrt{\alpha}}a\omega-m\sqrt{\alpha}-is},\qquad
    \mu_4=-(\mu_1+\mu_2+\mu_3+1),\\
    z_0=-\frac{i\pa{1+i\sqrt{\alpha}}^2}{4\sqrt{\alpha}},\qquad 
    \alpha_{\rm H}=\mu_1+\mu_2+s+1,\qquad
    \beta=\mu_1+\mu_2+2\mu_3+s+1,\\
    q=\frac{i}{4\sqrt{\alpha}}\bigg\{\lambda_{\omega\ell m}^{(s)}+s-2i\sqrt{\alpha}+2a\omega(1+\alpha)(m+s)-\frac{m^2}{2}\br{4\alpha+\pa{1+i\sqrt{\alpha}}^2}\notag\\
    \qquad\qquad\qquad\quad+\frac{s^2}{2}\pa{1-i\sqrt{\alpha}}^2+2ims\sqrt{\alpha}\pa{1+i\sqrt{\alpha}}-\pa{1+i\sqrt{\alpha}}^2\pa{\mu_1+\mu_2+2\mu_1\mu_2}\notag\\
    -2i\sqrt{\alpha}\br{(2\mu_3+1)(2\mu_2+1)-1}\bigg\}.
\end{gather}
\end{subequations}

The derivation of these modes is presented in App.~\ref{app:AngularHeun} below, where we also describe a method for determining the eigenvalues $\lambda_{\omega\ell m}^{(s)}$ from the general Heun function.
As in Kerr, the $\lambda_{\omega\ell m}^{(s)}$ do not admit an analytical representation and must always be computed numerically.

For fixed $\omega$, $m$, and $s$, the hatted modes \eqref{eq:AngularModes} are orthogonal (but not quite orthonormal):
\begin{align}
    \label{eq:Orthogonality}
    \int_0^\pi\hat{S}_{\omega\ell m}^{(s)}(\theta)\hat{S}_{\omega\ell'm}^{(s)}(\theta)\sin{\theta}\ed\theta=\delta_{\ell\ell'}I_{\omega\ell m}^{(s)},
\end{align}
where the constants $I_{\omega\ell m}^{(s)}$ may be obtained numerically but also admit a closed-form expression in terms of the general Heun function and its derivatives, which is given in App.~\ref{app:AngularNormalization} below.

If the frequency $\omega$ is real, then these modes are complete over $\theta\in[0,\pi]$.\footnote{For real $\omega$, the angular ODE can be written in self-adjoint Sturm-Liouville form. 
It is likely possible to generalize the results of Stewart \cite{Stewart1975} for Kerr to show at least weak completeness for $\omega$ within a complex disk.}
Thus, by Eq.~\eqref{eq:Orthogonality},
\begin{align}
    \sum_{\ell=\max\pa{|s|,|m|}}^\infty\frac{\hat{S}_{\omega\ell m}^{(s)}(\theta)\hat{S}_{\omega\ell m}^{(s)}(\theta')}{I_{\omega\ell m}^{(s)}}=\delta\pa{\cos{\theta}-\cos{\theta'}}.
\end{align}
The ``spin-weighted spheroidal harmonics'' are traditionally defined in asymptotically flat space \cite{Breuer1977}.
The hatted modes \eqref{eq:AngularModes} generalize them to the case of a cosmological constant $\Lambda\neq0$ via
\begin{align}
    Z_{\omega\ell m}^{(s)}(\theta,\phi)=\frac{1}{\sqrt{2\pi I_{\omega\ell m}^{(s)}}}\hat{S}_{\omega\ell m}^{(s)}(\theta)e^{im\phi}.
\end{align}
For real frequency $\omega$, these harmonics form a complete, orthonormal set over the 2-sphere:
\begin{subequations}
\begin{gather}
    \label{eq:Orthnormality}
    \int_0^\pi\int_0^{2\pi}Z_{\omega\ell m}^{(s)}(\theta,\phi)\ol{Z}_{\omega\ell'm'}^{(s)}(\theta,\phi)\sin{\theta}\ed\theta\ed\phi=\delta_{\ell\ell'}\delta_{mm'},\\
    \sum_{\ell=|s|}^\infty\sum_{m=-\ell}^{+\ell}Z_{\omega\ell m}^{(s)}(\theta,\phi)\ol{Z}_{\omega\ell m}^{(s)}(\theta',\phi')=\delta\pa{\cos{\theta}-\cos{\theta'}}\delta\pa{\phi-\phi'}.
\end{gather}
\end{subequations}
As $a\to0$ and spherical symmetry is restored, the spin-weighted spheroidal harmonics $Z_{\omega\ell m}^{(s)}(\theta,\phi)$ reduce (up to signs) to their standard spherical counterparts $Y_{\ell m}^{(s)}(\theta,\phi)=P_{\ell m}^{(s)}(\cos{\theta})e^{im\phi}$, where
\begin{align}
    P_{\ell m}^{(s)}(\cos{\theta})=N_{\ell m}^{(s)}\sin^{2\ell}\pa{\frac{\theta}{2}}\sum_{k=0}^{\ell-s}(-1)^{k}\binom{\ell-s}{k}\binom{\ell+s}{k+s-m}\cot^{2k+s-m}\pa{\frac{\theta}{2}},
\end{align}
are ``spin-weighted associated Legendre polynomials'' with normalization\footnote{This choice follows from the orthonormality condition \eqref{eq:Orthnormality}, that is, $\int_{S^2}Y^{(s)}_{\ell m}(\Omega)\ol{Y}^{(s)}_{\ell'm'}(\Omega)\ed\Omega=\delta_{\ell\ell'}\delta_{mm'}$.}
\begin{align}
    N_{\ell m}^{(s)}=(-1)^{\ell+m-s}\sqrt{\frac{2\ell+1}{4\pi}\frac{(\ell+m)!}{(\ell+s)!}\frac{(\ell-m)!}{(\ell-s)!}}.
\end{align}
As can be seen from the ODE \eqref{eq:AngularODE}, the dependence on $\Lambda\neq0$ only enters through the product $\Lambda a^2$ and therefore drops out in the limit $a\to0$.
At the mode level, this limit is quite complicated and requires the use of Eq.~\eqref{eq:HeunGToHeunC} because multiple parameters of $\HeunG$
blow up in Eq.~\eqref{eq:AngularHeun}.
As $a\to0$, the eigenvalues $\lambda_{\omega\ell m}^{(s)}$ lose their dependence on both $\Lambda$ and $\omega$, and degenerate to
\begin{align}
    \lambda_{0\ell m}^{(s)}=\ell(\ell+1)-s(s+1).
\end{align}

Finally, the symmetries of the angular ODE \eqref{eq:AngularODE} imply that our hatted angular eigenmodes \eqref{eq:AngularModes} and their eigenvalues obey again the following identities:\footnote{The proportionality factor in the last identity is set by the specific normalization of our hatted modes \eqref{eq:AngularModes}.}
\begin{align}
    \label{eq:AngularSymmetries}
    \lambda_{\omega\ell m}^{(-s)}=\lambda_{\omega\ell m}^{(s)}+2s,\qquad
    \ol{\lambda}_{\omega\ell m}^{(s)}=\lambda_{-\ol{\omega},\ell,-m}^{(s)},\qquad
    \ol{\hat{S}}_{\omega\ell m}^{(s)}=\hat{S}_{-\ol{\omega},\ell,-m}^{(-s)}.
\end{align}

\subsection{Radial modes}
\label{subsec:RadialModes}

As in Kerr,  for every frequency $\omega\in\mathbb{C}$ and integer harmonics $(\ell,m)$ with $\ell\ge|s|$ and $-\ell\le m\le\ell$, the second-order radial ODE \eqref{eq:RadialODE} admits a two-dimensional space of solutions.
In this paper, we focus on the ``in'' and ``out'' solutions, which correspond to modes that are purely ingoing or purely outgoing at the outer black hole horizon $(r=r_1)$, respectively:
\begin{subequations}
\label{eq:RadialModes}
\begin{align}
    \hat{R}_{\omega\ell m}^{(s)\,{\rm in}}(r)&=\frac{\pa{r-r_1}^{-\xi_1-s}}{\br{\Delta'(r_1)}^s}\pa{\frac{r-r_2}{r_1-r_2}}^{\xi_2}\pa{\frac{r-r_3}{r_1-r_3}}^{\xi_3}\pa{\frac{r-r_4}{r_1-r_4}}^{\xi_4}H^{\rm in}(z),\\
    \hat{R}_{\omega\ell m}^{(s)\,{\rm out}}(r)&=\pa{r-r_1}^{\xi_1}\pa{\frac{r-r_2}{r_1-r_2}}^{\xi_2}\pa{\frac{r-r_3}{r_1-r_3}}^{\xi_3}\pa{\frac{r-r_4}{r_1-r_4}}^{\xi_4}  
    H^{\rm out}(z).
\end{align}
\end{subequations}
Here, the hat indicates that we are referring to the modes with this specific normalization, and
\begin{subequations}
\begin{align}
    H^{\rm in}(z)&=\HeunG\pa{z_0,q+\pa{\epsilon+\delta z_0}(1-\gamma),1+\alpha_{\rm H}-\gamma,1+\beta-\gamma,2-\gamma,\delta;z},\\
    H^{\rm out}(z)&=\HeunG\pa{z_0,q,\alpha_{\rm H},\beta,\gamma,\delta;z},
\end{align}
\end{subequations}
are functions of a new radial coordinate
\begin{align}
    z=-\pa{\frac{r_2-r_4}{r_1-r_2}}\pa{\frac{r-r_1}{r-r_4}},
\end{align}
with parameters [recall that the $C_i$ were defined in Eq.~\eqref{eq:TortoiseCoordinate}]
\begin{subequations}
\label{eq:RadialModeParameters}
\begin{gather}
    \xi_1=iC_1\pa{\omega-m\Omega_1},\qquad
    \xi_2=iC_2\pa{\omega-m\Omega_2},\\
    \xi_3=iC_3\pa{\omega-m\Omega_3},\qquad
    \xi_4=-(\xi_1+\xi_2+\xi_3+2s+1),\\
    z_0=z\vert_{r=r_3}=-\pa{\frac{r_2-r_4}{r_1-r_2}}\pa{\frac{r_3-r_1}{r_1-r_2}},\qquad
    \alpha_{\rm H}=2s+1,\qquad
    \beta=-\frac{2i\Xi K_4}{\Delta'(r_4)}+s+1,\\
    \gamma=2\xi_1+s+1,\qquad 
    \delta=2\xi_2+s+1,\qquad
    \epsilon=\alpha_{\rm H}+\beta-\gamma-\delta+1,\\
    q=\frac{18\Xi^2}{\Lambda^2\mc{D}}\frac{(r_2-r_3)^2(r_2-r_4)^2(r_1-r_4)(r_3-r_4)}{r_1-r_2}\tilde{q}+\frac{6is\Xi}{\Lambda}\frac{\omega(r_1r_4+a^2)-am}{(r_1-r_2)(r_1-r_4)(r_3-r_4)}\notag\\
    \quad+\frac{3}{\Lambda(r_1-r_2)(r_3-r_4)}\pa{\lambda_{\omega\ell m}^{(s)}+\frac{s\Lambda a^2}{3}}+\frac{(s+1)(2s+1)}{r_1-r_2}\pa{r_4-r_2+\frac{2r_4^2}{r_3-r_4}}\notag\\
    +2\xi_1\pa{z_0\xi_2+\xi_3}+(s+1)\br{(z_0+1)\xi_1+z_0\xi_2+\xi_3},\\
    \tilde{q}=-\omega^2r_1^3\pa{r_1r_2+r_1r_3-2r_2r_3}-2a\omega(m-a\omega)r_1\pa{r_2r_3-r_1^2}-a^2(m-a\omega)^2(2r_1-r_2-r_3),\notag
\end{gather}
\end{subequations}
where $K_4=K\vert_{r=r_4}$ [with $K$ defined in Eq.~\eqref{eq:Potentials}], and $\triangle$ is the discriminant of $-\frac{3}{\Lambda}\Delta(r)$,
\begin{align}
    \triangle=\prod_{1\leq i<j\leq 4}(r_i-r_j)^2.
\end{align}

The derivation of these modes is presented in App.~\ref{app:RadialHeun} below.
Their behavior near the horizon is best described in terms of the tortoise coordinate \eqref{eq:TortoiseCoordinate}:
\begin{subequations}
\label{eq:HorizonBehavior}
\begin{align}
    \hat{R}_{\omega\ell m}^{(s)\,{\rm in}}(r)&\stackrel{r\to r_1}{\approx}\br{\Delta'(r_1)}^{-s}\pa{r-r_1}^{-iC_1k-s}
    \approx c^{\rm in}\Delta^{-s}e^{-ikr_*},\\
    \hat{R}_{\omega\ell m}^{(s)\,{\rm out}}(r)&\stackrel{r\to r_1}{\approx}\pa{r-r_1}^{iC_1k}
    \approx c^{\rm out}e^{ik r_*},
\end{align}
\end{subequations}
where $k=\omega-m\Omega_1$ and $c^{\rm in/out}$ are constants that are given explicitly in App.~\ref{app:RTSDerivation} below.

Finally, the symmetries of the radial ODE \eqref{eq:RadialODE} imply that the hatted radial eigenmodes \eqref{eq:RadialModes} again obey the following identities (with proportionality factors set by normalization):
\begin{align}
    \label{eq:RadialSymmetries}
    \ol{\hat{R}}_{\omega\ell m}^{(s)\,{\rm in/out}} 
    = \hat{R}_{-\ol{\omega},\ell,-m}^{(s)\,{\rm in/out}} 
    = \Delta^{-s}\hat{R}_{\ol{\omega}\ell m}^{(-s)\,{\rm out/in}}.
\end{align}

\subsection{Teukolsky--Starobinsky constants}
\label{subsec:TeukolskyStarobinskyConstants}

The Teukolsky--Starobinsky constants are derived in Sec.~\ref{subsec:TeukolskyStarobinsky} below; here, we merely quote them.
For some authors, the term refers to the quantities\footnote{\label{fn:TS}We have expressed these constants in terms of $\lambda_{\omega\ell m}^{(+2)}$, but could have also used $\lambda_{\omega\ell m}^{(-2)}=\lambda_{\omega\ell m}^{(+2)}+4$; see Eq.~\eqref{eq:AngularSymmetries}.}
\begin{align}
    \label{eq:TeukolskyStarobinskyD}
    \mc{D}_{\omega\ell m}&=\br{\pa{\lambda_{\omega\ell m}^{(+2)}+4-\frac{2}{3}\Lambda a^2}\pa{\lambda_{\omega\ell m}^{(+2)}+6-\frac{4}{3}\Lambda a^2}+4\Lambda a^2}^2\notag\\
    &\phantom{=}+8a\omega\Xi^2\pa{m-a\omega}\pa{\lambda_{\omega\ell m}^{(+2)}+4-\frac{2}{3}\Lambda a^2}\pa{5\lambda_{\omega\ell m}^{(+2)}+26-\frac{16}{3}\Lambda a^2}\notag\\
    &\phantom{=}+96a^2\Xi^2\pa{\lambda_{\omega\ell m}^{(+2)}+4-\frac{2}{3}\Lambda a^2}\br{\omega^2-\frac{1}{3}\Lambda\pa{m-a\omega}^2}\notag\\
    &\phantom{=}+144a^2\omega\Xi^2(m-a\omega)\br{\Xi^2\omega(m-a\omega)-\frac{2}{3}a\Lambda}.
\end{align}
For other authors, the term instead refers to the frequency-shifted version
\begin{align}
    \label{eq:RadialConstant}
    \mc{C}_{\omega\ell m}=\mc{D}_{\omega\ell m}+\pa{12\omega\Xi M}^2.
\end{align}
Both versions factorize into products of other quantities with important properties described in Sec.~\ref{subsec:TeukolskyStarobinsky}.
In particular, $\mc{D}_{\omega\ell m}$ is the product of ``angular Teukolsky--Starobinsky constants''
\begin{align}
    \label{eq:Factorization}
    \mc{D}_{\omega\ell m}=\hat{D}_{\omega\ell m}\hat{D}_{\omega\ell m}',
\end{align}
which take the explicit form\footref{fn:TS}
\begin{align}
    \label{eq:ConstantsD}
    \hat{D}_{\omega\ell m}=
    \begin{cases}
        \frac{(m-2)!}{(m+2)!}\Xi^{-2}\mc{D}_{\omega\ell m}
        &m\geq2,\\
        -\frac{1}{6\Xi}p_\omega^{(3)}\pa{\lambda_{\omega\ell m}^{(+2)}}
        &m=1,\\
        p_{-\omega}^{(2)}\pa{\lambda_{\omega\ell m}^{(+2)}}
        &m=0,\\
        -6\Xi p_{-\omega}^{(1)}\pa{\lambda_{\omega\ell m}^{(+2)}}
        &m=-1,\\
        \frac{(m+1)!}{(m-3)!}\Xi^2
        &m\leq-2,
    \end{cases}\qquad
    \hat{D}_{\omega\ell m}'=
    \begin{cases}
        \frac{(m+2)!}{(m-2)!}\Xi^2
        &m\geq2,\\
        -6\Xi p_\omega^{(1)}\pa{\lambda_{\omega\ell m}^{(+2)}}
        &m=1,\\
        p_\omega^{(2)}\pa{\lambda_{\omega\ell m}^{(+2)}}
        &m=0,\\
        -\frac{1}{6\Xi}p_{-\omega}^{(3)}\pa{\lambda_{\omega\ell m}^{(+2)}}
        &m=-1,\\
        \frac{(m-3)!}{(m+1)!}\Xi^{-2}\mc{D}_{\omega\ell m}
        &m\leq-2,
    \end{cases}
\end{align}
where $\frac{(m-3)!}{(m+1)!}=\frac{1}{(m+1)m(m-1)(m-2)}$, etc., and in terms of $c\equiv a\omega\Xi$ and the previously defined $\alpha=\frac{\Lambda a^2}{3}$, we introduced the polynomials\footnote{The factorization \eqref{eq:Factorization} is nontrivial for $|m|\le1$ and relies on the constants \eqref{eq:TeukolskyStarobinskyD} and polynomials \eqref{eq:Polynomials} obeying, for $x=\lambda_{\omega\ell m}^{(+2)}$, the identities $\mc{D}_{\omega\ell,1}=p_\omega^{(1)}(x)p_\omega^{(3)}(x)$, $\mc{D}_{\omega\ell,0}=p_\omega^{(2)}(x)p_{-\omega}^{(2)}(x)$, and $\mc{D}_{\omega\ell,-1}=p_{-\omega}^{(1)}(x)p_{-\omega}^{(3)}(x)$.}%
\begin{subequations}
\label{eq:Polynomials}
\begin{align}
    p_\omega^{(1)}(x)&=x+6c+4(1-2\alpha),\\
    p_\omega^{(2)}(x)&=x^2+2\pa{4c+5-3\alpha}x+4\br{3c^2+4\pa{2-\alpha}c+2\pa{3-2\alpha+\alpha^2}},\\
    p_\omega^{(3)}(x)&=x^3-2\pa{3c-8+2\alpha}x^2-4\br{c^2+4\pa{2-\alpha}c-21+2\alpha-5\alpha^2}x\notag\\
    &\phantom{=}+8\br{3c^3-2\pa{4+\alpha}c^2-\pa{1-10\alpha+\alpha^2}c+2\pa{9+\alpha^2-2\alpha^3}}.
\end{align}
\end{subequations}
Likewise, $\mc{C}_{\omega\ell m}$ is the product of ``radial Teukolsky--Starobinsky constants''
\begin{align}
    \mc{C}_{\omega\ell m}=\hat{\msc{C}}_{\omega\ell m}^{\rm in}\hat{\msc{C}}_{\omega\ell m}^{{\rm in}\,\prime}
    =\hat{\msc{C}}_{\omega\ell m}^{\rm out}\hat{\msc{C}}_{\omega\ell m}^{{\rm out}\,\prime},
\end{align}
which are separately defined for the ``in'' and ``out'' solutions as
\begin{align}
    \label{eq:ConstantsC}
    \hat{\msc{C}}_{\omega\ell m}^{\rm in}=\Gamma,\qquad
    \hat{\msc{C}}_{\omega\ell m}^{{\rm in}\,\prime}=\frac{\mc{C}_{\omega\ell m}}{\Gamma},\qquad
    \hat{\msc{C}}_{\omega\ell m}^{\rm out}=\frac{\mc{C}_{\omega\ell m}}{\wt{\Gamma}},\qquad
    \hat{\msc{C}}_{\omega\ell m}^{{\rm out}\,\prime}=\wt{\Gamma}.
\end{align}
Here, mirroring Paper I, we introduced the quantities
\begin{subequations}
\label{eq:GammaSigmaW}
\begin{gather}
    \Gamma=(w+2i\varsigma)(w+i\varsigma)w(w-i\varsigma),\qquad
    \wt{\Gamma}=(w-2i\varsigma)(w-i\varsigma)w(w+i\varsigma),\\
    \varsigma=\Delta'(r_1),\qquad
    w=2k\Xi(r_1^2+a^2).
\end{gather}
\end{subequations}
As in Kerr, the precise form of the angular and radial Teukolsky--Starobinsky constants depends on the specific normalization of our angular modes \eqref{eq:AngularModes} and radial modes \eqref{eq:RadialModes}, which is why these constants are hatted, but this normalization choice drops out of the products that define the Teukolsky--Starobinsky constants $\mc{D}_{\omega\ell m}$ or $\mc{C}_{\omega\ell m}$, which are therefore unhatted.

We will henceforth suppress the labels ``in'' or ``out'' on  $\hat{\msc{C}}_{\omega\ell m}$, $\hat{\msc{C}}_{\omega\ell m}'$, and $\hat{R}_{\omega\ell m}^{(s)}(r)$, as our equations will hold for either choice, though not for a linear combination thereof---caveat lector!

Finally, by careful inspection and use of the symmetries of $\lambda_{\omega\ell m}^{(s)}$ given in Eq.~\eqref{eq:AngularSymmetries}, one has%
\begin{subequations}
\label{eq:TeukolskyStarobinskySymmetries}
\begin{gather}
    \ol{\hat{D}}_{\ol{\omega}\ell m}=\hat{D}_{\omega\ell m},\qquad
    \ol{\hat{D}_{}'}_{\!\!\ol{\omega}\ell m}=\hat{D}_{\omega\ell m}',\qquad
    \hat{D}_{-\omega,\ell,-m}=\hat{D}_{\omega\ell m}',\\
    \ol{\hat{\msc{C}}}_{\omega\ell m}=\hat{\msc{C}}_{-\ol{\omega},\ell,-m},\qquad
    \ol{\hat{\msc{C}}_{}'}_{\!\!\!\omega\ell m}=\hat{\msc{C}}_{-\ol{\omega},\ell,-m}'.
\end{gather}
\end{subequations}
The first two identities simply reflect the fact that $\hat{D}_{\omega\ell m}$ and $\hat{D}_{\omega\ell m}'$ are real if the frequency $\omega$ is real.
On the other hand, $\hat{\msc{C}}_{\omega\ell m}$ and $\hat{\msc{C}}_{\omega\ell m}'$ are generally complex, even for real $\omega$.

\subsection{Radial and angular Teukolsky--Starobinsky identities}
\label{subsec:TeukolskyStarobinsky}

Using the radial ``potential'' $K$ defined in Eq.~\eqref{eq:Potentials}, we now define radial operators
\begin{align}
    \label{eq:Dn}
    \msc{D}_n=\pd_r-\frac{i\Xi K}{\Delta}+n\frac{\Delta'}{\Delta},\qquad
    \msc{D}_n^\dag=\pd_r+\frac{i\Xi K}{\Delta}+n\frac{\Delta'}{\Delta}.
\end{align}
Likewise, using an angular ``potential'' $Q=-a\omega\sin{\theta}+\frac{m}{\sin{\theta}}$, we also define angular operators
\begin{subequations}
\begin{align}
    \label{eq:Ln}
    \msc{L}_n&=\sqrt{\Upsilon}\pd_\theta+\frac{\Xi Q}{\sqrt{\Upsilon}}+n\pa{\sqrt{\Upsilon}\cot{\theta}+\frac{\Upsilon'}{2\sqrt{\Upsilon}}},\\
    \msc{L}_n^\dag&=\sqrt{\Upsilon}\pd_\theta-\frac{\Xi Q}{\sqrt{\Upsilon}}+n\pa{\sqrt{\Upsilon}\cot{\theta}+\frac{\Upsilon'}{2\sqrt{\Upsilon}}}.
\end{align}
\end{subequations}
These operators are the ``mode versions'' of $l$, $\frac{\Sigma}{\Delta}n$, $\mc{L}_n$, and $\ol{\mc{L}}_n$ of Eqs.~\eqref{eq:KinnersleyTetrad} and \eqref{eq:CurlyLn}, meaning%
\begin{subequations}
\label{eq:ModeVersions}
\begin{align}
    l\br{f(r,\theta)e^{-i\omega t+im\phi}}&=\msc{D}_0f(r,\theta)e^{-i\omega t+im\phi},\\
    \frac{\Sigma}{\Delta}n\br{f(r,\theta)e^{-i\omega t+im\phi}}&=-\frac{1}{2}\msc{D}_0^\dag f(r,\theta)e^{-i\omega t+im\phi},\\
    \mc{L}_n\br{f(r,\theta)e^{-i\omega t+im\phi}}&=\msc{L}_nf(r,\theta)e^{-i\omega t+im\phi},\\
    \ol{\mc{L}}_n\br{f(r,\theta)e^{-i\omega t+im\phi}}&=\msc{L}_n^\dag f(r,\theta)e^{-i\omega t+im\phi}.
\end{align}
\end{subequations}
In terms of these differential operators, the radial and angular ODEs \eqref{eq:AngularODE} and \eqref{eq:RadialODE} are just
\begin{subequations}
\begin{align} 
    \pa{\Delta\msc{D}_1\msc{D}_2^\dag+6i\omega\Xi r-2\Lambda r^2+\frac{2}{3}\Lambda a^2-\lambda_{\omega\ell m}^{(+2)}-4}R_{\omega\ell m}^{(+2)}&=0,\\
    \pa{\Delta\msc{D}_{-1}^\dag\msc{D}_0-6i\omega\Xi r-2\Lambda r^2+\frac{2}{3}\Lambda a^2-\lambda_{\omega\ell m}^{(-2)}}R_{\omega\ell m}^{(-2)}&=0,\\
    \pa{\msc{L}_{-1}^\dag\msc{L}_2-6a\omega\Xi\cos{\theta}-2\Lambda a^2\cos^2{\theta}-\frac{2}{3}\Lambda a^2+\lambda_{\omega\ell m}^{(+2)}+4}S_{\omega\ell m}^{(+2)}&=0,\\
    \pa{\msc{L}_{-1}\msc{L}_2^\dag+6a\omega\Xi\cos{\theta}-2\Lambda a^2\cos^2{\theta}-\frac{2}{3}\Lambda a^2+\lambda_{\omega\ell m}^{(-2)}}S_{\omega\ell m}^{(-2)}&=0.
\end{align}
\end{subequations}

By direct computation, one can show that there exists a certain fourth-order differential operator whose application to any solution $S_{\omega\ell m}^{(\pm2)}$ of the angular ODE \eqref{eq:AngularODE} with $s=\pm2$ yields another solution $S_{\omega\ell m}^{(\mp2)}$ of the same ODE but with opposite spin $s=\mp2$:
\begin{subequations}
\begin{align}
    \msc{L}_{-1}\msc{L}_0\msc{L}_1\msc{L}_2 S_{\omega\ell m}^{(+2)}&\propto S_{\omega\ell m}^{(-2)},\\
    \msc{L}_{-1}^\dag\msc{L}_0^\dag\msc{L}_1^\dag\msc{L}_2^\dag S_{\omega\ell m}^{(-2)}&\propto S_{\omega\ell m}^{(+2)}.
\end{align}
\end{subequations}
There is also a fourth-order differential operator whose action on any solution $R_{\omega\ell m}^{(\pm2)}$ of the radial ODE \eqref{eq:RadialODE} with $s=\pm2$ yields a solution $R_{\omega\ell m}^{(\mp2)}$ of the ODE with opposite spin $s=\mp2$:
\begin{subequations}
\begin{align}
    \msc{D}_0^4R_{\omega\ell m}^{(-2)}&\propto R_{\omega\ell m}^{(+2)},\\
    \Delta^2\pa{\msc{D}_0^\dag}^4\Delta^2R_{\omega\ell m}^{(+2)}&\propto R_{\omega\ell m}^{(-2)}.
\end{align}
\end{subequations}
These properties follow from the assumption that $R_{\omega\ell m}^{(\pm2)}$ and $S_{\omega\ell m}^{(\pm2)}$ obey the radial and angular ODEs \eqref{eq:RadialODE} and \eqref{eq:AngularODE} and hold regardless of the linear combination of modes being considered or their normalization.
However, if one considers specific modes with a particular normalization, then the proportionality factors become fixed.
For the angular modes $\hat{S}_{\omega\ell m}^{(\pm2)}$ defined in Eq.~\eqref{eq:AngularModes},%
\begin{subequations}
\label{eq:ATSI1}
\begin{align}
    \label{eq:ATSI1a}
    \msc{L}_{-1}\msc{L}_0\msc{L}_1\msc{L}_2\hat{S}_{\omega\ell m}^{(+2)}&=\hat{D}_{\omega\ell m}\hat{S}_{\omega\ell m}^{(-2)},\\
    \label{eq:ATSI1b}
    \msc{L}_{-1}^\dag\msc{L}_0^\dag\msc{L}_1^\dag\msc{L}_2^\dag\hat{S}_{\omega\ell m}^{(-2)}&=\hat{D}_{\omega\ell m}'\hat{S}_{\omega\ell m}^{(+2)},
\end{align}
\end{subequations}
where the proportionality constants $\hat{D}_{\omega\ell m}$ and $\hat{D}'_{\omega\ell m}$ are given in Eq.~\eqref{eq:ConstantsD}.
As for the radial modes $\hat{R}_{\omega\ell m}^{(\pm2)}$ defined in Eq.~\eqref{eq:RadialModes}, the ``in'' and ``out'' modes do not mix:
\begin{subequations}
\label{eq:RTSI1}
\begin{align}
    \label{eq:RTSI1a}
    \msc{D}_0^4\hat{R}_{\omega\ell m}^{(-2)\,{\rm in/out}}&=\hat{\msc{C}}_{\omega\ell m}^{\rm in/out}\hat{R}_{\omega\ell m}^{(+2)\,{\rm in/out}},\\
    \label{eq:RTSI1b}
    \Delta^2\pa{\msc{D}_0^\dag}^4\Delta^2\hat{R}_{\omega\ell m}^{(+2)\,{\rm in/out}}&=\hat{\msc{C}}_{\omega\ell m}^{{\rm in/out}\,\prime}
    \hat{R}_{\omega\ell m}^{(-2)\,{\rm in/out}},
\end{align}
\end{subequations}
where the proportionality constants $\hat{\msc{C}}_{\omega\ell m}$ and $\hat{\msc{C}}_{\omega\ell m}'$ are given in Eq.~\eqref{eq:ConstantsC}.
We will refer to Eqs.~\eqref{eq:ATSI1} and \eqref{eq:RTSI1} as the angular and radial Teukolsky--Starobinsky identities in first form.
We review their relation to the Teukolsky--Starobinsky identities in first form \eqref{eq:TeukolskyStarobinskyIdentities} in Sec.~\ref{subsec:ConsistencyChecks}, and give a slick derivation of the Teukolsky--Starobinsky constants in Eq.~\eqref{eq:RTSI1} in App.~\ref{app:RadialHeun}.

Plugging Eqs.~\eqref{eq:ATSI1a} and \eqref{eq:ATSI1b} into each other yields an eighth-order differential relation for each mode $\hat{S}_{\omega\ell m}^{(\pm2)}$.
These are the angular Teukolsky--Starobinsky identities in second form:
\begin{subequations}
\label{eq:ATSI2}
\begin{align}
    \msc{L}_{-1}^\dag\msc{L}_0^\dag\msc{L}_1^\dag\msc{L}_2^\dag\msc{L}_{-1}\msc{L}_0\msc{L}_1\msc{L}_2\hat{S}_{\omega\ell m}^{(+2)}&=\mc{D}_{\omega\ell m}\hat{S}_{\omega\ell m}^{(+2)},\\
    \msc{L}_{-1}\msc{L}_0\msc{L}_1\msc{L}_2\msc{L}_{-1}^\dag\msc{L}_0^\dag\msc{L}_1^\dag\msc{L}_2^\dag\hat{S}_{\omega\ell m}^{(-2)}&=\mc{D}_{\omega\ell m}\hat{S}_{\omega\ell m}^{(-2)}.
\end{align}
\end{subequations}
Likewise, plugging Eqs.~\eqref{eq:RTSI1a} and \eqref{eq:RTSI1b} into each other results in an eighth-order differential relation for each $\hat{R}_{\omega\ell m}^{(\pm2)}$.
These are the radial Teukolsky--Starobinsky identities in second form:
\begin{subequations}
\label{eq:RTSI2}
\begin{align}
    \msc{D}_0^4\Delta^2\pa{\msc{D}_0^\dag}^4\Delta^2 \hat{R}_{\omega\ell m}^{(+2)\,{\rm in/out}}
    &=\mc{C}_{\omega\ell m}\hat{R}^{(+2)\,{\rm in/out}}_{\omega\ell m},\\
    \Delta^2\pa{\msc{D}_0^\dag}^4\Delta^2\msc{D}_0^4 \hat{R}_{\omega\ell m}^{(-2)\,{\rm in/out}} 
    &=\mc{C}_{\omega\ell m}\hat{R}_{\omega\ell m}^{(-2)\,{\rm in/out}}.
\end{align}
\end{subequations}
The entire discussion at the end of Sec.~2.5 in Paper I applies here.
In short, the first-form identities \eqref{eq:ATSI1} and \eqref{eq:RTSI1} hold for our particular choice of angular and radial modes $\hat{S}_{\omega\ell m}^{(\pm2)}$ and $\hat{R}_{\omega\ell m}^{(\pm2)}$, while their second-form analogues \eqref{eq:ATSI2} and \eqref{eq:RTSI2} apply to any solution of their respective ODEs.
Thus, the hatted constants $\hat{D}_{\omega\ell m}$, $\hat{D}_{\omega\ell m}'$, $\hat{\msc{C}}_{\omega\ell m}$, and $\hat{\msc{C}}_{\omega\ell m}'$ depend on our choice of hatted modes, while their products $\mc{D}_{\omega\ell m}=\hat{D}_{\omega\ell m}\hat{D}_{\omega\ell m}'$ and $\mc{C}_{\omega\ell m}=\hat{\msc{C}}_{\omega\ell m}\hat{\msc{C}}_{\omega\ell m}'$ are independent of this choice of modes (which is why they are unhatted). Furthermore, as in Paper I, the first forms of the identities imply the second forms, whereas the converse is not true.

\subsection{Relation between modes of the Weyl scalars \texorpdfstring{$\psi_0$}{ψ0} and \texorpdfstring{$\psi_4$}{ψ4}}
\label{subsec:WeylRelations}

After mode inversion and elimination of the Hertz potentials in Sec.~\ref{subsec:ModeInversion} below, we find the (two modes of) $\psi_4$ associated with a given (single mode of) $\psi_0$, and vice versa.

Given a single mode of $\psi_0$ of the form
\begin{align}
    \label{eq:Mode0}
    \psi_0=e^{-i\omega t+im\phi}\hat{R}_{\omega\ell m}^{(+2)}\hat{S}_{\omega\ell m}^{(+2)},
\end{align}
the corresponding $\psi_4$ is given by
\begin{align}
    \label{eq:CorrespondingMode4}
    \zeta^4\psi_4=\frac{\hat{D}_{\omega\ell m}}{4\hat{\msc{C}}_{\omega\ell m}}e^{-i\omega t+im\phi}\hat{R}_{\omega\ell m}^{(-2)}\hat{S}_{\omega\ell m}^{(-2)}-\frac{3i\ol{\omega}M\Xi}{\ol{\hat{\msc{C}}}_{\omega\ell m}}e^{i\ol{\omega}t-im\phi}\hat{R}_{-\ol{\omega},\ell,-m}^{(-2)}\hat{S}_{-\ol{\omega},\ell,-m}^{(-2)}.
\end{align}
Conversely, given a single mode of $\zeta^4\psi_4$ of the form
\begin{align}
    \label{eq:Mode4}
    \zeta^4\psi_4=e^{-i\omega t+im\phi}\hat{R}_{\omega\ell m}^{(-2)}\hat{S}_{\omega\ell m}^{(-2)},
\end{align}
the corresponding $\psi_0$ is
\begin{align}
    \label{eq:CorrespondingMode0}
    \psi_0=\frac{4\hat{D}_{\omega\ell m}'}{\hat{\msc{C}}_{\omega\ell m}'}e^{-i\omega t+im\phi}\hat{R}_{\omega\ell m}^{(+2)}\hat{S}_{\omega\ell m}^{(+2)}+\frac{48i\ol{\omega}M\Xi}{\ol{\hat{\msc{C}}_{}'}_{\!\!\!\omega\ell m}}e^{i\ol{\omega}t-im\phi}\hat{R}_{-\ol{\omega},\ell,-m}^{(+2)}\hat{S}_{-\ol{\omega},\ell,-m}^{(+2)}.
\end{align}
Consistency requires that each of these pairs of $(\psi_0,\psi_4)$---namely, either the pair \eqref{eq:Mode0}--\eqref{eq:CorrespondingMode4}, or the pair \eqref{eq:Mode4}--\eqref{eq:CorrespondingMode0}---satisfies the Teukolsky--Starobinsky identities \eqref{eq:TeukolskyStarobinskyIdentities}.
This consistency check can be explicitly carried out using the symmetry properties \eqref{eq:AngularSymmetries} and \eqref{eq:RadialSymmetries}, together with the radial and angular Teukolsky--Starobinsky identities \eqref{eq:ATSI1} and \eqref{eq:RTSI1}.

\subsection{Metric reconstruction in ingoing radiation gauge}
\label{subsec:MetricReconstructionIRG}

Here, we explicitly reconstruct the metric perturbation $h_{\mu\nu}$ in ingoing radiation gauge \eqref{eq:IRG} that is associated with a single mode \eqref{eq:Mode0} of $\psi_0$ or with a single mode \eqref{eq:Mode4} of $\zeta^4\psi_4$.

To treat both cases simultaneously, we express the components of $h_{\mu\nu}^{\rm IRG}$ in terms of constants $A^{\rm IRG}$ and $B^{\rm IRG}$.
These must take different values according to whether one reconstructs the metric perturbation from a single mode \eqref{eq:Mode0} of $\psi_0$, in which case one must set
\begin{align}
\label{eq:IRGMode0Solution}
    A^{\rm IRG}=0,\qquad
    B^{\rm IRG}=\frac{4}{\ol{\hat{\msc{C}}}_{\omega\ell m}},
\end{align}
or from a single mode \eqref{eq:Mode4} of $\zeta^4\psi_4$, in which case one must set
\begin{align}
\label{eq:IRGMode4Solution}
    A^{\rm IRG}=-\frac{192i\omega M\Xi}{\mc{C}_{\omega\ell m}},\qquad
    B^{\rm IRG}=\frac{16\ol{\hat{D}_{}'}_{\!\!\omega\ell m}}{\ol{\mc{C}}_{\omega\ell m}}.
\end{align}
The Boyer-Lindquist components of the real metric perturbation $h_{\mu\nu}^{\rm IRG}$ are then given by\footnote{\label{fn:Trivial}These expressions are in some sense trivial, as they are directly obtained by inverting Eq.~\eqref{eq:TetradProjection}.
The nontrivial part of metric reconstruction comes in Eqs.~\eqref{eq:HIRG} and \eqref{eq:HORG}, whose derivation is sketched in Sec.~3.4 of Paper I.}
\begin{subequations}
\label{eq:MetricComponentsIRG}
\begin{align}
    h_{tt}^{\rm IRG}&=\frac{1}{\Xi^2}\br{-a^2\Upsilon\sin^2{\theta}\,\mc{M}_+-2a\sqrt{\Upsilon}\sin{\theta}\,\mc{N}_-+h_{nn}},\\
    h_{rr}^{\rm IRG}&=\frac{\Sigma^2}{\Delta^2}h_{nn},\\
    h_{\theta\theta}^{\rm IRG}&=\frac{\Sigma^2}{\Upsilon}\mc{M}_+,\\
    h_{\phi\phi}^{\rm IRG}&=-\frac{\sin^2{\theta}}{\Xi^2}\br{\pa{r^2+a^2}^2\Upsilon\mc{M}_++2a\pa{r^2+a^2}\sqrt{\Upsilon}\sin{\theta}\,\mc{N}_--a^2\sin^2{\theta}\,h_{nn}},\\
    h_{tr}^{\rm IRG}&=-\frac{\Sigma}{\Xi\Delta}\br{h_{nn}-a\sqrt{\Upsilon}\sin{\theta}\,\mc{N}_-},\\
    h_{t\theta}^{\rm IRG}&=\frac{\Sigma}{\Xi}\br{\frac{\mc{N}_+}{\sqrt{\Upsilon}}-a\sin{\theta}\,\mc{M}_-},\\
    h_{t\phi}^{\rm IRG}&=\frac{a\sin^2{\theta}}{\Xi^2}\br{\pa{r^2+a^2}\Upsilon\mathcal{M}_++\pa{\frac{\Sigma}{a\sin{\theta}}+2a\sin{\theta}}\sqrt{\Upsilon}\mathcal{N}_--h_{nn}},\\
    h_{r\theta}^{\rm IRG}&=-\frac{\Sigma^2}{\Delta\sqrt{\Upsilon}}\mathcal{N}_+,\\
    h_{r\phi}^{\rm IRG}&=-\frac{\Sigma\sin{\theta}}{\Xi\Delta}\br{\pa{r^2+a^2}\sqrt{\Upsilon}\mathcal{N}_--a\sin{\theta}\,h_{nn}},\\
    h_{\theta\phi}^{\rm IRG}&=\frac{\Sigma\sin{\theta}}{\Xi}\br{\pa{r^2+a^2}\mathcal{M}_--\frac{a\sin{\theta}}{\sqrt{\Upsilon}} \mathcal{N}_+},
\end{align}
\end{subequations}
where we introduced the manifestly real projections $\mc{M}_+\equiv\re\pa{\zeta^{-2}h_{mm}}$, $\mc{M}_-\equiv\im\pa{\zeta^{-2}h_{mm}}$, $\mc{N}_+\equiv\sqrt{2}\re\pa{\zeta^{-1}h_{nm}}$, and $\mc{N}_-\equiv\sqrt{2}\im\pa{\zeta^{-1}h_{nm}}$: more explicitly, recalling Eq.~\eqref{eq:TetradProjection},
\begin{subequations}
\label{eq:RealProjections}
\begin{gather}
    \mc{M}_+=\frac{1}{2}\pa{\frac{h_{mm}}{\zeta^2}+\frac{h_{\ol{m}\ol{m}}}{\ol{\zeta}^2}},\qquad
    \mc{M}_-=\frac{1}{2i}\pa{\frac{h_{mm}}{\zeta^2}-\frac{h_{\ol{m}\ol{m}}}{\ol{\zeta}^2}},\\
    \mc{N}_+=\frac{1}{\sqrt{2}}\pa{\frac{h_{nm}}{\zeta}+\frac{h_{n\ol{m}}}{\ol{\zeta}}},\qquad
    \mc{N}_-=\frac{1}{\sqrt{2}i}\pa{\frac{h_{nm}}{\zeta}-\frac{h_{n\ol{m}}}{\ol{\zeta}}}.
\end{gather}
\end{subequations}
The projections $h_{nn}$, $h_{nm}$, $h_{n\ol{m}}$, $h_{mm}$, and $h_{\ol{m}\ol{m}}$ (the only ones needed in IRG) decompose as
\begin{align}
    \label{eq:TetradComponentsIRG}
    h_{ab}=e^{-i\omega t+im\phi}h_{ab}^{(+)}(r,\theta)+e^{i\ol{\omega}t-im\phi}h_{ab}^{(-)}(r,\theta),
\end{align}
where each $h_{ab}^{(\pm)}(r,\theta)$ is expressible in terms of a single function $H_{\omega\ell m}^{ab}$.

In fact, only three such functions are needed, since
\begin{subequations}
\label{eq:TetradProjectionsIRG}
\begin{align}
    h_{nn}^{(+)}&=A^{\rm IRG}H_{\omega\ell m}^{nn}+\ol{B}^{\rm IRG}\ol{H}_{-\ol{\omega},\ell,-m}^{nn},\\
    h_{nn}^{(-)}&=\ol{h}_{nn}^{(+)},\\
    h_{nm}^{(+)}&=\ol{B}^{\rm IRG}\ol{H}_{-\ol{\omega},\ell,-m}^{nm},\\
    h_{nm}^{(-)}&=\ol{A}^{\rm IRG}\ol{H}_{\omega\ell m}^{nm},\\
    h_{n\ol{m}}^{(\pm)}&=\ol{h}_{nm}^{(\mp)},\\
    h_{mm}^{(+)}&=\ol{B}^{\rm IRG}\ol{H}_{-\ol{\omega},\ell,-m}^{mm},\\
    h_{mm}^{(-)}&=\ol{A}^{\rm IRG}\ol{H}_{\omega\ell m}^{mm},\\
    h_{\ol{m}\ol{m}}^{(\pm)}&=\ol{h}_{mm}^{(\mp)}.
\end{align}
\end{subequations}
Specifying the three functions $H_{\omega\ell m}^{nn}$, $H_{\omega\ell m}^{nm}$, and $H_{\omega\ell m}^{mm}$ finally determines the metric components:%
\begin{subequations}
\label{eq:HIRG}
\begin{align}
    H_{\omega\ell m}^{nn}&\equiv-\frac{\epsilon_g}{4\ol{\zeta}^2}\pa{\msc{L}_1^\dag-\frac{2ia\sqrt{\Upsilon}\sin{\theta}}{\zeta}}\msc{L}_2^\dag\hat{R}_{\omega\ell m}^{(-2)}\hat{S}_{\omega\ell m}^{(-2)},\\
    H_{\omega\ell m}^{nm}&\equiv-\frac{\epsilon_g}{2\sqrt{2}\ol{\zeta}}\pa{\msc{D}_0\msc{L}^\dag_2+\frac{a^2\sin{2\theta}}{\Sigma}\msc{D}_0-\frac{2r}{\Sigma}\msc{L}^\dag_2}\hat{R}_{\omega\ell m}^{(-2)}\hat{S}_{\omega\ell m}^{(-2)},\\
    H_{\omega\ell m}^{mm}&\equiv-\frac{\epsilon_g}{2}\pa{\msc{D}_0-\frac{2}{\zeta}}\msc{D}_0\hat{R}_{\omega\ell m}^{(-2)}\hat{S}_{\omega\ell m}^{(-2)}.
\end{align}
\end{subequations}

\subsection{Metric reconstruction in outgoing radiation gauge}
\label{subsec:MetricReconstructionORG}

Here, we explicitly reconstruct the metric perturbation $h_{\mu\nu}$ in outgoing radiation gauge \eqref{eq:ORG} that is associated with a single mode \eqref{eq:Mode0} of $\psi_0$ or with a single mode \eqref{eq:Mode4} of $\zeta^4\psi_4$.

To handle both cases simultaneously, we write the components of $h_{\mu\nu}^{\rm ORG}$ in terms of constants $A^{\rm ORG}$ and $B^{\rm ORG}$.
These must take different values according to whether one reconstructs the metric perturbation from a single mode \eqref{eq:Mode0} of $\psi_0$, in which case one must set
\begin{align}
\label{eq:ORGMode0Solution}
    A^{\rm ORG}=\frac{192i\omega M\Xi}{\mc{C}_{\omega\ell m}},\qquad
    B^{\rm ORG}=\frac{16\ol{\hat{D}}_{\omega\ell m}}{\ol{\mc{C}}_{\omega\ell m}},
\end{align}
or from a single mode \eqref{eq:Mode4} of $\zeta^4\psi_4$, in which case one must set
\begin{align}
\label{eq:ORGMode4Solution}
    A^{\rm ORG}=0,\qquad
    B^{\rm ORG}=\frac{64}{\ol{\hat{\msc{C}}_{}'}_{\!\!\!\omega\ell m}}.
\end{align}
The Boyer-Lindquist components of the real metric perturbation $h_{\mu\nu}^{\rm ORG}$ are then given by\footref{fn:Trivial}
\begin{subequations}
\label{eq:MetricComponentsORG}
\begin{align}
    h_{tt}^{\rm ORG}&=\frac{1}{\Xi^2}\br{-a^2\Upsilon\sin^2{\theta}\,\mathcal{M}_+-\frac{a\Delta\sqrt{\Upsilon}\sin{\theta}}{\Sigma}\mathcal{L}_-+\frac{\Delta^2}{4\Sigma^2}h_{ll}},\\
    h_{rr}^{\rm ORG}&=\frac{h_{ll}}{4},\\
    h_{\theta\theta}^{\rm ORG}&=\frac{\Sigma^2}{\Upsilon}\mc{M}_+,\\
    h_{\phi\phi}^{\rm ORG}&=-\frac{\sin^2{\theta}}{\Xi^2}\br{\pa{r^2+a^2}^2\Upsilon\mc{M}_++\frac{a\Delta\pa{r^2+a^2}\sqrt{\Upsilon}\sin{\theta}}{\Sigma}\mc{L}_--\frac{a^2\Delta^2\sin^2{\theta}}{4\Sigma^2}h_{ll}},\\
    h_{tr}^{\rm ORG}&=\frac{1}{2\Xi}\br{\frac{\Delta}{2\Sigma}h_{ll}-a\sqrt{\Upsilon}\sin{\theta}\,\mc{L}_-},\\
    h_{t\theta}^{\rm ORG}&=\frac{1}{\Xi}\br{\frac{\Delta}{2\sqrt{\Upsilon}}\mc{L}_+-a\Sigma\sin{\theta}\,\mc{M}_-},\\
    h_{t\phi}^{\rm ORG}&=\frac{a\sin^2{\theta}}{\Xi^2}\br{\pa{r^2+a^2}\Upsilon\mc{M}_++\frac{\Delta\sqrt{\Upsilon}}{2\Sigma}\pa{\frac{\Sigma}{a\sin{\theta}}+2a\sin{\theta}}\mc{L}_--\frac{\Delta^2}{4\Sigma^2}h_{ll}},\\
    h_{r\theta}^{\rm ORG}&=\frac{\Sigma}{2\sqrt{\Upsilon}}\mc{L}_+,\\
    h_{r\phi}^{\rm ORG}&=\frac{\sin{\theta}}{2\Xi}\br{\pa{r^2+a^2}\sqrt{\Upsilon}\mc{L}_--\frac{a\Delta\sin{\theta}}{2\Sigma}h_{ll}},\\
    h_{\theta\phi}^{\rm ORG}&=\frac{\Sigma\sin{\theta}}{\Xi}\br{\pa{r^2+a^2}\mc{M}_--\frac{a\Delta\sin{\theta}}{2\Sigma\sqrt{\Upsilon}}\mc{L}_+}.
\end{align}
\end{subequations}
Here, the manifestly real projections $\mc{M}_\pm$ were defined in Eq.~\eqref{eq:RealProjections}, and we also introduced
\begin{gather}
    \label{eq:RealProjectionsBis}
    \mc{L}_+\equiv\sqrt{2}\re\frac{h_{lm}}{\zeta}
    =\frac{1}{\sqrt{2}}\pa{\frac{h_{lm}}{\zeta}+\frac{h_{l\ol{m}}}{\ol{\zeta}}},\qquad
    \mc{L}_-\equiv\sqrt{2}\im\frac{h_{lm}}{\zeta}
    =\frac{1}{\sqrt{2}i}\pa{\frac{h_{lm}}{\zeta}-\frac{h_{l\ol{m}}}{\ol{\zeta}}}.
\end{gather}
The projections $h_{ll}$, $h_{lm}$, $h_{l\ol{m}}$, $h_{mm}$, and $h_{\ol{m}\ol{m}}$ (the only ones needed in ORG) decompose as
\begin{align}
    \label{eq:TetradComponentsORG}
    h_{ab}=e^{-i\omega t+im\phi}h_{ab}^{(+)}(r,\theta)+e^{i\ol{\omega}t-im\phi}h_{ab}^{(-)}(r,\theta),
\end{align}
where once again each $h_{ab}^{(\pm)}(r,\theta)$ is expressible in terms of a single function $H_{\omega\ell m}^{ab}$:
\begin{subequations}
\label{eq:TetradProjectionsORG}
\begin{align}
    h_{ll}^{(+)}&=A^{\rm ORG}H_{\omega\ell m}^{ll}+\ol{B}^{\rm ORG}\ol{H}_{-\ol{\omega},\ell,-m}^{ll},\\
    h_{ll}^{(-)}&=\ol{h}_{ll}^{(+)},\\
    h_{lm}^{(+)}&=A^{\rm ORG}H_{\omega\ell m}^{lm},\\
    h_{lm}^{(-)}&=B^{\rm ORG}H_{-\ol{\omega},\ell,-m}^{lm},\\
    h_{l\ol{m}}^{(\pm)}&=\ol{h}_{lm}^{(\mp)},\\
    h_{mm}^{(+)}&=A^{\rm ORG}H_{\omega\ell m}^{mm},\\
    h_{mm}^{(-)}&=B^{\rm ORG}H_{-\ol{\omega},\ell,-m}^{mm},\\
    h_{\ol{m}\ol{m}}^{(\pm)}&=\ol{h}_{mm}^{(\mp)}.
\end{align}
\end{subequations}
Specifying the three functions $H_{\omega\ell m}^{nn}$, $H_{\omega\ell m}^{nm}$, and $H_{\omega\ell m}^{mm}$ finally determines the metric components:%
\begin{subequations}
\label{eq:HORG}
\begin{align}
    H_{\omega\ell m}^{ll}&\equiv-\frac{\epsilon_g}{4}\zeta^2\pa{\msc{L}_1-\frac{2ia\sqrt{\Upsilon}\sin{\theta}}{\zeta}}\msc{L}_2 \hat{R}_{\omega\ell m}^{(+2)}\hat{S}_{\omega\ell m}^{(+2)},\\
    H_{\omega\ell m}^{lm}&\equiv\frac{\epsilon_g}{4\sqrt{2}}\frac{\zeta^2}{\ol{\zeta}\Delta}\pa{\msc{D}_0^\dag\msc{L}_2+\frac{a^2\sqrt{\Upsilon}\sin{2\theta}}{\Sigma}\msc{D}_0^\dag-\frac{2r}{\Sigma}\msc{L}_2}\Delta^2\hat{R}_{\omega\ell m}^{(+2)}\hat{S}_{\omega\ell m}^{(+2)},\\
    H_{\omega\ell m}^{mm}&\equiv-\frac{\epsilon_g}{8}\frac{\zeta^2}{\ol{\zeta}^2}\pa{\msc{D}_0^\dag-\frac{2}{\zeta}}\msc{D}_0^\dag\Delta^2\hat{R}_{\omega\ell m}^{(+2)}\hat{S}_{\omega\ell m}^{(+2)}.
\end{align}
\end{subequations}

\subsection{Consistency checks}
\label{subsec:ConsistencyChecks}

As we did for Kerr in Paper I, we have reconstructed the metric perturbation $h_{\mu\nu}$ corresponding to the modes of a given Weyl scalar $\psi_0$ or $\psi_4$, in either ingoing radiation gauge (in Sec.~\ref{subsec:MetricReconstructionIRG}) or outgoing radiation gauge (in Sec.~\ref{subsec:MetricReconstructionORG}).
The derivation of the metric components relies on the use of the Newman--Penrose and Geroch--Held--Penrose formalisms, which we apply to the Kerr--(anti-)de Sitter metric \eqref{eq:KerrdS} in Sec.~\ref{subsec:KerrdS}.
However, the results of this section can be directly checked without the use of these formalisms.

First, one can plug either of the metric perturbations $h_{\mu\nu}^{\rm IRG}$ or $h_{\mu\nu}^{\rm ORG}$ given in Eqs.~\eqref{eq:MetricComponentsIRG} and \eqref{eq:MetricComponentsORG} into Eq.~\eqref{eq:LinearizedEE}---or the simpler version in Eq.~\eqref{eq:TracelessEE}---and verify that they satisfy the vacuum linearized Einstein equations.
One can also plug $h_{\mu\nu}^{\rm IRG/ORG}$ into the equations for the Weyl scalars in Eq.~\eqref{eq:WeylScalars}, using Eq.~\eqref{eq:LinearizedWeyl} to express the linearized Weyl tensor $C_{\mu\nu\rho\sigma}^{(1)}$ in terms of the metric perturbation, and then verify that this recovers the expected modes of $\psi_0$ and $\psi_4$, thereby ``closing the loop''.
Finally, one can also check that the resulting Weyl scalars are related by the equations of Sec.~\ref{subsec:WeylRelations}.

\subsection{Different forms of the Teukolsky--Starobinsky identities}
\label{subsec:TeukolskyStarobinskyForms}

Finally, as in Paper I, we present other forms of the Teukolsky--Starobinsky identities \eqref{eq:TeukolskyStarobinskyIdentities}. 

For the sake of brevity, we adopt our notation from Paper I.
For instance, we will refer to the first form \eqref{eq:TeukolskyStarobinskyIdentities} of the Teukolsky--Starobinsky identities as the TSI$_1$.
These can be derived from Eqs.~\eqref{eq:IngoingPotential} and \eqref{eq:OutgoingPotential} by eliminating the Hertz potentials $\Psi_{\rm H}$ and $\Psi_{\rm H}'$.
Eliminating the Weyl scalars $\psi_0$ and $\psi_4$ instead yields the TSI$_1$ for the Hertz potentials,
\begin{subequations}
\begin{align}
    l^4\Psi_{\rm H}&=\frac{1}{4}\mc{L}_{-1}\mc{L}_0\mc{L}_1\mc{L}_2\frac{\Psi_{\rm H}'}{\zeta^4}+3M\Xi\pd_t\frac{\ol{\Psi}_{\rm H}'}{\ol{\zeta}^4},\\
    \Delta^2\pa{\frac{\Sigma}{\Delta}n}^4\Delta^2\frac{\Psi_{\rm H}'}{\zeta^4}&=\frac{1}{4}\ol{\mc{L}}_{-1}\ol{\mc{L}}_0\ol{\mc{L}}_1\ol{\mc{L}}_2\Psi_{\rm H}-3M\Xi\pd_t\ol{\Psi}_{\rm H}.
\end{align}
\end{subequations}
One can also plug the TSI$_1$ into each other to obtain eighth-order differential relations involving only one Weyl scalar.
These are the Teukolsky--Starobinsky identities in second form: the TSI$_2$,%
\begin{subequations}
\label{eq:TSI2}
\begin{align}
    \Delta^2\pa{\frac{\Sigma}{\Delta}n}^4\Delta^2l^4\zeta^4\psi_4&=\frac{1}{16}\mc{L}_{-1}\mc{L}_0\mc{L}_1\mc{L}_2\ol{\mc{L}}_{-1}\ol{\mc{L}}_0\ol{\mc{L}}_1\ol{\mc{L}}_2\zeta^4\psi_4-9M^2\Xi^2\pd_t^2\zeta^4\psi_4,\\
    l^4\Delta^2\pa{\frac{\Sigma}{\Delta}n}^4\Delta^2\psi_0&=\frac{1}{16}\ol{\mc{L}}_{-1}\ol{\mc{L}}_0\ol{\mc{L}}_1\ol{\mc{L}}_2\mc{L}_{-1}\mc{L}_0\mc{L}_1\mc{L}_2\psi_0-9M^2\Xi^2\pd_t^2\psi_0,
\end{align}
\end{subequations}
The analogous relations for the Hertz potentials are given by replacing $(\zeta^4\psi_0,\zeta^4\psi_4)\to(\Psi_{\rm H}',\Psi_{\rm H})$.

As in Kerr, the Teukolsky--Starobinsky identities come in multiple versions.
The TSI$_1$ are unseparated in $(r,\theta)$ but couple opposite spins, while the TSI$_2$ are also unseparated but decouple opposite spins.
These also have separated angular and radial counterparts from Sec.~\ref{subsec:TeukolskyStarobinsky}, namely: angular Teukolsky--Starobinsky identities in first (ATSI$_1$) form \eqref{eq:ATSI1} and second (ATSI$_2$) form \eqref{eq:ATSI2}, and radial Teukolsky--Starobinsky identities in first (RTSI$_1$) form \eqref{eq:RTSI1} and second (RTSI$_2$) form \eqref{eq:RTSI2}.
The logical interrelations between these identities are explained in Sec.~2.10 of Paper I.
The only change in that discussion is that Eqs.~(2.61a) and (2.61b) therein become
\begin{subequations}
\begin{align}
    \frac{\Delta^2\pa{\msc{D}_0^\dag}^4\Delta^2\msc{D}_0^4\Delta^2R_{\omega\ell m}^{(-2)}}{R_{\omega\ell m}^{(-2)}}&=\frac{\msc{L}_{-1}\msc{L}_0\msc{L}_1\msc{L}_2\msc{L}_{-1}^\dag\msc{L}_0^\dag\msc{L}_1^\dag\msc{L}_2^\dag S_{\omega\ell m}^{(-2)}}{S_{\omega\ell m}^{(-2)}}+(12\omega M\Xi)^2,\\
    \frac{\msc{D}_0^4\Delta^2\pa{\msc{D}_0^\dag}^4\Delta^2R_{\omega\ell m}^{(+2)}}{R_{\omega\ell m}^{(+2)}}&=\frac{\msc{L}_{-1}^\dag\msc{L}_0^\dag\msc{L}_1^\dag\msc{L}_2^\dag\msc{L}_{-1}\msc{L}_0\msc{L}_1\msc{L}_2S_{\omega\ell m}^{(+2)}}{S_{\omega\ell m}^{(+2)}}+(12\omega M\Xi)^2.
\end{align}
\end{subequations}

\section{Derivation of results}
\label{sec:Derivation}

This section derives the identities presented in Sec.~\ref{sec:Results}, following the structure of Paper I and using the formalisms developed by Newman and Penrose \cite{Newman1962}, and by Geroch, Held, and Penrose \cite{Geroch1973} (which are reviewed, for instance, by Whiting and Price \cite{Whiting2005} and in Price's thesis \cite{PriceThesis}).
In Sec.~3.1 of Paper I, we reviewed the metric reconstruction operators ${\mc{S}_0^\dag}_{\mu\nu}$ and ${\mc{S}_4^\dag}_{\mu\nu}$, together with the relations between the Weyl scalars $(\psi_0,\psi_4)$ and the Hertz potentials $(\Psi_{\rm H},\Psi_{\rm H}')$.
Surprisingly, the presence of a nonzero cosmological constant does not change any of these equations, even though of course the tetrad vectors, differential operators, and spin coefficients all implicitly depend on $\Lambda$.
In Sec.~\ref{subsec:KerrdS}, we apply these general formulas to the special case of a Kerr--(anti-)de Sitter black hole to obtain the formulas presented in Sec.~\ref{sec:Results}.
Then in Sec.~\ref{subsec:ModeInversion}, we derive the mode inversion formula that leads to the relations in Sec.~\ref{subsec:WeylRelations}, and briefly explain how to obtain the metric reconstruction formulas in Secs.~\ref{subsec:MetricReconstructionIRG} and \ref{subsec:MetricReconstructionORG}.

\subsection{Application to the Kerr--(anti-)de Sitter black hole}
\label{subsec:KerrdS}

The formulas Sec.~3.1 of Paper I are applicable to any spacetime of Petrov type D that is also non-accelerating (or equivalently, that admits a rank-2 Killing tensor).
Since this class includes both Kerr and Kerr--(anti-)de Sitter, the relevant equations are identical to those in Paper I, so we do not repeat them here.
Instead, we directly specialize them to the particular case of the Kerr--(anti-)de Sitter metric \eqref{eq:KerrdS}, where the spin coefficients are
\begin{subequations}
\begin{gather} 
    \rho=-\frac{1}{\zeta},\qquad
    \rho'=\frac{\Delta}{2\zeta\Sigma},\qquad
    \tau=-\frac{ia\sqrt{\Upsilon}\sin{\theta}}{\sqrt{2}\Sigma},\qquad
    \tau'=-\frac{ia\sqrt{\Upsilon}\sin{\theta}}{\sqrt{2}\zeta^2},\\
    \beta=\frac{\sqrt{\Upsilon}}{2\sqrt{2}\,\ol{\zeta}}\pa{\cot{\theta}+\frac{\Upsilon'}{2\Upsilon}},\qquad
    \beta'=\frac{\sqrt{\Upsilon}}{2\sqrt{2}\zeta} \pa{\cot{\theta}-\frac{2ia\sin{\theta}}{\zeta}+\frac{\Upsilon'}{2\Upsilon}},\\
    \epsilon=0,\qquad
    \epsilon'=\frac{1}{2\Sigma}\pa{\frac{\Delta}{\zeta}-\frac{\Delta'}{2}},
\end{gather}
\end{subequations}
with all others coefficients zero.
The single, nonzero Weyl scalar of this background is
\begin{align} 
    \Psi_2=-M\zeta^{-3}.
\end{align}

The operators ${\mc{S}_0^\dag}_{\mu\nu}$ and ${\mc{S}_4^\dag}_{\mu\nu}$ take the general GHP forms given in Eqs.~(3.4) of Paper I, which are well-defined only when acting on quantities with definite GHP weights.
If ${\mc{S}_0^\dag}_{\mu\nu}$ acts on the IRG Hertz potential $\Psi_{\rm H}$ (or any quantity of weight $\cu{-4,0}$), or if ${\mc{S}_4^\dag}_{\mu\nu}$ acts on the ORG Hertz potential $\Psi_{\rm H}'$ (or any quantity of weight $\cu{4,0}$), then in the Kerr--(anti-)de Sitter background, these operators take the explicit forms
\begin{subequations}
\label{eq:KerrAdjointOperators}
\begin{align}
    \label{eq:AdjointOperator0}
    {\mc{S}_0^\dag}_{\mu\nu}\Psi_{\rm H}&=\Bigg[-\frac{1}{4\ol{\zeta}^2}l_\mu l_\nu\pa{\ol{\mc{L}}_1-\frac{2ia\sqrt{\Upsilon}\sin{\theta}}{\zeta}}\ol{\mc{L}}_2-\frac{1}{2}m_\mu m_\nu\pa{l-\frac{2}{\zeta}}l\notag\\
    &\phantom{=}\quad\,
    +\frac{1}{\sqrt{2}\,\ol{\zeta}}l_{(\mu}m_{\nu)}\pa{l\ol{\mc{L}}_2+\frac{a^2\sqrt{\Upsilon}\sin{2\theta}}{\Sigma}l-\frac{2r}{\Sigma}\ol{\mc{L}}_2}\Bigg]\Psi_{\rm H},\\
    \label{eq:AdjointOperator4}
    {\mc{S}_4^\dag}_{\mu\nu}\Psi_{\rm H}'&=\zeta^2\Bigg[-\frac{1}{4}n_\mu n_\nu\pa{\mc{L}_1-\frac{2ia\sqrt{\Upsilon}\sin{\theta}}{\zeta}}\mc{L}_2-\frac{1}{2\ol{\zeta}^2}\ol{m}_\mu\ol{m}_\nu\pa{\frac{\Sigma}{\Delta}n+\frac{1}{\zeta}}\frac{\Sigma}{\Delta}n\Delta^2\notag\\
    &\phantom{=}\qquad\,+\frac{1}{\sqrt{2}\,\ol{\zeta}\Delta}n_{(\mu}\ol{m}_{\nu)}\pa{\frac{\Sigma}{\Delta}n\mc{L}_2+\frac{a^2\sqrt{\Upsilon}\sin{2\theta}}{\Delta}n+\frac{r}{\Sigma}\mc{L}_2}\Delta^2\bigg]\frac{\Psi_{\rm H}'}{\zeta^4}.
\end{align}
\end{subequations}
One can then check that Eqs.~(3.5), (3.6), and (3.7) of Paper I directly reduce to Eqs.~\eqref{eq:IngoingPotential}, \eqref{eq:OutgoingPotential}, and \eqref{eq:TeukolskyStarobinskyIdentities}, respectively.

\subsection{Mode inversion and metric reconstruction}
\label{subsec:ModeInversion}

In this section, we put everything together to derive the metric reconstruction formulas presented in Secs.~\ref{subsec:MetricReconstructionIRG} and \ref{subsec:MetricReconstructionORG}.
But first, we derive the formula given in Sec.~\ref{subsec:WeylRelations} that directly relates modes of the Weyl scalars $\psi_0$ and $\psi_4$ associated with the same metric perturbation $h_{\mu\nu}$.
To do so, we must perform the mode inversion procedure: that is, we must invert Eqs.~\eqref{eq:IngoingPotential} and \eqref{eq:OutgoingPotential} to obtain the Hertz potentials $\Psi_{\rm H}$ and $\Psi_{\rm H}'$ that correspond to these modes.

This requires us to consider a linear combination of two modes in the Hertz potentials:
\begin{subequations}
\label{eq:HertzModes}
\begin{align}
    \label{eq:HertzModeIRG}
    \Psi_{\rm H}&=A^{\rm IRG}e^{-i\omega t+im\phi}\hat{R}_{\omega\ell m}^{(-2)}(r)\hat{S}_{\omega\ell m}^{(-2)}(\theta)+B^{\rm IRG}e^{i\ol{\omega}t-im\phi}\hat{R}_{-\ol{\omega},\ell,-m}^{(-2)}(r)\hat{S}^{(-2)}_{-\ol{\omega},\ell,-m}(\theta),\\
    \label{eq:HertzModeORG}
    \zeta^{-4}\Psi_{\rm H}'&=A^{\rm ORG}e^{-i\omega t+im\phi}\hat{R}_{\omega\ell m}^{(+2)}(r)\hat{S}_{\omega\ell m}^{(+2)}(\theta)+B^{\rm ORG}e^{i\ol{\omega}t-im\phi}\hat{R}_{-\ol{\omega},\ell,-m}^{(+2)}(r)\hat{S}_{-\ol{\omega},\ell,-m}^{(+2)}(\theta).
\end{align}
\end{subequations}
Here, $A^{\rm IRG/ORG}$ and $B^{\rm IRG/ORG}$ are yet-to-be-determined constants.
Plugging these expressions into Eqs.~\eqref{eq:IngoingPotential0} and \eqref{eq:OutgoingPotential0} and then applying the angular and radial Teukolsky--Starobinsky identities \eqref{eq:ATSI1b} and \eqref{eq:RTSI1a} yields 
\begin{subequations}
\begin{align}
    \label{eq:Weyl0ModeIRG}
    \psi_0&=\frac{1}{4}\pa{\hat{\msc{C}}_{\omega\ell m}\ol{B}^{\rm IRG}e^{-i\omega t+im\phi}\hat{R}_{\omega\ell m}^{(+2)}\hat{S}_{\omega\ell m}^{(+2)}+\ol{\hat{\msc{C}}}_{\omega\ell m}\ol{A}^{\rm IRG}e^{i\ol{\omega}t-im\phi}\hat{R}_{-\ol{\omega},\ell,-m}^{(+2)}\hat{S}_{-\ol{\omega},\ell,-m}^{(+2)}},\\
    \label{eq:Weyl0ModeORG}
    \psi_0&=\frac{1}{16}\Big[\pa{\ol{\hat{D}}_{\omega\ell m}\ol{A}^{\rm ORG}+12i\ol{\omega}M\Xi B^{\rm ORG}}e^{i\ol{\omega}t-im\phi}\hat{R}_{-\ol{\omega},\ell,-m}^{(+2)}\hat{S}_{-\ol{\omega},\ell,-m}^{(+2)}\notag\\
    &\phantom{=}\qquad+\pa{\hat{D}_{\omega\ell m}'\ol{B}^{\rm ORG}-12i\omega M\Xi A^{\rm ORG}}e^{-i\omega t+im\phi}\hat{R}_{\omega\ell m}^{(+2)}\hat{S}_{\omega\ell m}^{(+2)}\Big].
\end{align}
\end{subequations}
Setting these two expressions equal to the single mode \eqref{eq:Mode0} of $\psi_0$ determines the coefficients $A^{\rm IRG/ORG}$ and $B^{\rm IRG/ORG}$ to take the form given in Eqs.~\eqref{eq:IRGMode0Solution} and \eqref{eq:ORGMode0Solution}.
This completes the mode inversion procedure starting from a single mode \eqref{eq:Mode0} of $\psi_0$: to find its corresponding Weyl scalar $\psi_4$, one can either plug Eq.~\eqref{eq:HertzModeIRG} into Eq.~\eqref{eq:IngoingPotential4}, or else, one can equivalently plug Eq.~\eqref{eq:HertzModeORG} into Eq.~\eqref{eq:OutgoingPotential4}.
Either way, one recovers Eq.~\eqref{eq:CorrespondingMode4}.

Likewise, plugging the two Hertz potentials \eqref{eq:HertzModeIRG} and \eqref{eq:HertzModeORG} into Eqs.~\eqref{eq:IngoingPotential4} and \eqref{eq:OutgoingPotential4} and applying the angular and radial Teukolsky--Starobinsky identities \eqref{eq:ATSI1a} and \eqref{eq:RTSI1b} yields%
\begin{subequations}
\begin{align}
    \label{eq:Weyl4ModeIRG}
    \zeta^4\psi_4&=\frac{1}{16}\bigg[\pa{\ol{\hat{D}}_{\omega\ell m}'\ol{A}^{\rm IRG}-12i\ol{\omega}M\Xi B^{\rm IRG}}e^{i\ol{\omega}t+im\phi}\hat{R}_{-\ol{\omega},\ell,-m}^{(-2)}\hat{S}_{-\ol{\omega},\ell,-m}^{(-2)}\notag\\
    &\phantom{=}\qquad+\pa{\hat{D}_{\omega\ell m}\ol{B}^{\rm IRG}+12i\omega M\Xi A^{\rm IRG}}e^{-i\omega t+im\phi}\hat{R}_{\omega\ell m}^{(-2)}\hat{S}_{\omega\ell m}^{(-2)}\bigg],\\
    \label{eq:Weyl4ModeORG}
    \zeta^4\psi_4&=\frac{1}{64}\pa{\hat{\msc{C}}_{\omega\ell m}'\ol{B}^{\rm ORG}e^{-i\omega t+im\phi}\hat{R}_{\omega\ell m}^{(-2)}\hat{S}_{\omega\ell m}^{(-2)}+\ol{\hat{\msc{C}}'_{}}_{\!\!\!\omega\ell m}\ol{A}^{\rm ORG}e^{i\ol{\omega}t-im\phi}\hat{R}_{-\ol{\omega},\ell,-m}^{(-2)}\hat{S}_{-\ol{\omega},\ell,-m}^{(-2)}}.
\end{align}
\end{subequations}
Setting these two expressions equal to the single mode \eqref{eq:Mode4} of $\zeta^4\psi_4$ determines the coefficients $A^{\rm IRG/ORG}$ and $B^{\rm IRG/ORG}$ to take the form given in Eqs.~\eqref{eq:IRGMode4Solution} and \eqref{eq:ORGMode4Solution}.
This completes the mode inversion procedure starting from a single mode \eqref{eq:Mode4} of $\zeta^4\psi_4$: to find its corresponding Weyl scalar $\psi_0$, one can either plug Eq.~\eqref{eq:HertzModeIRG} into Eq.~\eqref{eq:IngoingPotential0}, or else, one can equivalently plug Eq.~\eqref{eq:HertzModeORG} into Eq.~\eqref{eq:OutgoingPotential0}.
Either way, one recovers Eq.~\eqref{eq:CorrespondingMode0}.

Just as in Paper I, the short length of this section reveals that the fundamental difficulty of mode inversion lies in obtaining the Teukolsky--Starobinsky identities in their various forms. Mode inversion follows easily after those are found.

Finally, we put everything together to derive the metric reconstruction formulas presented in Secs.~\ref{subsec:MetricReconstructionIRG} and \ref{subsec:MetricReconstructionORG}.
The metric perturbations $h_{\mu\nu}$ in IRG and ORG are given by Eqs.~\eqref{eq:MetricReconstructionIRG} and \eqref{eq:MetricReconstructionORG}, respectively, with the operators ${\mc{S}_0^\dag}_{\mu\nu}$ and ${\mc{S}_4^\dag}_{\mu\nu}$ taking the forms given in Eqs.~\eqref{eq:KerrAdjointOperators}.
We apply the operators in this form to the mode expansions \eqref{eq:HertzModes} of the IRG and ORG Hertz potentials.
Since each operator has three terms, we see that the metric components $h_{\mu\nu}$ can all be expressed in terms of three functions, which we called $H_{\omega\ell m}^{ab}(r,\theta)$ in Secs.~\ref{subsec:MetricReconstructionIRG} and \ref{subsec:MetricReconstructionORG}.
The computation of these functions is essentially the main nontrivial step in metric reconstruction.\footref{fn:Trivial}

As in Paper I, we omit these computations for the sake of brevity; for an explicit example, we direct the reader to Sec.~3.3 therein.

\section*{Acknowledgments}

This work was supported by NSF grant AST-2307888, the NSF CAREER award PHY-2340457, and the Simons Foundation award SFI-MPS-BH-00012593.

\clearpage
\appendix
\addtocontents{toc}{\protect\setcounter{tocdepth}{1}}

\section{Linearized gravity}
\label{app:LinearizedGravity}

This appendix collects known results in linearized gravity, repeating much of App.~A from Paper I but allowing the background spacetime to have a nonzero cosmological constant $\Lambda$. 
In App.~\ref{subsec:LinearizedEE}, we re-derive the linearized Einstein equations for metric perturbations $h_{\mu\nu}$ around an arbitrary background $g_{\mu\nu}$. 
We then specialize to a background which obeys the Einstein equations with a nonzero cosmological constant, recovering the form \eqref{eq:LinearizedEE} of the equations. 
Then, in App.~\ref{subsec:LinearizedWeyl}, we derive an explicit formula for the linearized Weyl tensor $C_{\mu\nu\rho\sigma}^{(1)}$ in terms of $h_{\mu\nu}$.
We use the following linearized identities, which are derived for any background in App.~A.3 of Paper I:
\begin{subequations}
\label{eq:Linearizations}
\begin{align}
    \label{eq:LinearizedRiemann}
    R_{\mu\nu\rho\sigma}^{(1)}&=\nabla_{\nu}\nabla_{[\rho}h_{\sigma]\mu}-\nabla_\mu\nabla_{[\rho} h_{\sigma]\nu}-{h^\tau}_{[\rho} R_{\sigma]\tau\mu\nu},\\
    \label{eq:LinearizedRicci}
    R_{\mu\nu}^{(1)}&=\nabla^\rho\nabla_{(\mu}h_{\nu)\rho}-\frac{1}{2}\nabla^2h_{\mu\nu}-\frac{1}{2}\nabla_\mu \nabla_\nu h,\\
    R^{(1)}&=\nabla^\mu\nabla^\nu h_{\mu\nu}-\nabla^2h-h^{\mu\nu}R_{\mu\nu}.
\end{align}
\end{subequations}

\subsection{Linearized Einstein equations and radiation gauge}
\label{subsec:LinearizedEE}

In the presence of a nonzero cosmological constant, the vacuum Einstein equations are
\begin{align}
    \label{eq:EE}
    G_{\mu\nu}+\epsilon_g\Lambda g_{\mu\nu}=0.
\end{align}
Taking $g_{\mu\nu}\to g_{\mu\nu}+h_{\mu\nu}$ and expanding to first order yields the linearized Einstein equations
\begin{align}
    \label{eq:LinearizedEEFullForm}
    G_{\mu\nu}^{(1)}+\epsilon_g\Lambda h_{\mu\nu}=0,
\end{align}
where $G_{\mu\nu}^{(1)}=R_{\mu\nu}^{(1)}-\frac{1}{2}g_{\mu\nu}R^{(1)}-\frac{1}{2}h_{\mu\nu}R$ is the linearized Einstein tensor.
Explicitly, we have
\begin{align}
    -\nabla^2h_{\mu\nu}+2\nabla^\rho\nabla_{(\mu}h_{\nu)\rho}-\nabla_\mu\nabla_\nu h-h_{\mu\nu}R+g_{\mu\nu}\pa{-\nabla^\rho\nabla^\sigma h_{\rho\sigma}+\nabla^2h+R^{\rho\sigma}h_{\rho\sigma}}+2\epsilon_g \Lambda h_{\mu\nu}=0,
\end{align}
where we used Eqs.~\eqref{eq:Linearizations} and $h=g^{\mu\nu}h_{\mu\nu}$ denotes the trace of the metric perturbation.
If the background metric obeys the Einstein equations \eqref{eq:EE}, as is the case with the Kerr--(anti-)de Siter spacetime \eqref{eq:KerrdS}, then $R_{\mu\nu}=\epsilon_g \Lambda g_{\mu\nu}$, and so this simplifies to
\begin{align}
    \label{eq:LinearizedEERepeat}
    -\nabla^2h_{\mu\nu} + 2\nabla^\rho\nabla_{(\mu}h_{\nu)\rho} - \nabla_\mu\nabla_\nu h + g_{\mu\nu}\pa{-\nabla^\rho\nabla^\sigma h_{\rho\sigma} + \nabla^2h} + \epsilon_g\Lambda \pa{g_{\mu\nu} h - 2h_{\mu\nu}} = 0,
\end{align}
which is precisely Eq.~\eqref{eq:LinearizedEE} in Sec.~\ref{sec:Introduction}. In this paper we work in either ingoing \eqref{eq:IRG} or outgoing \eqref{eq:ORG} radiation gauge. Both impose tracelessness ($h=0$) and simplify the equations to
\begin{align}
    \label{eq:TracelessEE}
    -\nabla^2h_{\mu\nu} + 2\nabla^\rho\nabla_{(\mu}h_{\nu)\rho} - g_{\mu\nu}\nabla^\rho\nabla^\sigma h_{\rho\sigma} - 2\epsilon_g\Lambda h_{\mu\nu} = 0.
\end{align}
The discussion about counting degrees of freedom from App.~A.1 in Paper I also applies here.

\subsection{Linearized Weyl tensor}
\label{subsec:LinearizedWeyl}

As reviewed in the Introduction, the linearized Weyl tensor $C_{\mu\nu\rho\sigma}^{(1)}$ contains two Weyl scalars \eqref{eq:WeylScalars} that encode the two degrees of freedom carried by a perturbation $h_{\mu\nu}$ of the background spacetime.
Here, we derive an expression for $C_{\mu\nu\rho\sigma}^{(1)}$ in terms of the metric perturbation $h_{\mu\nu}$, which is useful for performing the consistency checks in Sec.~\ref{subsec:ConsistencyChecks}.

The Weyl tensor is defined as the completely traceless part of the Riemann tensor,
\begin{align}
    C_{\mu\nu\rho\sigma}=R_{\mu\nu\rho\sigma}+\frac{1}{2}\pa{R_{\mu\sigma}g_{\nu\rho}-R_{\mu\rho}g_{\nu\sigma}+R_{\nu\rho}g_{\mu\sigma}-R_{\nu\sigma}g_{\mu\rho}}+\frac{1}{6}R\pa{g_{\mu\rho}g_{\nu\sigma}-g_{\mu\sigma}g_{\nu\rho}}.
\end{align}
Taking $g_{\mu\nu}\to g_{\mu\nu}+h_{\mu\nu}$ and extracting the pieces linear in $h_{\mu\nu}$ yields the linearized Weyl tensor
\begin{align}
    C_{\mu\nu\rho\sigma}^{(1)}&=R_{\mu\nu\rho\sigma}^{(1)}+\frac{1}{2}\pa{R_{\mu\sigma}^{(1)}g_{\nu\rho}-R_{\mu\rho}^{(1)}g_{\nu\sigma}+R_{\nu\rho}^{(1)}g_{\mu\sigma}-R_{\nu\sigma}^{(1)}g_{\mu\rho}}\notag\\
    &\phantom{=}+\frac{1}{2}\pa{R_{\mu\sigma}h_{\rho\nu}-R_{\mu\rho}h_{\nu\sigma}+R_{\nu\rho}h_{\mu\sigma}-R_{\nu\sigma}h_{\mu\rho}}+\frac{1}{6}R^{(1)}\pa{g_{\mu\rho}g_{\nu\sigma}-g_{\mu\sigma}g_{\nu\rho}}\notag\\
    &\phantom{=}+\frac{1}{6}R\pa{g_{\mu\rho}h_{\nu\sigma}-g_{\mu\sigma}h_{\nu\rho}+g_{\nu\sigma}h_{\mu\rho}-g_{\nu\rho}h_{\mu\sigma}}.
\end{align}
If the background $g_{\mu\nu}$ obeys the Einstein equations, so $R_{\mu\nu}=\epsilon_g\Lambda g_{\mu\nu}$, and the linearized metric perturbation $h_{\mu\nu}$ solves the vacuum equations \eqref{eq:LinearizedEERepeat}, then 
\begin{align}
    R_{\mu\nu}^{(1)}=\frac{1}{2}g_{\mu\nu}R^{(1)}+\epsilon_g\Lambda h_{\mu\nu},
\end{align}
and the above expression simplifies to
\begin{align}
    C_{\mu\nu\rho\sigma}^{(1)}&=R_{\mu\nu\rho\sigma}^{(1)}-\frac{1}{3}R^{(1)}\pa{g_{\mu\rho}g_{\nu\sigma}-g_{\mu\sigma}g_{\nu\rho}}-\frac{\epsilon_g\Lambda}{3}\pa{g_{\mu\rho}h_{\nu\sigma}-g_{\nu\rho}h_{\mu\sigma}+g_{\nu\sigma}h_{\mu\rho}-g_{\mu\sigma}h_{\nu\rho}}.
\end{align}
Using the identities \eqref{eq:Linearizations}, this can be rewritten explicitly in terms of $h_{\mu\nu}$ as
\begin{align}
    \label{eq:LinearizedWeyl}
    C_{\mu\nu\rho\sigma}^{(1)}&=\nabla_{\nu}\nabla_{[\rho}h_{\sigma]\mu}-\nabla_\mu\nabla_{[\rho} h_{\sigma]\nu}-{h^\tau}_{[\rho} R_{\sigma]\tau\mu\nu}-\frac{1}{3}\pa{\nabla^\mu\nabla^\nu h_{\mu\nu}-\nabla^2h-\epsilon_g\Lambda h}\pa{g_{\mu\rho}g_{\nu\sigma}-g_{\mu\sigma}g_{\nu\rho}}\notag\\
    &\phantom{=}-\frac{\epsilon_g\Lambda}{3}\pa{g_{\mu\rho}h_{\nu\sigma}-g_{\nu\rho}h_{\mu\sigma}+g_{\nu\sigma}h_{\mu\rho}-g_{\mu\sigma}h_{\nu\rho}}.
\end{align}

\section{Metric reconstruction in ingoing and outgoing coordinates}
\label{app:IngoingOutgoing}

In this appendix, we take the metrics reconstructed in ingoing and outgoing radiation gauges (as given in Secs.~\ref{subsec:MetricReconstructionIRG} and \ref{subsec:MetricReconstructionORG}, respectively) and transform their Boyer-Lindquist components $h_{\mu\nu}^{\rm IRG/ORG}$ to ingoing coordinates (in App.~\ref{subsec:IngoingMetricReconstruction}) and to outgoing coordinates (in App.~\ref{subsec:OutgoingMetricReconstruction}).
These coordinate systems remain regular near the event horizon of the black hole, so the metric components in these coordinates are particularly useful for analyzing the perturbations near the horizon, where the usual Boyer-Lindquist coordinates become singular.
In addition, many components of $h_{\mu\nu}^{\rm ORG}$ (or $h_{\mu\nu}^{\rm IRG}$) vanish in ingoing (or resp., outgoing) coordinates.

\subsection{Metric reconstruction in ingoing coordinates}
\label{subsec:IngoingMetricReconstruction}

Ingoing Kerr--(anti-)de Sitter coordinates $(v,r,\theta,\psi)$ are related to Boyer-Lindquist coordinates $(t,r,\theta,\phi)$ via
\begin{align}
    \label{eq:IngoingCoordinates}
    v=t+r_*,\qquad
    \psi=\phi+r_\sharp,
\end{align}
where the tortoise coordinate $r_*$ and the radius $r_\sharp$ are defined in Eqs.~\eqref{eq:TortoiseCoordinate} and \eqref{eq:SharpCoordinate}.

Under this transformation, the line element \eqref{eq:KerrdS} becomes
\begin{align}
    \frac{ds^2}{\epsilon_g}&=-\pa{\frac{\Delta-a^2\Upsilon\sin^2{\theta}}{\Xi^2\Sigma}}\ed v^2+\frac{2}{\Xi}\ed v\ed r+\frac{\Sigma}{\Upsilon}\ed\theta^2-\frac{2a\sin^2{\theta}}{\Xi}\br{\frac{\pa{r^2+a^2}\Upsilon-\Delta}{\Xi\Sigma}\ed v+\ed r}\ed\psi\notag\\ 
    &\phantom{=}+\frac{\sin^2{\theta}}{\Xi^2\Sigma}\br{\pa{r^2+a^2}^2\Upsilon-a^2\Delta\sin^2{\theta}}\ed\psi^2.
\end{align}
Since the $(r,\theta)$ coordinates are untouched, single modes transform simply as follows:
\begin{align}
    \label{eq:IngoingModes}
    e^{-i\omega t+im\phi}R(r)S(\theta)\to e^{i\omega r_*-imr_\sharp}e^{-i\omega v+im\psi}R(r)S(\theta).
\end{align}

In ingoing coordinates \eqref{eq:IngoingCoordinates}, the IRG metric components \eqref{eq:MetricComponentsIRG} become
\begin{subequations}
\begin{align}
    h_{vv}^{\rm IRG}&=\frac{1}{\Xi^2}\br{-a^2\Upsilon\sin^2{\theta}\,\mathcal{M}_+-2a\sqrt{\Upsilon}\sin{\theta}\,\mathcal{N}_-+h_{nn}},\\ 
    h_{rr}^{\rm IRG}&=\frac{4\Sigma^2}{\Delta^2}h_{nn},\\
    h_{\theta\theta}^{\rm IRG}&=\frac{\Sigma^2}{\Upsilon}\mathcal{M}_+,\\
    h_{\psi\psi}^{\rm IRG}&=-\frac{\sin^2{\theta}}{\Xi^2}\br{\pa{r^2+a^2}^2\Upsilon\mathcal{M}_++2a\pa{r^2+a^2}\sqrt{\Upsilon}\sin{\theta}\,\mathcal{N}_--a^2\sin^2{\theta}\,h_{nn}},\\
    h_{vr}^{\rm IRG}&=\frac{2\Sigma}{\Xi\Delta}\br{a\sqrt{\Upsilon}\sin{\theta}\,\mathcal{N}_--h_{nn}},\\
    h_{v\theta}^{\rm IRG}&=\frac{\Sigma}{\Xi}\br{\frac{\mathcal{N}_+}{\sqrt{\Upsilon}}-a\sin{\theta}\,\mathcal{M}_-},\\
    h_{v\psi}^{\rm IRG}&=\frac{a\sin^2{\theta}}{\Xi^2}\br{\pa{r^2+a^2}\Upsilon\mathcal{M}_++\pa{\frac{\Sigma}{a\sin{\theta}}+2a\sin{\theta}}\sqrt{\Upsilon}\mathcal{N}_--h_{nn}},\\
    h_{r\theta}^{\rm IRG}&=-\frac{2\Sigma^2}{\Delta\sqrt{\Upsilon}}\mathcal{N}_+,\\ 
    h_{r\psi}^{\rm IRG}&=\frac{2\Sigma\sin{\theta}}{\Xi\Delta}\br{-\pa{r^2+a^2}\sqrt{\Upsilon}\mathcal{N}_-+a\sin{\theta}\,h_{nn}},\\
    h_{\theta\psi}^{\rm IRG}&=\frac{\Sigma\sin{\theta}}{\Xi}\br{\pa{r^2+a^2}\mc{M}_--\frac{a\sin{\theta}}{\sqrt{\Upsilon}}\mc{N}_+},
\end{align}
\end{subequations}
while the ORG metric components \eqref{eq:MetricComponentsORG} become (the $h_{r\mu}$ all vanish)
\begin{subequations}
\begin{align}
    h_{vv}^{\rm ORG}&=\frac{1}{\Xi^2}\br{-a^2\Upsilon\sin^2{\theta}\,\mc{M}_+-\frac{a\Delta\sqrt{\Upsilon}\sin{\theta}}{\Sigma}\mc{L}_-+\frac{\Delta^2}{4\Sigma^2}h_{ll}},\\
    h_{r\mu}^{\rm ORG}&=0,\\
    h_{\theta\theta}^{\rm ORG}&=\frac{\Sigma^2}{\Upsilon}\mc{M}_+,\\
    h_{\psi\psi}^{\rm ORG}&=-\frac{\sin^2{\theta}}{\Xi^2}\br{\pa{r^2+a^2}^2\Upsilon\mc{M}_++\frac{a\Delta\pa{r^2+a^2}\sqrt{\Upsilon}\sin{\theta}}{\Sigma}\mc{L}_--\frac{a^2\Delta^2\sin^2{\theta}}{4\Sigma^2}h_{ll}},\\
    h_{v\theta}^{\rm ORG}&=\frac{1}{\Xi}\br{\frac{\Delta}{2\sqrt{\Upsilon}}\mc{L}_+-a\Sigma\sin{\theta}\,\mc{M}_-},\\
    h_{v\psi}^{\rm ORG}&=\frac{a\sin^2{\theta}}{\Xi^2}\br{\pa{r^2+a^2}\Upsilon\mc{M}_++\frac{\Delta\sqrt{\Upsilon}}{2\Sigma}\pa{\frac{\Sigma}{a\sin{\theta}}+2a\sin{\theta}}\mc{L}_- - \frac{\Delta^2}{4\Sigma^2}h_{ll}},\\
    h_{\theta\psi}^{\rm ORG}&=\frac{\Sigma\sin{\theta}}{\Xi}\br{\pa{r^2+a^2}\mc{M}_- - \frac{a\Delta\sin{\theta}}{2\sqrt{\Upsilon}\Sigma}\mc{L}_+},
\end{align}
\end{subequations}
with $\mathcal{L}_\pm$, $\mathcal{M}_\pm$, and $\mathcal{N}_\pm$ as defined in Eqs.~\eqref{eq:RealProjections} and \eqref{eq:RealProjectionsBis}.
Now, we only need to specify the tetrad components \eqref{eq:TetradProjection} in these new coordinates.
Since the projections $h_{ab}=a^\mu b^\nu h_{\mu\nu}$ are spacetime scalars, they transform simply by substitution of \eqref{eq:IngoingCoordinates}.
By Eq.~\eqref{eq:IngoingModes}, the result is
\begin{align}
    h_{ab}=e^{i\omega r_*-imr_\sharp}e^{-i\omega v+im\psi}h_{ab}^{(+)}(r,\theta)+e^{-i\ol{\omega}r_*+imr_\sharp}e^{i\ol{\omega}v-im\psi}h_{ab}^{(-)}(r,\theta),
\end{align}
where the $h_{ab}^{(\pm)}(r,\theta)$ are the same as in Boyer-Lindquist coordinates---and are given in Eqs.~\eqref{eq:TetradProjectionsIRG} and \eqref{eq:TetradProjectionsORG}---since the coordinates $r$ and $\theta$ are unchanged under the transformation \eqref{eq:IngoingCoordinates}.

\subsection{Metric reconstruction in outgoing coordinates}
\label{subsec:OutgoingMetricReconstruction}

Outgoing Kerr--(anti-)de Sitter coordinates $(u,r,\theta,\psi)$ are related to Boyer-Lindquist coordinates $(t,r,\theta,\phi)$ via
\begin{align}
    \label{eq:OutgoingCoordinates}
    u=t-r_*,\qquad
    \psi=\phi-r_\sharp,
\end{align}
where the tortoise coordinate $r_*$ and the radius $r_\sharp$ are defined in Eqs.~\eqref{eq:TortoiseCoordinate} and \eqref{eq:SharpCoordinate}.

Under this transformation, the line element \eqref{eq:KerrdS} becomes
\begin{align}
    \frac{ds^2}{\epsilon_g}&=-\pa{\frac{\Delta-a^2\Upsilon\sin^2{\theta}}{\Xi^2\Sigma}}\ed u^2-\frac{2}{\Xi}\ed v\ed r+\frac{\Sigma}{\Upsilon}\ed\theta^2-\frac{2a\sin^2{\theta}}{\Xi}\br{\frac{\pa{r^2+a^2}\Upsilon-\Delta}{\Xi\Sigma}\ed u-\ed r}\ed\psi\notag\\
    &\phantom{=}+\frac{\sin^2{\theta}}{\Xi^2\Sigma}\br{\pa{r^2+a^2}^2\Upsilon-a^2\Delta\sin^2{\theta}}\ed\psi^2.
\end{align}
Since the $(r,\theta)$ coordinates are untouched, single modes transform simply as follows:
\begin{align}
    \label{eq:OutgoingModes}
    e^{-i\omega t+im\phi}R(r)S(\theta)\to e^{-i\omega r_*+imr_\sharp}e^{-i\omega u+im\psi}R(r)S(\theta).
\end{align}
In outgoing coordinates \eqref{eq:OutgoingCoordinates}, the IRG metric components \eqref{eq:MetricComponentsIRG} become (the $h_{r\mu}$ all vanish)
\begin{subequations}
\begin{align}
    h_{uu}^{\rm IRG}&=\frac{1}{\Xi^2}\br{-a^2\Upsilon\sin^2{\theta}\,\mc{M}_+-2a\sqrt{\Upsilon}\sin{\theta}\,\mc{N}_-+h_{nn}},\\
    h_{r\mu}^{\rm IRG}&=0,\\
    h_{\theta\theta}^{\rm IRG}&=\frac{\Sigma^2}{\Upsilon}\mc{M}_+,\\
    h_{\psi\psi}^{\rm IRG}&=-\frac{\sin^2{\theta}}{\Xi^2}\br{\pa{r^2+a^2}^2\Upsilon\mc{M}_++2a\pa{r^2+a^2}\sqrt{\Upsilon}\sin{\theta}\,\mc{N}_--a^2\sin^2{\theta}\,h_{nn}},\\
    h_{u\theta}^{\rm IRG}&=\frac{\Sigma}{\Xi}\br{\frac{1}{\sqrt{\Upsilon}}\mc{N}_+-a\sin{\theta}\,\mc{M}_-},\\
    h_{u\psi}^{\rm IRG}&=\frac{a\sin^2{\theta}}{\Xi^2}\br{\pa{r^2+a^2}\Upsilon\mc{M}_++\sqrt{\Upsilon}\pa{\frac{\Sigma}{a\sin{\theta}}+2a\sin{\theta}}\mc{N}_--h_{nn}},\\
    h_{\theta\psi}^{\rm IRG}&=\frac{\Sigma\sin{\theta}}{\Xi}\br{\pa{r^2+a^2}\mc{M}_--\frac{a\sin{\theta}}{\sqrt{\Upsilon}}\mc{N}_+},
\end{align}
\end{subequations}
while the ORG components \eqref{eq:MetricComponentsORG} become
{\allowdisplaybreaks
\begin{subequations}
\begin{align}
    h_{uu}^{\rm ORG}&=\frac{1}{\Xi^2}\br{-a^2\Upsilon\sin^2{\theta}\,\mc{M}_+-\frac{a\Delta\sqrt{\Upsilon}\sin{\theta}}{\Sigma}\mc{L}_-+\frac{\Delta^2}{4\Sigma^2}h_{ll}},\\
    h_{rr}^{\rm ORG}&=h_{ll},\\
    h_{\theta\theta}^{\rm ORG}&=\frac{\Sigma^2}{\Upsilon}\mc{M}_+,\\
    h_{\psi\psi}^{\rm ORG}&=-\frac{\sin^2{\theta}}{\Xi^2}\br{\pa{r^2+a^2}^2\Upsilon\mc{M}_++\frac{a\Delta\pa{r^2+a^2}\sqrt{\Upsilon}\sin{\theta}}{\Sigma}\mc{L}_--\frac{a^2\Delta^2\sin^2{\theta}}{4\Sigma^2}h_{ll}},\\
    h_{ur}^{\rm ORG}&=\frac{1}{\Xi}\br{-a\sqrt{\Upsilon}\sin{\theta}\,\mathcal{L}_-+\frac{\Delta}{2\Sigma}h_{ll}},\\
    h_{u\theta}^{\rm ORG}&=\frac{1}{\Xi}\br{\frac{\Delta}{2\sqrt{\Upsilon}}\mc{L}_+-a\Sigma\sin{\theta}\,\mc{M}_-},\\
    h_{u\psi}^{\rm ORG}&=\frac{a\sin^2{\theta}}{\Xi^2}\br{\pa{r^2+a^2}\Upsilon\mc{M}_++\frac{\Delta\sqrt{\Upsilon}}{2\Sigma}\pa{\frac{\Sigma}{a\sin{\theta}}+2a\sin{\theta}}\mc{L}_--\frac{\Delta^2}{4\Sigma^2}h_{ll}},\\
    h_{r\theta}^{\rm ORG}&=\frac{\Sigma}{\sqrt{\Upsilon}}\mathcal{L}_+,\\
    h_{r\psi}^{\rm ORG}&=\frac{\sin{\theta}}{\Xi}\br{\pa{r^2+a^2}\sqrt{\Upsilon}\mathcal{L}_--\frac{a\Delta\sin{\theta}}{2\Sigma}h_{ll}},\\
    h_{\theta\psi}^{\rm ORG}&=\frac{\Sigma\sin{\theta}}{\Xi}\br{\pa{r^2+a^2}\mc{M}_--\frac{a\Delta\sin{\theta}}{2\sqrt{\Upsilon}\Sigma}\mc{L}_+},
\end{align}
\end{subequations}}
with $\mathcal{L}_\pm$, $\mathcal{M}_\pm$, and $\mathcal{N}_\pm$ as defined in Eqs.~\eqref{eq:RealProjections} and \eqref{eq:RealProjectionsBis}.
Now, we only need to specify the tetrad components \eqref{eq:TetradProjection} in these new coordinates.
Since the projections $h_{ab}=a^\mu b^\nu h_{\mu\nu}$ are spacetime scalars, they transform simply by substitution of \eqref{eq:OutgoingCoordinates}.
By Eq.~\eqref{eq:OutgoingModes}, the result is
\begin{align}
    h_{ab}=e^{-i\omega r_*+imr_\sharp}e^{-i\omega u+im\psi}h_{ab}^{(+)}(r,\theta)+e^{i\ol{\omega}r_*-imr_\sharp}e^{i\ol{\omega}u-im\psi}h_{ab}^{(-)}(r,\theta).
\end{align}
where the $h_{ab}^{(\pm)}(r,\theta)$ are the same as in Boyer-Lindquist coordinates---and are given in Eqs.~\eqref{eq:TetradProjectionsIRG} and \eqref{eq:TetradProjectionsORG}---since the coordinates $r$ and $\theta$ are unchanged under the transformation \eqref{eq:OutgoingCoordinates}.

\section{Roots of \texorpdfstring{$\Delta(r)$}{Δ(r)}}
\label{app:Roots}

In this appendix, we investigate the nature of the roots of $\Delta(r)$, which determine the locations of the horizons in the Kerr--(anti-)de Sitter spacetime.
We first review the roots of a depressed cubic polynomial in App.~\ref{subsec:CubicRoots} and then those of a generic quartic polynomial in App.~\ref{subsec:QuarticRoots}, with an emphasis on the necessary conditions for them to be real.
Next, in App.~\ref{subsec:ConstrainingDelta}, we write down explicit expressions for the roots $r_i$ and use their reality conditions to constrain the parameters of the Kerr--(anti-)de Sitter black hole.
Finally, in App.~\ref{subsec:RootsProof}, we use Vieta's formulas to prove that in Kerr--de Sitter, $\Delta(r)$ has three positive real roots and one negative real one, so that the spacetime has three physical horizons, whereas in Kerr--anti-de Sitter, $\Delta(r)$ has two positive real roots and two complex-conjugate roots, so that the spacetime has two physical horizons.

\subsection{Roots of a depressed cubic}
\label{subsec:CubicRoots}

Any cubic $x^3+Ax^2+Bx+C$ can be mapped, via a change of variables, to the depressed form
\begin{align}
    \label{eq:DepressedCubic}
    f(x)=x^3+px+q.
\end{align}
For $p$ and $q$ are real, the nature of the roots of $f(x)$ is determined by the sign of the discriminant
\begin{align}
    \triangle_3=-4p^3-27q^2.
\end{align}
If $x_1$, $x_2,$ and $x_3$ are the roots (with multiplicities) of $f(x)$, then the discriminant obeys
\begin{align}
    \triangle_3=\prod_{1\leq i<j\leq3}(x_i-x_j)^2.
\end{align}
If $\triangle_3>0$, then $f(x)$ has three distinct real roots, and if $\triangle_3<0$, then $f(x)$ has only one real root and two distinct complex-conjugate roots.
If $\triangle_3=0$, then at least two roots are equal.
In all three cases, the three roots can be expressed as
\begin{align}
    \label{eq:xRootsCubic}
    x_k=-\frac{1}{3}\pa{\xi^k Q-\frac{3p}{\xi^k Q}},\qquad
    \xi=\frac{-1+i\sqrt{3}}{2},\qquad 
    Q=\sqrt[3]{\frac{27q-3\sqrt{-3\triangle_3}}{2}}.
\end{align}
In this form, it is not manifestly clear which roots are real.\footnote{When $\triangle_3>0$, $Q^3$ is not real and thus $Q$ is not either, but all three $x_k$ are.
When $\triangle_3<0$, $Q^3$ is real, though $Q$ may not be.
If $Q\in\mathbb{R}$, then $Q^3>0$ and thus $p<0$ and $q\geq2\pa{\frac{|p|}{3}}^{3/2}$, so $x_0$ is the only real root in this case.
If $Q\notin\mathbb{R}$, then $Q^3<0$ and thus either $p>0$ or $p<0$ and $q\leq-2\pa{\frac{|p|}{3}}^{3/2}$, so $x_1$ is the only real root in this case.}
A more useful form (for $p\neq 0$) is
\begin{gather}
    \label{eq:TrigRoots}
    x_k=-2\,{\rm sign}(p)\sqrt{-\frac{p}{3}}\cos\br{\frac{1}{3}\arccos\pa{\frac{3q}{2p}\sqrt{-\frac{3}{p}}}-\frac{2\pi k}{3}},\qquad
    k\in\cu{0,1,2}.
\end{gather}
When $\triangle_3>0$, so that $p<0$ and $|q|<\frac{2|p|}{3}\sqrt{\frac{|p|}{3}}$, the arccosine in the above equation is manifestly real, resulting in three real roots.
When $\triangle_3<0$, this formula still holds but involves complex cosines and arccosines, resulting in only one real root expressible in the manifestly real form
\begin{align}
    \label{eq:HyperbolicTrigRoots}
    x_0=
    \begin{cases}
        -2\,{\rm{sign}}(q)\sqrt{\frac{|p|}{3}}\cosh\cu{\frac{1}{3}{\rm{arccosh}}\br{\frac{|q|}{2}\pa{\frac{3}{|p|}}^{3/2}}},
        &p<0\text{ and }|q|>2\pa{\frac{|p|}{3}}^{3/2},\\
        -2\,{\rm{sign}}(q)\sqrt{\frac{p}{3}}\sinh\cu{\frac{1}{3}{\rm{arcsinh}}\br{\frac{|q|}{2}\pa{\frac{3}{p}}^{3/2}}},
        &p>0.
    \end{cases}
\end{align}

\subsection{Roots of a quartic}
\label{subsec:QuarticRoots}

Given a generic quartic polynomial with real coefficients (we assume that $a\neq 0$)
\begin{align}
    \label{eq:Quartic}
    f(x) = ax^4+bx^3+cx^2+dx+e,
\end{align}
the nature of the four roots is determined by multiple signs, include that of the discriminant
\begin{align}
    \triangle_4&=256a^3e^3-192a^2bde^2-128a^2c^2e^2+144 a^2cd^2e-27a^2d^4\notag\\
    &\phantom{=}+144ab^2ce^-6ab^2d^2e-80abc^2de+18abcd^3+16ac^4e\notag\\
    &\phantom{=}-4ac^3d^2-27b^4e^2+18b^3cde-4b^3d^3-4b^2c^3e+b^2c^2d^2.
\end{align}
If $x_1$, $x_2$, $x_3$, and $x_4$ are the roots (with multiplicities) of $f(x)$, then this discriminant obeys
\begin{align}
    \triangle_4=a^6\prod_{1\leq i<j\leq4}(x_i-x_j)^2.
\end{align}
Further refinement on the nature of the roots comes from the quantities
\begin{align}
    P=8ac-3b^2,\qquad
    D=64a^3e-16a^2c^2+16ab^2c-16a^2bd-3b^4.
\end{align}
The possible cases for the roots of $f(x)$ are as follows:
\begin{itemize}
    \item If $\triangle_4<0$, then there are two distinct real roots and two complex-conjugate roots.
    \item If $\triangle_4>0$, then the roots are either all real or all non-real:
    \begin{itemize}
        \item If $P<0$ and $D<0$, then all four roots are real and distinct.
        \item If $P>0$ or $D>0$, then there are two pairs of complex-conjugate roots.
    \end{itemize}
    \item If $\triangle_4=0$, then there is a repeated root.
\end{itemize}

The four roots of the quartic can be written as\footnote{These expressions are derived from an analysis of the depressed quartic $y^4+py^2+qy+r$, to which Eq.~\eqref{eq:Quartic} can be converted (up to an overall factor of $a$ that does not affect the roots) by taking $x=y-\frac{b}{4a}$.}
\begin{subequations}
\label{eq:xRootsQuartic}
\begin{align}
    x_{1,2}&=-\frac{b}{4a}-\frac{\sqrt{2m}}{2}\pm\frac{1}{2}\sqrt{-\pa{2m+2p-\frac{\sqrt{2}q}{\sqrt{m}}}},\\
    x_{3,4}&=-\frac{b}{4a}+\frac{\sqrt{2m}}{2}\pm\frac{1}{2}\sqrt{-\pa{2m+2p+\frac{\sqrt{2}q}{\sqrt{m}}}},
\end{align}
\end{subequations}
where $m$ is any real solution of the ``resolvent cubic'' equation
\begin{align}
    \label{eq:ResolventCubic}
    m^3+pm^2+\pa{\frac{p^2}{4}-r}m-\frac{q^2}{8}=0,
\end{align}
with coefficients
\begin{subequations}
\begin{align}
    p&=\frac{P}{8a^2},\qquad
    q=\frac{b^3-4abc+8a^2d}{8a^3},\\
    r&=\frac{-3b^4+256a^3e-64a^2bd+16ab^2c}{256a^4}
    =\frac{1}{64a^4}\pa{D+\frac{P^2}{4}}.
\end{align}
\end{subequations}
The resolvent cubic \eqref{eq:ResolventCubic} always admits at least one real positive root.\footnote{Like any cubic, the resolvent \eqref{eq:ResolventCubic} has at least one real root.
By one of Vieta's formulas, the product of its roots must equal $\frac{q^2}{8}$ and hence be nonnegative.
If $q\neq0$, then when all three roots are real, at least one must be positive; when only one root is real, the product of the complex-conjugate roots is positive, so this real root must also be positive.
If $q=0$, then $m=0$ is a real solution and Eqs.~\eqref{eq:xRootsQuartic} fail, but the depressed quartic degenerates to a biquadratic that contains only even powers of $y$ and can then be solved via the quadratic formula for $z=y^2$.}
Such a root can be obtained by letting $m=t-\frac{p}{3}$ to convert the resolvent cubic \eqref{eq:ResolventCubic} to a depressed cubic
\begin{align}
    \label{eq:DepressedResolventCubic}
    t^3+p't+q'=0,\qquad
    p'=-\pa{\frac{p^2}{12}+r},\qquad
    q'=-\pa{\frac{p^3}{108}+\frac{q^2}{8}-\frac{pr}{3}},
\end{align}
with discriminant
\begin{align}
    \triangle_3=-4\pa{p'}^3-27\pa{q'}^2
    =\frac{\triangle_4}{64a^6}.
\end{align}
In summary, one can use the expressions from App.~\ref{subsec:CubicRoots}---namely, Eq.~\eqref{eq:xRootsCubic}, or alternatively Eqs.~\eqref{eq:TrigRoots} or \eqref{eq:HyperbolicTrigRoots}---to find a real solution to Eq.~\eqref{eq:DepressedResolventCubic} and hence a positive real root $m=t-\frac{p}{3}$ of the resolvent cubic \eqref{eq:ResolventCubic}.
Then the four roots of the quartic \eqref{eq:Quartic} are obtained by plugging that $m$ into Eqs.~\eqref{eq:xRootsQuartic}.

Lastly, we define for future purposes two quantities $\triangle_0$ and $\triangle_1$ such that $\triangle_1^2-4\triangle_0^3=-27\triangle_4$:%
\begin{subequations}
\begin{align}
    \triangle_0&=-12a^2p'
    =c^2-3bd+12ae,\\
    \triangle_1&=-216a^3q'
    =2c^3-9bcd+27b^2e+27ad^2-72ace.
\end{align}
\end{subequations}

\subsection{Constraining the parameters of $\Delta(r)$}
\label{subsec:ConstrainingDelta}

We now apply these results to the quartic polynomial that appears in the line element \eqref{eq:KerrdS},
\begin{align}
    \Delta(r)=-\frac{\Lambda}{3}r^4+\pa{1-\frac{\Lambda a^2}{3}}r^2-2Mr+a^2.
\end{align}
In the Kerr--de Sitter case ($\Lambda>0$), we require that all four roots of this polynomial be real.
In the Kerr--anti-de Sitter case ($\Lambda<0$), we only require that two of the roots be real.
To determine the constraints these two cases place on $\Lambda$ and $a$, we first compute the discriminant of $\Delta(r)$:
\begin{align}
    \label{eq:KerrdSDiscriminant}
    \triangle_4=-\frac{16\Lambda}{243}\big(&\Lambda^4a^{10}+12\Lambda^3a^8+54\Lambda^2a^6+3\Lambda^3a^6M^2+108\Lambda a^4+297\Lambda^2a^4M^2\notag\\
    &-891\Lambda a^2M^2+81a^2+729\Lambda M^4-81M^2\big).
\end{align}
For future reference, we compute the various parameters from Sec.~\ref{subsec:QuarticRoots}:
\begin{subequations}
\label{eq:DeltaParameters}
\begin{gather}
    P=-\frac{8\Lambda}{3}\pa{1-\frac{\Lambda a^2}{3}},\qquad
    D=-\frac{16\Lambda^2}{9}\pa{1+\frac{\Lambda a^2}{3}}^2,\\
    p=\frac{9P}{8\Lambda^2},\qquad
    q=\frac{6M}{\Lambda},\qquad
    r=-\frac{3a^2}{\Lambda},\\
    \triangle_0=\frac{\Lambda^2a^4}{9}-\frac{14\Lambda a^2}{3}+1,\qquad
    \triangle_1=-2\pa{\frac{\Lambda^3a^6}{27}+\frac{11\Lambda^2a^4}{3}-11\Lambda a^2+18\Lambda M^2-1}.
\end{gather}
\end{subequations}
The condition $\triangle_4 = 0$ can be represented as a set of curves in the plane $(a,\Lambda)$.
Since $\Delta(r)$ and thus $\triangle_4$ only depend on $a$ via $a^2$, this set of curves must be symmetric about the axis $a=0$. 

We first consider the Schwarzchild--(anti-)de Sitter black hole with zero spin.
When $a=0$, the discriminant \eqref{eq:KerrdSDiscriminant} degenerates to a quadratic in $\Lambda$,
\begin{align}
    \triangle_4=-48M^4\Lambda\pa{\Lambda-\Lambda_{\rm ext}},\qquad
    \Lambda_{\rm ext}=\frac{1}{9M^2},
\end{align}
which clearly has distinct real roots at $\Lambda=0$ and $\Lambda=\Lambda_{\rm ext}$.
It therefore follows that $\triangle_4>0$ for $0<\Lambda<\Lambda_{\rm ext}$.
In that range, Eqs.~\eqref{eq:DeltaParameters} also imply that $P<0$ and $D<0$, so all four roots are real, whereas outside $\triangle_4<0$, so there are two real roots and two complex roots.
As such, a nonrotating black hole can only fit in an asymptotically de Sitter spacetime with positive cosmological constant in the range $0<\Lambda<\Lambda_{\rm ext}$.
By contrast, there is no such restriction for a Schwarzschild black hole in asymptotically anti-de Sitter spacetime.

Next, we turn to the Kerr case of an asymptotically flat black hole with vanishing cosmological constant $\Lambda=0$.
In this case, $\triangle_4=0$ as $\Delta(r)$ degenerates from a quartic to a quadratic with discriminant $\triangle_2=4\pa{M^2-a^2}$.
To ensure the existence of two horizons (two real roots), we must demand that this discriminant be positive, leading to the usual Kerr bound $0\leq|a|\leq M$.

From here on, we shall consider the general case with both $\Lambda\neq0$ and $a\neq0$, in which $-\frac{243\triangle_4}{16\Lambda}$ is itself a quartic polynomial in $\Lambda$.
It has leading coefficient $a^{10}$ and parameters from App.~\ref{subsec:QuarticRoots}
\begin{subequations}
\begin{align}
    P_\Lambda&=27a^{12}M^2\pa{80a^2-M^2}, \\
    D_\Lambda&=81a^{24}M^2\pa{4096a^6-14592a^4M^2+48a^2M^4-3M^6},\\
    p_\Lambda&=\frac{P_\Lambda}{8a^{20}},\\
    q_\Lambda&=-\frac{27M^2}{8a^{12}}\pa{768a^4-96a^2M^2-M^4},\\
    r_\Lambda&=\frac{81}{256a^{16}}\pa{16384a^6M^2-768a^4M^4-1248a^2M^6-3M^8},\\
    \triangle_0^\Lambda&=243a^6M^2\pa{256a^4+288a^2M^2-27M^4},\\
    \triangle_1^\Lambda&=314928a^{10}M^4\pa{256a^4+27M^4}.
\end{align}
\end{subequations}
Its quartic discriminant $\triangle_\Lambda$ is manifestly positive for $|a|>a_{\rm max}=\frac{1}{4}\sqrt{9+6\sqrt{3}}M\approx1.101M$:
\begin{align}
    \frac{\triangle_\Lambda}{2^{26}3^{12}a^{18}M^6}=\pa{a^4-\frac{9}{8}a^2M^2-\frac{27}{256}M^4}^3
    =\br{a^2+\frac{3}{16}\pa{2\sqrt{3}-3}M^2}^3\pa{a^2-a_{\rm max}^2}^3.
\end{align}
We see that $P_\Lambda>0$ for $|a|>\frac{\sqrt{5}}{20}M\approx0.112M$, which is smaller than $a_{\rm max}$, so the quartic in $\Lambda$ has no real roots for $|a|>a_{\rm max}$.
Meanwhile, $\triangle_\Lambda<0$ for $0<|a|<a_{\rm max}$, so in this range the quartic in $\Lambda$ has two distinct real roots, and thus there are two values of $\Lambda$ (other than $\Lambda=0$) for which $\triangle_4$ vanishes.
Since $\triangle_\Lambda<0$ for $0<|a|<a_{\rm max}$, we must have $\pa{\triangle_1^\Lambda}^2>4\pa{\triangle_0^\Lambda}^3$ in this range, as we can also verify with the above expressions.
We further find that $\triangle_1^\Lambda>0$ for $|a|>0$.
To determine the two nonzero real roots, we can now use Eq.~\eqref{eq:xRootsQuartic}, once we determine the value of the real root $m_\Lambda$ of the resolvent cubic \eqref{eq:ResolventCubic}.
Using the results of App.~\ref{subsec:CubicRoots}, we find\footnote{This is a slightly different definition of $Q$ than the one used in App.~\ref{subsec:CubicRoots}.
Here, we took advantage of the arbitrary square root sign and the freedom to multiply by a cube root of $-1$ to ensure that $Q_\Lambda$ is real for $\triangle_\Lambda<0$.}
\begin{align}
    m_\Lambda=-\frac{p_\Lambda}{3}+\frac{1}{6a^{10}}\pa{Q_\Lambda+\frac{\Delta^\Lambda_0}{Q_\Lambda}},\qquad
    Q_\Lambda=\sqrt[3]{\frac{\triangle_1^\Lambda+\sqrt{\pa{\triangle_1^\Lambda}^2-4\pa{\triangle_0^\Lambda}^3}}{2}}.
\end{align}
As mentioned above, $\triangle_1^\Lambda>0$ for $|a|>0$, so $Q^3>0$ and thus $Q$ is real provided $0<|a|<a_{\rm max}$.

We can now use Eqs.~\eqref{eq:xRootsQuartic} to determine the two nonzero real roots $\Lambda$ of $\triangle_4$:
\begin{subequations}
\begin{align}
    \Lambda_{\rm max}&=
    \begin{cases}
         -\frac{3\pa{4a^2+M^2}}{4a^4}-\frac{\sqrt{2m_\Lambda}}{2}+\frac{1}{2}\sqrt{-\pa{2m_\Lambda+2p_\Lambda-\frac{\sqrt{2}q_\Lambda}{\sqrt{m_\Lambda}}}},
         &|a|<\frac{a_{\rm max}}{3}\\
         -\frac{3\pa{4a^2+M^2}}{4a^4}+\frac{\sqrt{2m_\Lambda}}{2}+\frac{1}{2}\sqrt{-\pa{2m_\Lambda+2p_\Lambda+\frac{\sqrt{2}q_\Lambda}{\sqrt{m}}}},
         &|a|>\frac{a_{\rm max}}{3}
    \end{cases}\\
    &=-\frac{3\pa{4a^2+M^2}}{4a^4}-{\rm sign}\pa{q_\Lambda}\frac{\sqrt{2m_\Lambda}}{2}+\frac{1}{2}\sqrt{-\pa{2m_\Lambda+2p_\Lambda-\frac{\sqrt{2}\ab{q_\Lambda}}{\sqrt{m_\Lambda}}}},\\
    \Lambda_{\rm min}&=
    \begin{cases}
         -\frac{3\pa{4a^2+M^2}}{4a^4}-\frac{\sqrt{2m_\Lambda}}{2}-\frac{1}{2}\sqrt{-\pa{2m_\Lambda+2p_\Lambda-\frac{\sqrt{2}q_\Lambda}{\sqrt{m_\Lambda}}}},
         &|a|<\frac{a_{\rm max}}{3}\\
         -\frac{3\pa{4a^2+M^2}}{4a^4}+\frac{\sqrt{2m_\Lambda}}{2}-\frac{1}{2}\sqrt{-\pa{2m_\Lambda+2p_\Lambda+\frac{\sqrt{2}q_\Lambda}{\sqrt{m_\Lambda}}}},
         &|a|>\frac{a_{\rm max}}{3}
    \end{cases}\\
    &=-\frac{3\pa{4a^2+M^2}}{4a^4}-{\rm sign}\pa{q_\Lambda}\frac{\sqrt{2m_\Lambda}}{2}-\frac{1}{2}\sqrt{-\pa{2m_\Lambda+2p_\Lambda-\frac{\sqrt{2}\ab{q_\Lambda}}{\sqrt{m_\Lambda}}}}.
\end{align}
\end{subequations}
When the spin is $a=\frac{a_{\rm max}}{3}$, these formulas cannot be used because both $q_\Lambda$ and $m_\Lambda$ vanish, but the two roots can be found using the quadratic formula:
\begin{subequations}
\begin{align}
    \Lambda_{\rm max}\big\vert_{|a|=\frac{a_{\rm max}}{3}}&=\frac{48}{M^2}\pa{-81+46\sqrt{3}+4\sqrt{6}\sqrt{774-447\sqrt{3}+8\sqrt{12786-7382\sqrt{3}}}}\\
    &\approx0.1146 M^{-2},\notag\\
    \Lambda_{\rm min}\big\vert_{|a|=\frac{a_{\rm max}}{3}}&=\frac{48}{M^2}\pa{-81+46\sqrt{3}-4\sqrt{6}\sqrt{774-447\sqrt{3}+8\sqrt{12786-7382\sqrt{3}}}}\\
    &\approx-127.4M^{-2}.\notag
\end{align}
\end{subequations}
Though it is certainly not apparent from the above expressions, the roots $\Lambda_{\rm max}$ and $\Lambda_{\rm min}$ are continuous functions of $a$, including across the special value of spin $a=\frac{a_{\rm max}}{3}$.

Plotted as functions of $a$, the roots $\Lambda_{\rm max}$ and $\Lambda_{\rm min}$ delineate boundaries in the $(a,\Lambda)$ plane.
Crossing either of these boundaries flips the sign of $\triangle_4$.
We find that $\Lambda_{\rm max}>0$ for $|a|<a_{\rm max}$, $\Lambda_{\rm min}>0$ for $M<|a|<a_{\rm max}$, and $\Lambda_{\rm min}<0$ for $|a|<M$.
Lastly, we also find, consistent with the fact (noted above) that $\triangle_\Lambda=0$ at $|a|=a_{\rm max}$, that
\begin{align}
    \Lambda_{\rm max}\big\vert_{|a|=a_{\rm max}}=\Lambda_{\rm min}\big\vert_{|a|=a_{\rm max}}
    =\frac{16}{3M^2}\pa{-45+26\sqrt{3}}
    \approx0.1777M^{-2}.
\end{align}

\begin{figure}[h]
\includegraphics[width=6cm]{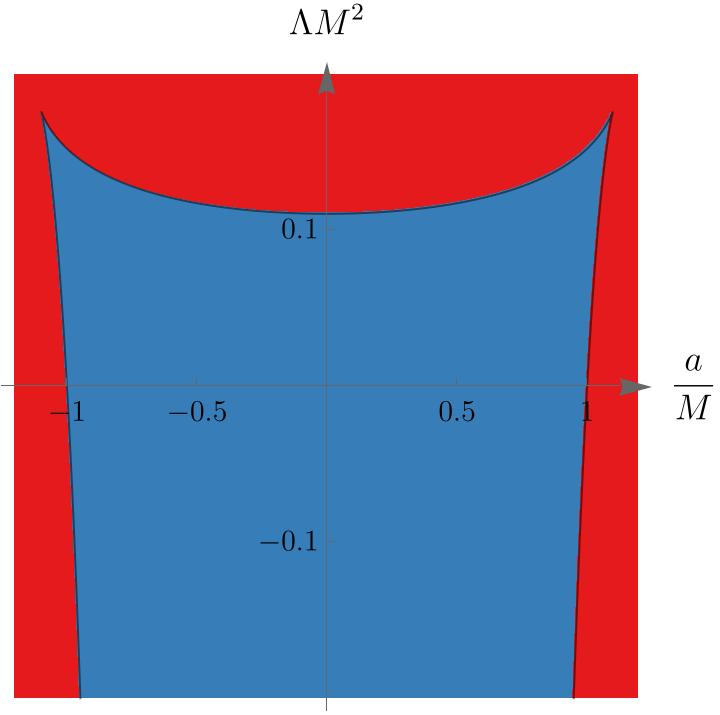}
\centering
\caption{The allowed region (blue) in the $(a,\Lambda)$ plane (shown in units of $M$).}
\label{fig:DiscriminantSign}
\end{figure}

We can now establish the nature of the roots of $\Delta(r)$ for all possible values of $a$ and $\Lambda$:
\begin{itemize}
    \item If $0<|a|<M$ and $0<\Lambda<\Lambda_{\rm max}(a)$, then $\triangle_4>0$.
    We see from Eq.~\eqref{eq:DeltaParameters} that $D$ is always negative, while $P$ is negative for $0<\Lambda<3/a^2$.
    We can show that $\Lambda_{\rm max}<\frac{3}{a^2}$ for all values of $|a|<a_{\rm max}$, so there are four real roots in this range.
    \item If $M<|a|<a_{\rm max}$ and $\Lambda_{\rm min}(a)<\Lambda<\Lambda_{\rm max}(a)$, then $\triangle_4>0$ and so by the previous logic, there are also four real roots.
    \item If $0<|a|<a_{\rm max}$ and either $0<\Lambda<\Lambda_{\rm min}(a)$ or $\Lambda>\Lambda_{\rm max}(a)$, then $\triangle_4<0$ so there are two real and two complex-conjugate roots.
    \item If $|a|>a_{\rm max}$ and $\Lambda>0$, then $\triangle_4<0$ so there are two real and two complex-conjugate roots.
    \item If $0<|a|<M$ and $\Lambda_{\rm min}(a)<\Lambda<0$, then $\triangle_4<0$ so there are two real and two complex-conjugate roots.
    \item If $0<|a|<a_{\rm max}$ and $\Lambda<\Lambda_{\rm min}(a)<0$, then $\triangle_4>0$ and $P>0$, so there are no real roots.
    \item If $|a|>a_{\rm max}$ and $\Lambda<0$, then $\triangle_4>0$ and $P>0$, so again there are no real roots.
\end{itemize}
The locus $\frac{\triangle_4}{\Lambda}=0$ in the $(a,\Lambda)$ plane is represented in Fig.~\ref{fig:DiscriminantSign}.
It delineates the (blue) allowed region of parameter space for Kerr--(anti-)de Sitter black holes, which corresponds to $\frac{\triangle_4}{\Lambda}>0$.

We can now write the four real roots of $\Delta(r)$ in the Kerr--de Sitter case $\Lambda>0$.
In this case, we must have $\triangle_4>0$, which implies that $\triangle_0>0$ and also that the resolvent cubic \eqref{eq:ResolventCubic} has three real roots, with at least one positive.
We can represent one such root using Eq. \eqref{eq:TrigRoots} as
\begin{align}
    m=\frac{1}{\Lambda}\cu{1-\frac{1}{3}\Lambda a^2+\sqrt{\triangle_0}\cos\br{\frac{1}{3}\arccos\pa{-\frac{\triangle_1}{2\triangle_0^{3/2}}}}}
    >0.
\end{align}
The roots are then
\begin{subequations}
\begin{align}
    r_1&=\frac{\sqrt{2m}}{2}-\frac{1}{2}\sqrt{-2m-2p-\frac{\sqrt{2}q}{\sqrt{m}}},
    &&r_2=-\frac{\sqrt{2m}}{2}+\frac{1}{2}\sqrt{-2m-2p+\frac{\sqrt{2}q}{\sqrt{m}}},\\
    r_3&=\frac{\sqrt{2m}}{2}+\frac{1}{2}\sqrt{-2m-2p-\frac{\sqrt{2}q}{\sqrt{m}}},
    &&r_4=-\frac{\sqrt{2m}}{2}-\frac{1}{2}\sqrt{-2m-2p+\frac{\sqrt{2}q}{\sqrt{m}}}.
\end{align}
\end{subequations}
Using these explicit expressions, one can directly verify that these roots are ordered as
\begin{align}
    \label{eq:dSOrdering}
    r_4<0<r_2<r_1<r_3.
\end{align}
We also give a more elegant proof in App.~\ref{subsec:RootsProof}.
At leading order in $\Lambda=\frac{3}{L^2}$, we have
\begin{subequations}
\begin{align}
    r_{1,2}&=M\pm\sqrt{M^2-a^2}+\mc{O}\pa{\frac{1}{L^2}},\\
    r_{3,4}&=\pm L-M+\mc{O}\pa{\frac{1}{L}}.
\end{align}
\end{subequations}
It follows that we can think of $r_1$ and $r_2$ as the outer and inner black hole horizons, respectively, and $r_3$ as the cosmological horizon.
Since $r_4<0$, it is non-physical.

Next, in the Kerr--anti-de Sitter case $\Lambda<0$, we write the two real and two complex-conjugate roots of $\Delta(r)$.
Since we must have $\triangle_4<0$, the resolvent cubic has one real, positive root
\begin{align}
    m=-\frac{p}{3}-\frac{1}{2\Lambda}\pa{Q+\frac{\triangle_0}{Q}}
    >0,\qquad
    Q=\sqrt[3]{\frac{\triangle_1+\sqrt{\triangle_1^2-4\triangle_0^3}}{2}}
    >0,
\end{align}
and it can also be shown that $\triangle_1>0$, so that $Q^3>0$ and thus $Q>0$.
The real roots are then
\begin{align}
    r_1=\frac{\sqrt{2m}}{2}+\frac{1}{2}\sqrt{-2m-2p-\frac{\sqrt{2}q}{\sqrt{m}}},\qquad
    r_2=\frac{\sqrt{2m}}{2}-\frac{1}{2}\sqrt{-2m-2p-\frac{\sqrt{2}q}{\sqrt{m}}},
\end{align}
while the complex-conjugate roots are
\begin{align}
    r_3=-\frac{\sqrt{2m}}{2}+\frac{1}{2}\sqrt{-2m-2p+\frac{\sqrt{2}q}{\sqrt{m}}},\qquad
    r_4=-\frac{\sqrt{2m}}{2}-\frac{1}{2}\sqrt{-2m-2p+\frac{\sqrt{2}q}{\sqrt{m}}}.
\end{align}
Using their explicit expressions, one can directly verify that the real roots are ordered as
\begin{align}
    \label{eq:AdSOrdering}
    0<r_2<r_1.
\end{align}
Again, for a proof of this, see App.~\ref{subsec:RootsProof}.
At leading order in $\Lambda=-\frac{3}{L^2}$, we have
\begin{subequations}
\begin{align}
    r_{1,2}&=M\pm\sqrt{M^2-a^2}+\mc{O}\pa{\frac{1}{L^2}}.\\
    r_{3,4}&=\pm iL-M+\mc{O}\pa{\frac{1}{L}}.
\end{align}
\end{subequations}
We can still think of $r_1$ and $r_2$ as the outer and inner black hole horizons, respectively, but now $r_3$ and $r_4$ are complex.

\subsection{Ordering of the roots of \texorpdfstring{$\Delta(r)$}{Δ(r)}}
\label{subsec:RootsProof}

In this section, we prove that $\Delta(r)$ has three positive roots and one negative root in the Kerr--de Sitter spacetime, but two positive roots and two complex-conjugate non-real roots in the Kerr--anti-de Sitter spacetime.

We apply Vieta's formulas to $\Delta(r)$:
\begin{subequations}
\label{eq:Vieta}
\begin{align}
    \sum_{i=1}^4r_i&=0,\\
    \sum_{1\leq i<j\leq4}r_ir_j&=a^2-\sigma^2L^2,\\
    \sum_{1\leq i<j<k\leq4}r_ir_jr_k&=-2\sigma^2ML^2,\\
    r_1r_2r_3r_4&=-\sigma^2a^2L^2,
\end{align}
\end{subequations}
where, as before, we have $\Lambda=\frac{3\sigma^2}{L^2}$ with
\begin{align}
    \sigma= 
    \begin{cases}
        1,
        &\Lambda>0,\\
        i,
        &\Lambda<0.
    \end{cases}
\end{align}

We first consider the Kerr--de Sitter spacetime.
As shown in App.~\ref{subsec:ConstrainingDelta}, $a$ and $\Lambda$ must obey certain constraints $\Delta(r)$ to have four real roots.
The relevant one here are solely $|a|>0$ and $\Lambda>0$.
Assuming that we have four real roots, Vieta's fourth formula in Eqs.~\eqref{eq:Vieta} implies that none of the roots can be zero, and that either one is negative or three of them are.
Since in either case, we have at least one positive and at least one negative root, we can without loss of generality let these roots be $r_3$ and $r_4$, respectively.
It then follows that $r_1r_2>0$, so that $r_1$ and $r_2$ have the same sign.
We then use the first of Vieta's formulas, in the form
\begin{align}
    r_3+r_4=-(r_1+r_2),
\end{align}
to rewrite the third as
\begin{align}
    \pa{r_1+r_2}\pa{r_3r_4-r_1r_2}=-2ML^2
    <0.
\end{align}
Since $r_3r_4<0$ and $r_1r_2>0$, we see that this implies $r_1+r_2>0$.
Since $r_1$ and $r_2$ have the same sign, they must both be positive.

We next consider the Kerr--anti-de Sitter spacetime.
Once again, as shown in App.~\ref{subsec:ConstrainingDelta}, $a$ and $\Lambda$ must obey certain constraints for there to be two real roots and two complex-conjugate roots of $\Delta(r)$, but here we only need $|a|>0$ and $\Lambda<0$.
Without loss of generality, we take $r_1$ and $r_2$ to be the real roots and $r_3$ and $r_4$ to be the complex-conjugate roots.
We see that $r_3r_4=|r_3|^2>0$, so once again Vieta's fourth formula tells us that $r_1r_2>0$ and thus that $r_1$ and $r_2$ have the same sign.
We again use Vieta's first formula to rewrite the second and third as
\begin{subequations}
\begin{align}
    r_1r_2+r_3r_4-\pa{r_1+r_2}^2&=a^2+L^2
    >0,\\
    \pa{r_1+r_2}\pa{r_3r_4-r_1r_2}&=2ML^2
    >0.
\end{align}
\end{subequations}
We can also rewrite the first equation as
\begin{align}
    r_3r_4-r_1r_2>r_1^2+r_2^2
    >0.
\end{align}
The second equation then implies that $r_1+r_2>0$, so once again since $r_1$ and $r_2$ have the same sign, they must both be positive.
This concludes the proof of the inequalities \eqref{eq:dSOrdering} and \eqref{eq:AdSOrdering}.

\section{Fuchsian differential equations}
\label{app:Fuchsian}

This appendix reviews the basic theory of Fuchsian differential equations and their singularities.

A homogeneous linear ODE whose coefficients are rational functions is called \textit{Fuchsian} if all of its singular points (including at infinity if there is one) are regular.
Any second-order Fuchsian equation with $n\geq 2$ distinct regular singular points $d_1, \ldots, d_{n-1}$, and $\infty$ can be written as \cite{Maier2007}
\begin{align}
    \label{eq:Fuchsian}
    Df\equiv\frac{d^2f}{dz^2}+\br{\sum_{i=1}^{n-1}\frac{1-\rho_i-\widehat{\rho}_i}{z-d_i}}\frac{df}{dz}+\br{\sum_{i=1}^{n-1}\frac{\rho_i\widehat{\rho}_i}{\pa{z-d_i}^2}+\frac{p_{n-3}(z)}{\prod_{i=1}^{n-1}\pa{z-d_i}}}f(z)=0,
\end{align}
where $p_{n-3}(z)=\sum_{i=0}^{n-3}c_iz^i$ is a polynomial of degree $n-3$.
For each regular singular point $d_i$, $\rho_i$ and $\widehat{\rho}_i$ are its characteristic exponents, that is, the roots of its indicial equation.
The leading coefficient of $p_{n-3}(z)$ is (here, $c_i=0$ for $i<0$)
\begin{align}
    \label{eq:LeadingCoefficient}
    c_{n-3}=-\sum_{i=1}^{n-1}\rho_i\widehat{\rho}_i+\rho_\infty\widehat{\rho}_\infty,
\end{align}
where $\rho_\infty$ and $\widehat{\rho}_\infty$ are the characteristic exponents for the point $z = \infty$.
The characteristic exponents obey the Fuchs relation:
\begin{align}
    \label{eq:FuchsRelation}
    \sum_{i=1}^{n-1}\pa{\rho_i+\widehat{\rho}_i}+\rho_\infty+\widehat{\rho}_\infty=n-2.
\end{align}
The $n-3$ subleading coefficients of $p_{n-3}(x)$ are independent of the characteristic exponents, and are conventionally called the accessory parameters.
Characteristic exponents can be defined for ordinary points as well as singular points.
Any ordinary point will have exponents 0 and 1, though the converse is not true: a finite point $d_i$ with exponents 0 and 1 is an ordinary point if and only if $p_{n-3}(d_i)=0$.
The point $z=\infty$ is an ordinary point if and only if it has exponents 0 and 1 and the first subleading coefficient of $p_{n-3}(z)$ is
\begin{align}
    \label{eq:OrdinaryInfinity}
    c_{n-4}=-\sum_{i=1}^{n-1}\rho_i\widehat{\rho}_i\pa{d_i-\sum_{j\neq i}d_j}.
\end{align}

Fuchsian equations have two important types of automorphisms.
The first consists of M\"obius transformations of the variable $z$.
Such transformations take the form 
\begin{align}
    Pz=\frac{Az+B}{Cz+D},\qquad
    AD-BC\neq0.
\end{align}
If $z=Pw$, then the equation $Df(z)=0$ becomes an equation $\wt{D}\wt{f}(w)=0$ for some function $\wt{f}(w)\equiv f(Pw)$, which has regular singular points at $w=P^{-1}d_i$ and $w=P^{-1}\infty$.
Characteristic exponents are preserved under M\"obius transformations, but accessory parameters are not.

The second type consists of the so-called index transformations, which are specified by $k$ complex numbers $e_i$ and $k$ ``index shifts'' $s_i$.
The transformed equation is $\wt{D}\wt{f}(z)=0$, where $\wt{f}(z)\equiv S(z)f(z)$ and $\wt{D}=SDS^{-1}$, with 
\begin{align}
    S(z)=\prod_{i=1}^k(z-e_i)^{s_i}.
\end{align}
The exponents associated with each point $x=e_i$ are shifted by $-s_i$ relative to the corresponding exponents in the original differential equation.
More precisely, if one of the $e_i$ is an ordinary point in the original equation, then it becomes a regular singular point in the transformed equation with exponents $-s_i$ and $1-s_i$.
Meanwhile, if one of the $e_i$ is a regular singular point with exponents $\rho_i$ and $\widehat{\rho}_i$ in the original equation, then its new exponents with respect to the transformed equation are $\rho_i-s_i$ and $\widehat{\rho}_i-s_i$, and the nature of the singularity depends upon these new exponents.
Consistent with Eq.~\eqref{eq:FuchsRelation}, the exponents of $z=\infty$ are shifted by $\underset{i=1}{\overset{k}{\sum}}s_i$. In general, the accessory parameters are also changed.

Fuchsian equations can be greatly simplified by applying index transformations that change the characteristic exponents in a judicious way: there is always a choice of $k=n-1$ shifts with
\begin{align}
    e_i=d_i,\qquad
    s_i=\widehat{\rho}_i,\qquad
    1\leq1\leq n-1,
\end{align}
that shifts one of the characteristic exponents to zero for each of the finite regular singular points (excluding infinity).
This transformation yields the ``reduced form'' for a Fuchsian equation,
\begin{align}
    \label{eq:ReducedFuchsian}
    \frac{d^2f}{dz^2}+\br{\sum_{i=1}^{n-1}\frac{1-\rho_i+\widehat{\rho}_i}{z-d_i}}\frac{df}{dz}+\br{\frac{p_{n-3}^0(z)}{\prod_{i=1}^{n-1}\pa{z-d_i}}}f(z)=0.
\end{align}
The standard forms of the hypergeometric and Heun equations are of this type.
Furthermore, via an index transformation, we can in some cases turn a regular singular point into an ordinary point: if the characteristic exponents of any regular singular point differ by exactly 1, then they can be shifted to precisely 0 and 1, though as explained above, this is a necessary but not sufficient condition for the point to become ordinary after the transformation.

\section{Heun functions}
\label{app:Heun}

In this appendix, we review the special function $\HeunG(z)$ that we use to express both the angular modes \eqref{eq:AngularModes} and the radial modes \eqref{eq:RadialModes}. 
Throughout, we closely follow Becker's study of the general Heun equation \cite{Becker1997}, directly quoting his results without derivation in many places.

First, we present the general Heun equation and its local Frobenius solutions in App.~\ref{app:CanonicalHeun}. 
In App.~\ref{app:Wronskian}, we quote Becker's functional form of the Wronskian associated with any two local Frobenius solutions around the two regular singular points $z=0$ and $z=1$.
Then, in App.~\ref{app:HeunFunctions}, we establish the condition that a solution of the general Heun equation must satisfy to be a general Heun \textit{function}.
In App.~\ref{app:HeunOrthogonality} and \ref{app:HeunNormalization}, we reproduce from Becker \cite{Becker1997} the orthogonality relations between these Heun functions together with a formula for the normalization integral of a Heun function.
Lastly, App.~\ref{app:Confluence} shows how the confluent Heun equation and its solution $\HeunC$ can be recovered from a limit of the general Heun equation and its solution $\HeunG$.

\subsection{Canonical form of the general Heun equation}
\label{app:CanonicalHeun}

The canonical form of the general Heun equation with parameters $(q,z_0,\alpha,\gamma,\delta)$ is
\begin{align}
    \label{eq:HeunEq}
    \frac{d^2f}{dz^2}+\pa{\frac{\gamma}{z}+\frac{\delta}{z-1}+\frac{\epsilon}{z-z_0}}\frac{df}{dz}+\frac{\alpha\beta z-q}{z(z-1)(z-z_0)}f(z)=0.
\end{align}
Here, $\epsilon$ is not an independent parameter but must obey Riemann's relation:
\begin{align}
    \label{eq:RiemannRelation}
    \epsilon=\alpha+\beta-\gamma-\delta+1
\end{align}
This is a Fuchsian equation of the form \eqref{eq:Fuchsian} with $n=4$ regular singular points located at $z=0$, $1$, $z_0$, and $\infty$.
The characteristic exponents for these points are
\begin{align}
     z=0:\quad1,\,1-\gamma,\qquad
     z=1:\quad0,\,1-\delta,\qquad
     z=z_0:\quad0,\,1-\epsilon,\qquad
     z=\infty:\quad\alpha,\,\beta,
\end{align}
so Riemann's relation \eqref{eq:RiemannRelation} is just the Fuchs relation \eqref{eq:FuchsRelation}, and indeed Eq.~\eqref{eq:LeadingCoefficient} holds.

As a second-order ODE, Eq.~\eqref{eq:HeunEq} admits two independent solutions.
The lowest powers of $z$ in the series expansions of the two solutions about $z=0$ are 0 and $1-\gamma$.
As for the two series expansions about $z=1$, their lowest powers of $1-z$ are 0 and $1-\delta$.
The function $\HeunG(z_0,q,\alpha,\gamma,\delta;z)$ denotes the unique solution whose power series expansion around $z=0$ has the leading term $z^0$ with coefficient 1.
This function is implemented in \textsc{Mathematica} as \texttt{HeunG}$[z_0,q,\alpha,\gamma,\delta,z]$ and may be regarded as a ``special function'' of mathematical physics.

In terms of this special function, the local Frobenius solutions around $z=0$ are\footnote{\label{fn:Log}When $1-\gamma$ is integer, one of the solutions is replaced by a more complicated expression involving a different power series plus a piece proportional to the other solution multiplied by $\ln(z)$.
If $1-\gamma>0$, then the leading term in the series is $z^0$, and the solution multiplying the logarithm has leading term $z^{1-\gamma}$.
The opposite holds for $1-\gamma<0$.
Similar statements can be made for the solutions around $z=1$ when $1-\delta$ is integer.}
\begin{subequations}
\label{eq:Frobenius0}
\begin{align}
    \label{eq:Frobenius0a}
    &\HeunG(z_0,q,\alpha,\gamma,\delta;z)=1+\frac{q}{\gamma z_0}z-\frac{\alpha\beta-\frac{q}{\gamma z_0}\br{\epsilon+q+\gamma+z_0(\gamma+\delta)}}{2z_0(\gamma+1)}z^2+\mc{O}\pa{z^3},\\
    \label{eq:Frobenius0b}
    &z^{1-\gamma}\HeunG\pa{z_0,q+(1-\gamma)(\epsilon+\delta z_0),1+\beta-\gamma,1+\alpha-\gamma,2-\gamma,\delta;z}.
\end{align}
\end{subequations}
Meanwhile, the local Frobenius solutions around $z=1$ are
\begin{subequations}
\label{eq:Frobenius1}
\begin{align}
    \label{eq:Frobenius1a}
    &\HeunG(1-z_0,\alpha\beta-q,\alpha,\beta,\delta,\gamma;1-z),\\
    \label{eq:Frobenius1b}
    &(1-z)^{1-\delta}\HeunG\pa{1-z_0,\alpha\beta-q+(1-\delta)(\epsilon+\gamma(1-z_0)),1+\beta-\delta,1+\alpha-\delta,2-\delta,\gamma;1-z}.
\end{align}
\end{subequations}

\subsection{Functional form of the Wronskian}
\label{app:Wronskian}

Fixing $z_0$, $\alpha$, $\beta$, $\gamma$, and $\delta$, but allowing the accessory parameter $q$ to vary, we let $f_0(q,z)$ denote either one of the local Frobenius solutions around $z=0$ given in Eq.~\eqref{eq:Frobenius0}.
Likewise, we let $f_1(q,z)$ denote either one of the local Frobenius solutions around $z=1$ given in Eq.~\eqref{eq:Frobenius1}.
Their Wronskian is
\begin{align} 
\label{eq:Wronskian}
    W(q,z)\equiv f_0(q,z)\frac{\pd f_1}{\pd z}(q,z)-f_1(q,z)\frac{\pd f_0}{\pd z}(q,z),
\end{align}
a definition which also holds for any two solutions more generally.
We quote from  Becker \cite{Becker1997} the following result for the functional form of this Wronskian. With the definition
\begin{align}
    p(z)\equiv z^\gamma(z-1)^\delta (z-z_0)^\epsilon,
\end{align}
the Wronskian obeys
\begin{align}
    \frac{d}{dz}\pa{p(z)W(q,z)}=0,
\end{align}
which implies the existence of some function $D(q)$ such that
\begin{align}
    \label{eq:WronskianDecomposition}
    W(q,z)=\frac{D(q)}{p(z)}.
\end{align}

\subsection{General Heun functions}
\label{app:HeunFunctions}

Any Frobenius solution around $z=0$ is a linear combination of the two Frobenius solutions around $z=1$, and vice versa.
A solution that is simultaneously Frobenius around both points is a (general) Heun \textit{function}.
Such functions are classified by their exponents at $z=0$ and $1$:
\begin{table}[h]
\centering
\begin{tabular}{| c | c |}
    \hline
        Class I & $(0,0)$ \\
        Class II & $(1-\gamma,0)$ \\
        Class III & $(0,1-\delta)$ \\
        Class IV & $(1-\gamma,1-\delta)$ \\
    \hline
\end{tabular}
\caption{Classes of general Heun functions}
\label{tbl:Classes}
\end{table}

Given a choice of parameters $z_0$, $\alpha$, $\beta$, $\gamma$, and $\delta$, there is a discrete but infinite set of values $\cu{q_n}$ of the accessory parameter for which a Frobenius solution becomes a general Heun function.
There is no closed form for these values, but they can be numerically determined as follows.
When $q=q_n$, the local Frobenius $f_0(q_n,z)$ and $f_1(q_n,z)$ are linearly dependent, so we have
\begin{align}
    f_0(q_n,z)=A(q_n)f_1(q_n,z),
\end{align}
and as a result, the Wronskian \eqref{eq:Wronskian} vanishes:
\begin{align}
    \label{eq:Quantization}
    W(q_n,z)=0.
\end{align}
The $n^\text{th}$ Heun function associated with the Heun equation \eqref{eq:HeunEq} is 
\begin{align}
    \label{eq:HeunFunctions}
    H_n(z)\equiv f_0(q_n,z).
\end{align}

\subsection{Orthogonality of the general Heun functions}
\label{app:HeunOrthogonality}

The general Heun functions \eqref{eq:HeunFunctions} satisfy orthogonality relations, which we again quote from Becker \cite{Becker1997}. He derives the following relation:
\begin{align} 
    (q_n-q_m)\int_0^1w(z)H_n(z)H_m(z)\ed z=p(z)\left.\pa{H_m(z)\frac{dH_n}{dz}-H_n(z)\frac{dH_m}{dz}}\right|_0^1,
\end{align}
where the relevant weight function in this general case is
\begin{align}
    w(z)\equiv z^{\gamma-1}(z-1)^{\delta-1}(z-z_0)^{\epsilon-1}.
\end{align}
By examining the behavior of the general Heun function near the regular singular points $z=0$ and $z=1$, we can see that both the $z=0$ and $z=1$ pieces of the right-hand side vanish, provided that the following class-dependent \textit{existence conditions} are met:
\begin{table}[h]
\centering
\begin{tabular}{| c | c c |}
    \hline
        Class I & $\re[\gamma]>0$, & $\re[\delta]>0$ \\
        Class II & $\re[\gamma]<2$, & $\re[\delta]>0$ \\
        Class III & $\re[\gamma]>0$, & $\re[\delta]<2$ \\
        Class IV & $\re[\gamma]<2$, & $\re[\delta]<2$ \\
    \hline
\end{tabular}
\caption{Existence conditions for orthogonal Heun functions}
\label{tbl:ExistenceConditions}
\end{table}

If the existence conditions are met, then we obtain the following orthogonality relation:
\begin{align} 
    (q_n-q_m)\int_0^1w(z)H_n(z)H_m(z)\ed z=0.
\end{align}

\subsection{Normalization of the Heun functions}
\label{app:HeunNormalization}

Becker also derives a formula for the normalization integral
\begin{align} 
    I_n\equiv\int_0^1w(z)\br{H_n(z)}^2\ed z.
\end{align}
We omit the details of derivation here and simply quote the result.
Let $q$ denote an arbitrary accessory parameter, and let $q_n$ satisfy the quantization condition \eqref{eq:Quantization}.
Then $f_0(q_n,z)$ is a general Heun \textit{function}, while $f_0(q,z)$ is generically not, even though they both solve the Heun equation \eqref{eq:HeunEq} with respect to their individual accessory parameters.
We also know that $f_0(q,z)$ must be some linear combination of the Frobenius solutions around $z=1$:
\begin{align}
    \label{eq:FrobeniusSolutions}
    f_0(q,z)=A(q)f_1(q,z)+B(q)\wt{f}_1(q,z),
\end{align}
where $f_1(q,z)$ denotes the Frobenius solution \eqref{eq:Frobenius1a} with exponent 0, and $\wt{f}_1(q,z)$ denotes the Frobenius solution \eqref{eq:Frobenius1b} with exponent $1-\delta$.
For the time being, we now restrict our attention to Heun functions of class I or II, for which $B(q_n)=0$. 
Then it can be shown that
\begin{align} 
    \label{eq:HeunNormalization}
    I_n=-p(z)\frac{\pd W}{\pd q}(q_n,z)\frac{f_0(q_n,z)}{f_1(q_n,z)},
\end{align}
Recalling Eqs.~\eqref{eq:WronskianDecomposition} and \eqref{eq:FrobeniusSolutions}, this can also be expressed as
\begin{align} 
    I_n=-D'(q_n)A(q_n),
\end{align}
which confirms that each piece is a constant (that is, independent of $z$).

This argument can be extended to general Heun functions of class III or IV by essentially exchanging $A$ and $B$ in the above derivation.
This results in an expression identical to Eq.~\eqref{eq:HeunNormalization}, but with $f_1$ now representing the local Frobenius solution with exponent $1-\delta$ (so long as the existence conditions are satisfied).

\subsection{Confluence of the general Heun equation}
\label{app:Confluence}

A confluence of a differential equation occurs when two or more regular singular points ``flow together'' to form an irregular singular point.
The relevant case in this work is the confluence of the general Heun equation \eqref{eq:HeunEq} in which the regular singular points at $z = z_0$ and $z=\infty$ merge to form an irregular singular point at $z=\infty$.
This yields the confluent Heun equation,
\begin{align}
    \label{eq:ConfluentHeun}
    f''(z)+\pa{\frac{\gamma}{z}+\frac{\delta}{z-1}+\epsilon}f'(z)+\frac{\alpha z-q}{z(z-1)}f(z)=0,
\end{align}
where $\epsilon$ is now unconstrained.
Its local Frobenius solutions can be expressed in terms of the special function $\HeunC$ reviewed in App.~E of Paper I.
This function admits the expansion
\begin{align}
    \label{eq:ConfluentFrobenius0}
    \HeunC(q,\alpha,\gamma,\delta,\epsilon;z)=1-\frac{q}{\gamma}z+\frac{1}{2(\gamma+1)}\pa{\alpha+\frac{q}{\gamma}(q-\gamma-\delta+\epsilon)}z^2+\mc{O}\pa{z^3}.
\end{align}
To transform the general Heun equation \eqref{eq:HeunEq} into the confluent Heun equation \eqref{eq:ConfluentHeun}, we take $q\to-qz_0$, $\alpha \to \alpha/\epsilon$, $\beta\to-\epsilon z_0$, and then let $z_0\to\infty$.
a, we have the identity
\begin{align}
    \label{eq:HeunGToHeunC}
    \HeunC(q,\alpha,\gamma,\delta,\epsilon;z)=\lim_{z_0\to\infty}\HeunG(z_0,-qz_0,\alpha/\epsilon,-\epsilon z_0,\gamma,\delta;z).
\end{align}
We shall use this identity in App.~\ref{app:KerrLimit}.

\section{Angular modes as general Heun functions}
\label{app:AngularHeun}

In this appendix, we recast the angular modes \eqref{eq:AngularModes} as general Heun functions and apply some of the results derived in App.~\ref{app:Heun}.
First, we map the angular ODE \eqref{eq:AngularODE} to the general Heun equation in App.~\ref{app:HeunAngularODE}.
Then, in App.~\ref{app:AngularBoundaryConditions}, we impose physical boundary conditions on its solutions to derive the spectrum of angular modes \eqref{eq:AngularModes} and their separation constants.
Finally, we determine their normalization in App.~\ref{app:AngularNormalization} using the formulas of App.~\ref{app:HeunNormalization}.

\subsection{Mapping the angular ODE to the Heun equation}
\label{app:HeunAngularODE}

This section adapts the techniques of Borissov and Fiziev \cite{Borissov2009} to map the angular ODE \eqref{eq:AngularODE} to the Heun equation.
First, we change to a variable $u=\cos{\theta}$, such that Eq.~\eqref{eq:AngularODE} becomes
\begin{subequations}
\label{eq:AngularVariableChange}
\begin{gather}
    \cu{\frac{d}{du}\br{\Upsilon\pa{1-u^2}\frac{d}{du}}-\frac{V_\theta(u)}{\Upsilon}-\frac{2\Lambda a^2}{3}\pa{2s^2+1}u^2+s+\lambda_{\omega\ell m}^{(s)}}S_{\omega\ell m}^{(s)}(u)=0,\\
    V_\theta(u)=\pa{a\omega\Xi}^2\pa{1-u^2}-2a\omega\Xi^2\pa{m-su}+\frac{\br{m\Xi+s(2\Upsilon-\Xi)u}^2}{1-u^2}.
\end{gather}
\end{subequations}
This equation has five regular singular points at $u=\pm1$, $u=\pm\frac{i}{\sqrt{\alpha}}$ with $\alpha=\frac{\Lambda a^2}{3}$, and $u=\infty$.
In particular, the singular point at infinity has characteristic exponents $\rho_\infty=1$ and $\widehat{\rho}_\infty=2$.

To recast this equation into a standard form, we first perform the redefinition
\begin{align}
    \label{eq:AngularCosineMode}
    S_{\omega\ell m}^{(s)}(u)=(1-u)^{\mu_1}(1+u)^{\mu_2}\pa{u+\frac{i}{\sqrt{\alpha}}}^{\mu_3}\pa{u-\frac{i}{\sqrt{\alpha}}}^{\mu_4}H(u),
\end{align}
where the $\mu_i$ are yet to be fixed.
This transformation turns Eq.~\eqref{eq:AngularVariableChange} into
\begin{align}
    \label{eq:AngularIntermediate}
    H''(u)+\pa{\frac{2\mu_1+1}{u-1}+\frac{2\mu_2+1}{u+1}+\frac{2\mu_3+1}{u+\frac{i}{\sqrt{\alpha}}}+\frac{2\mu_4+1}{u-\frac{i}{\sqrt{\alpha}}}}H'(u)+\mathcal{V}(u)H(u)=0, 
\end{align}
where we introduced a potential
\begin{align}
    \mathcal{V}(u)&=\frac{\mu_1^2-\frac{1}{4}(m+s)^2}{(u-1)^2}+\frac{\mu_2^2-\frac{1}{4}(m-s)^2}{(u+1)^2}+\frac{c_2u^2+2c_1u+c_0}{\alpha(u-1)(u+1)\pa{u+\frac{i}{\sqrt{\alpha}}}\pa{u-\frac{i}{\sqrt{\alpha}}}}\notag\\
    &\phantom{=}+\frac{\mu_3^2+\frac{1}{4}\pa{\frac{1+\alpha}{\sqrt{\alpha}}a\omega-m\sqrt{\alpha}-is}^2}{\pa{u+\frac{i}{\sqrt{\alpha}}}^2}+\frac{\mu_4^2+\frac{1}{4}\pa{\frac{1+\alpha}{\sqrt{\alpha}}a\omega-m\sqrt{\alpha}+is}^2}{\pa{u-\frac{i}{\sqrt{\alpha}}}^2},
\end{align}
with coefficients
\begin{subequations}
\begin{align}
    c_2&=-\frac{1}{2}\br{a\omega-\alpha(m-a\omega)}^2+\alpha\pa{\frac{m^2}{2}+s^2+2+3\sum_{i=1}^4\mu_i+2\sum_{1\leq i<j\leq4}\mu_i\mu_j},\\
    c_1&=sa\omega-i\sqrt{\alpha}(\mu_1+\mu_2+1)(\mu_3-\mu_4)+\alpha\br{sa\omega+(\mu_1-\mu_2)(\mu_3+\mu_4+1)},\\
    c_0&=\frac{1}{2}\br{a\omega-\alpha(m-a\omega)}^2-\lambda_{\omega\ell m}^{(s)}+\frac{m^2-s^2}{2}-s-2am\omega+\mu_1+\mu_2+2\mu_1\mu_2\notag\\
    &\phantom{=}-2i\sqrt{\alpha}(\mu_1-\mu_2)(\mu_3-\mu_4)+\alpha\br{2m(m-a\omega)+\frac{s^2}{2}-\mu_3-\mu_4-2\mu_3\mu_4}.
\end{align}
\end{subequations}
By carefully choosing the $\mu_i$, we can remove all the second-order poles in $\mc{V}(u)$ and put Eq.~\eqref{eq:AngularIntermediate} into the ``reduced form'' \eqref{eq:ReducedFuchsian} of a Fuchsian equation with five regular singular points (with one at infinity).
However, we can do even better than this.
By imposing the constraint 
\begin{align}
    \label{eq:MuConstraint}
    \sum_{i=1}^4\mu_i=-1,
\end{align}
we can shift the characteristic exponents of the point $u=\infty$ to $\rho_\infty=0$ and $\widehat{\rho}_\infty=1$.
Remarkably, when this constraint is imposed, Eq.~\eqref{eq:OrdinaryInfinity} is also satisfied, so $u=\infty$ becomes an ordinary point!
The removability of this singular point by an index transformation, leaving only four (finite and regular) singular points, is what allows us to map the angular ODE to the general Heun equation.
This property of the angular ODE was discovered by Suzuki, Takasugi, and Umetsu \cite{Suzuki1998}.

After imposing Eq.~\eqref{eq:MuConstraint}, only three of the $\mu_i$ can still be fixed independently: we set
\begin{align}
    \label{eq:MuConditions}
    \mu_1=\pm\frac{s+m}{2},\qquad
    \mu_2=\pm\frac{s-m}{2},\qquad
    \mu_3=\pm\frac{i}{2}\pa{\frac{1+\alpha}{\sqrt{\alpha}}a\omega-m\sqrt{\alpha}-is},
\end{align}
and then the value of $\mu_4$ is fixed by Eq.~\eqref{eq:MuConstraint}.
This index transformation sets one of the characteristic exponents for each of the singular points $u=\pm1$ and $u=-\frac{i}{\sqrt{\alpha}}$ to zero, while the characteristic exponents for $u=\frac{i}{\sqrt{\alpha}}$ both remain nonzero.
For later use, we also define
\begin{align}
    \widetilde{\mu}_3=\mp\frac{i}{2}\pa{\frac{1+\alpha}{\sqrt{\alpha}}a\omega-m\sqrt{\alpha}+is},
\end{align}
which for real $\omega$ and $\alpha > 0$ is simply the complex conjugate $\ol{\mu}_3$.

To precisely recover the canonical form \eqref{eq:HeunEq} of the Heun equation, we still have to perform a M\"obius transformation $u\to z$ that maps the regular singular points $u\in\cu{\pm1,\pm\frac{i}{\sqrt{\alpha}}}$ to their canonical positions $z\in\cu{0,1,z_0,\infty}$.
This transformation must send $u=\frac{i}{\sqrt{\alpha}}$ to $z=\infty$, as the characteristic exponents of the point $z=\infty$ are both nonzero in the general Heun equation.
We also wish for the Kerr limit $\alpha\to0$ to land on the confluent Heun equation \eqref{eq:ConfluentHeun}, so the transformation should send $u=-\frac{i}{\sqrt{\alpha}}$ to $z=z_0$.
These two desiderata leave only two choices:
\begin{align}
    u\to z_+
    =\frac{1}{2}\pa{1-\frac{i}{\sqrt{\alpha}}}\frac{u+1}{u-\frac{i}{\sqrt{\alpha}}},\qquad
    u\to z_-
    =\frac{1}{2}\pa{1+\frac{i}{\sqrt{\alpha}}}\frac{u-1}{u-\frac{i}{\sqrt{\alpha}}}.
\end{align}
The map to $z_+$ sends the north pole $\theta=0$ to $z_+=1$ and the south pole $\theta=\pi$ to $z_+=0$, while the map to $z_-$ does the reverse.
Both choices transform Eq.~\eqref{eq:AngularIntermediate} into Eq.~\eqref{eq:HeunEq}, namely
\begin{align}
    \label{eq:zAngularHeun}
    \frac{d^2H}{dz_\pm^2}+\pa{\frac{\gamma_\pm}{z_\pm}+\frac{\delta_\pm}{z_\pm-1}+\frac{\epsilon_\pm}{z-z_0^\pm}}\frac{dH}{dz_\pm}+\frac{\alpha_\pm \beta_\pm z_\pm-q_\pm}{z_\pm(z_\pm-1)(z-z_0^\pm)}H(z_\pm)=0,
\end{align}
but with different parameters:
\begin{subequations}
\label{eq:AngularHeunParameters}
\begin{gather}
    z_0^\pm=\mp\frac{i\pa{1\pm i\sqrt{\alpha}}^2}{4\sqrt{\alpha}},\qquad
    \gamma_+=\delta_-
    =2\mu_2+1,\qquad
    \delta_+=\gamma_-
    =2\mu_1+1,\\
    \epsilon_\pm=2\mu_3+1,\qquad
    \alpha_\pm=1+\mu_1+\mu_2+\mu_3+\widetilde{\mu}_3,\qquad
    \beta_\pm=1+\mu_1+\mu_2+\mu_3-\widetilde{\mu}_3,\\
    q_\pm=\pm\frac{i}{4\sqrt{\alpha}}\bigg\{\lambda_{\omega\ell m}^{(s)}+s\mp2i\sqrt{\alpha}+2a\omega(1+\alpha)(m\pm s)-\frac{m^2}{2}\br{4\alpha+\pa{1\pm i\sqrt{\alpha}}^2}+\frac{s^2}{2}\pa{1\mp i\sqrt{\alpha}}^2\notag\\
    \phantom{=}+2ims\sqrt{\alpha}\pa{1\pm i\sqrt{\alpha}}-\pa{1\pm i\sqrt{\alpha}}^2\pa{\mu_1+\mu_2+2\mu_1\mu_2}\mp2i\sqrt{\alpha}\br{(2\mu_3+1)\gamma_\pm-1}\bigg\}.
\end{gather}
\end{subequations}

\subsection{Boundary conditions and quantization of the mode spectrum}
\label{app:AngularBoundaryConditions}

The variables $z_+$ and $z_-$ are related by the joint transformations $u\to-u$ and $\sqrt{\alpha}\to-\sqrt{\alpha}$, or equivalently, $\theta\to\pi-\theta$ with $L\to-L$.
This is a discrete symmetry of the Kerr--(anti-)de Sitter metric \eqref{eq:KerrdS}, so the choice of either $z_+$ or $z_-$ is arbitrary---from now on, we consider only the mapping to $z_+$ and drop the $+$ subscript.

As in Paper I, the angular modes $S_{\omega\ell m}^{(s)}(\theta)$ must be regular on the sphere at both the north pole ($\theta=0$ or $z=1$) and south pole ($\theta=\pi$ or $z=0$).
The addition of a cosmological constant does not change the structure of the angular ODE at these points, so the process of imposing regularity there is nearly identical to that in App.~F.2 of Paper I.
The two regularity conditions again force the $H(u)$ appearing in Eq.~\eqref{eq:AngularCosineMode} to be a general Heun \textit{function}, which we choose to be of class I as in Paper I, requiring
\begin{align}
    \mu_1=\frac{|s+m|}{2},\qquad
    \mu_2=\frac{|s-m|}{2}.
\end{align}
The Heun parameters $\gamma$ and $\delta$ given in Eq.~\eqref{eq:AngularHeunParameters} then obey the existence conditions given in Table~\ref{tbl:ExistenceConditions}, so the angular modes are orthogonal and obey the normalization given in App.~\ref{app:AngularNormalization} below, which is derived from the one in App.~\ref{app:HeunNormalization}.

The choice of sign for $\mu_3$ in Eq.~\eqref{eq:MuConditions} changes the behavior at the regular singular point $z=z_0$, which does not lie on the sphere.
It is therefore arbitrary, and we choose the $+$ sign.

As in Paper I, and as discussed in App.~\ref{app:HeunFunctions}, $H(u)$ is only a general Heun \textit{function} if the accessory parameter $q$ belongs to the infinite but discrete set of values $\cu{q_n}$ such that the Wronskian \eqref{eq:Wronskian} vanishes.
This then imposes a quantization condition on the separation constant $\lambda_{\omega\ell m}^{(s)}$, whose discrete values we label using $\ell$ rather than $n$.

\subsection{Normalization of the angular modes}
\label{app:AngularNormalization}

Since the Heun functions $H(z)$ in the angular modes $S_{\omega\ell m}^{(s)}(z)$ obey the existence conditions in Table~\ref{tbl:ExistenceConditions}, according to Sec.~\ref{app:HeunNormalization}, the normalization constants $I_{\omega\ell m}^{(s)}$ appearing in Eq.~\eqref{eq:Orthogonality} are
\begin{align} 
    I_{\omega\ell m}^{(s)}\equiv\int_0^\pi\br{\hat{S}_{\omega\ell m}^{(s)}(\theta)}^2\sin{\theta}\ed\theta
    =C\int_0^1w(z)[H(z)]^2\ed z,
\end{align}
where $H(z)$ is given in Eq.~\eqref{eq:AngularHeun}, while
\begin{align}
    C=(-1)^{\mu_3}2^{|s+m|+|s-m|+2\mu_3+1}\pa{1-i\sqrt{\alpha}}^{4\mu_2+4\mu_3+2}(1+\alpha)^{-2\mu_2-4\mu_3-1}\alpha^{\mu_3}.
\end{align}
The results of App.~\ref{app:HeunNormalization} enable us to express this normalization as
\begin{subequations}
\begin{align}
    I_{\omega\ell m}^{(s)}&=-Cp(z)\frac{dW}{dq}(q_n,z)\frac{f_0(q_n,z)}{f_1(q_n,z)}\\
    &=-Cp(z)\left.\frac{d}{dq}\pa{f_0\frac{\pd f_1}{\pd z}-f_1\frac{\pd f_0}{\pd z}}\right|_{q=q_n}\frac{f_0(q_n,z)}{f_1(q_n,z)},
\end{align}
\end{subequations}
where $f_0(q,z)$ and $f_1(q,z)$ are given in Eqs.~\eqref{eq:Frobenius0a} and \eqref{eq:Frobenius1a}, and a $q$ without an $n$ subscript is \textit{not} a zero of the Wronskian (otherwise the derivative with respect to $q$ would be meaningless).

\section{Radial modes as general Heun functions}
\label{app:RadialHeun}

In this appendix, we recast the radial modes \eqref{eq:RadialModes} as solutions of the general Heun equation \eqref{eq:HeunEq}, though these solutions are generically \textit{not} general Heun \textit{functions} as defined in App.~\ref{app:HeunFunctions}.
After mapping the radial ODE \eqref{eq:RadialODE} to the Heun equation in App.~\ref{app:HeunRadialODE}, we adapt an elegant method of Ori's \cite{Ori2003} to derive the radial Teukolsky--Starobinsky constants \eqref{eq:ConstantsC} in App.~\ref{app:RTSDerivation}.

\subsection{Mapping the radial ODE to the Heun equation}
\label{app:HeunRadialODE}

In this section, we adapt the techniques of Borissov and Fiziev \cite{Borissov2009} to map the radial ODE \eqref{eq:RadialODE} to the Heun equation.
It is useful to rewrite the equation as
\begin{align}
    \label{eq:RadialODEAlt}
    \br{\frac{d^2}{dr^2}+(s+1)\frac{\Delta'}{\Delta}\frac{d}{dr}+\frac{f_4(r)}{\Delta^2}+\frac{f_2(r)}{\Delta}}R_{\omega\ell m}^{(s)}(r)=0,
\end{align}
where $f_4(r)$ and $f_2(r)$ are polynomials of degree 4 and 2, respectively, which are given by
\begin{subequations}
\begin{align}
    f_4(r)&=\Xi^2K^2-is\Xi K\Delta',\\
    f_2(r)&=-2(2s+1)(s+1)\frac{\Lambda}{3}r^2+4is\omega\Xi r-\frac{s\Lambda a^2}{3}-\lambda_{\omega\ell m}^{(s)}.
\end{align}
\end{subequations}
This equation admits exactly five singular points (all regular) located at the four roots of $\Delta(r)$ and $r=\infty$.
Analysis of the characteristic exponents reveals that $\rho_\infty = 2s+1$ and $\widehat{\rho}_\infty = 2s+2$. 

We will need the following partial fraction decomposition:
\begin{align}
    \label{eq:PartialFractions}
    \frac{f_4(r)}{\Delta^2(r)}=\sum_{i=1}^4\pa{\frac{f_4(r_i)}{\br{\Delta'(r_i)}^2(r-r_i)^2}+\frac{a_i}{r-r_i}},\qquad
    a_i\equiv\frac{f_4'(r_i)\Delta'(r_i)-f_4(r_i)\Delta''(r_i)}{\Delta'(r_i)^3}.
\end{align}
The coeffcients $a_i$ satisfy the useful relation 
\begin{align}
    \sum_{i=1}^4a_i=0,
\end{align}
which can be proved by applying the residue theorem to the left hand side of Eq.~\eqref{eq:PartialFractions} on a sufficiently large circular contour.
We will also need the identity
\begin{align}
    \sum_{i=1}^4\frac{a_i}{r-r_i}=\frac{b_2r^2+b_1r+b_0}{\prod_{i=1}^4(r-r_i)}, 
\end{align}
where we introduced new coefficients
\begin{align}
    b_2&=-\sum_{i=1}^4a_ir_i,\qquad
    b_1=\sum_{i=1}^4a_i\sum_{\substack{1\leq j<k\leq4\\j,k\neq i}}r_jr_k,\qquad
    b_0=-\sum_{i=1}^4a_i\sum_{\substack{1\leq j<k<l\leq4\\j,k,l\neq i}}r_jr_kr_l.
\end{align}
We now have all the ingredients needed to put Eq.~\eqref{eq:RadialODEAlt} into standard form.
We first transform
\begin{align}
    R_{\omega\ell m}^{(s)}(r)=\pa{r-r_1}^{\xi_1}\pa{r-r_2}^{\xi_2}\pa{r-r_3}^{\xi_3}\pa{r-r_4}^{\xi_4} H(r),
\end{align}
where the $\xi_i$ are yet to be fixed.
This transformation turns Eq.~\eqref{eq:RadialODE} into
\begin{align}
    \label{eq:RadialIntermediate}
    H''(r)+\pa{\sum_{i=1}^4\frac{2\xi_i+s+1}{r-r_i}}H'(r)+\br{\sum_{i=1}^4\frac{p_i(\xi_i)}{(r-r_i)^2}+\frac{c_2r^2+c_1r+c_0}{\prod_{i=1}^4(r-r_i)}}H(r)=0.
\end{align}
In this last equation, we introduced four polynomials
\begin{align}
    p_i(\xi)=\br{\xi-iC_i\pa{\omega-m\Omega_i}}\br{\xi+s+iC_i\pa{\omega-m\Omega_i}},
\end{align}
with constants
\begin{align}
    C_i=\frac{\Xi\pa{r_i^2+a^2}}{\Delta'(r_i)},\qquad
    \Omega_i=\frac{a}{r_i^2+a^2},
\end{align}
and new coefficients
\begin{subequations}
\begin{align}
    c_2&=b_2+2(s+1)(2s+1)+3(s+1)\sum_{i=1}^4\xi_i+2\sum_{1\leq i<j\leq4}\xi_i\xi_j,\\
    c_1&=b_1-\frac{12is\omega\Xi}{\Lambda}+2(s+1)\sum_{i=0}^4\xi_ir_i-2\sum_{i=1}^4r_i\sum_{\substack{1\leq j<k\leq4\\j,k\neq i}}\xi_j\xi_k,\\
    c_0&=b_0+\frac{3}{\Lambda}\pa{\lambda_{\omega\ell m}^{(s)}+s}+s\sum_{1\leq i<j\leq4}r_ir_j+(s+1)\sum_{i=1}^4\xi_i\sum_{\substack{1\leq j<k\leq4\\j,k\neq i}}r_jr_k.
\end{align}
\end{subequations}
Here, we used the first and second relations in Eq.~\eqref{eq:Vieta} to simplify $c_1$ and $c_0$, respectively.
As in the angular case, we see that by carefully choosing the $\xi_i$, it is possible to remove all the second-order poles in the coefficient of $H(r)$.
This would put the differential equation into the ``reduced form'' \eqref{eq:ReducedFuchsian} of a Fuchsian equation with five regular singular points (including one at infinity).
As before, however, we can do even better than this.
By imposing the constraint
\begin{align}
    \label{eq:XiConstraint}
    \sum_{i=1}^4\xi_i=-(2s+1),
\end{align}
we can shift the characteristic exponents of the point $r=\infty$ to $\rho_\infty=0$ and $\widehat{\rho}_\infty=1$.
Remarkably, when this constraint is imposed (and the first relation in Eq.~\eqref{eq:Vieta} is used), Eq.~\eqref{eq:OrdinaryInfinity} is also satisfied, so $r=\infty$ becomes an ordinary point!
Once again, it is the removability of this singular point by an index transformation, leaving only four (finite and regular) singular points, that allows us to map the radial ODE to the general Heun equation \cite{Suzuki1998}.

After imposing Eq.~\eqref{eq:XiConstraint}, only three of the $\xi_i$ can still be fixed independently: we set
\begin{gather}
    \label{eq:XiConditions}
    \xi_i=\frac{i\Xi K_i}{\Delta'(r_i)},
    =iC_i\pa{\omega-m\Omega_i}
    \qquad\text{or}\qquad 
    \xi_i=-s-\frac{i\Xi K_i}{\Delta'(r_i)}
    =-s-iC_i\pa{\omega-m\Omega_i},
\end{gather}
where $K_i=K\vert_{r=r_i}$, and then the value of $\xi_4$ is fixed by Eq.~\eqref{eq:XiConstraint}.
This index transformation sets one of the characteristic exponents for each of the singular points $r=r_i$ with $i\in\cu{1,2,3}$ to zero, while the characteristic exponents for $r=r_4$ both remain nonzero.
With this choice, $p_i(\xi_i)=0$ for $i\in\cu{1,2,3}$, but $p_4(\xi_4)\neq0$.
Thus, all second-order poles have been removed except for the one at $r=r_4$, which we will next push to infinity via a M\"obius transformation.

Indeed, to precisely recover the canonical form \eqref{eq:HeunEq} of the Heun equation, we still have perform a M\"obius transformation $r\to z$ that maps the regular singular points $r\in\cu{r_1,r_2,r_3,r_4}$ to their canonical positions $z\in\cu{0,1,z_0,\infty}$.
This transformation must send $r=r_4$ to $z=\infty$, as the characteristic exponents of the point $z=\infty$ are both nonzero in the Heun equation.
We also wish for the Kerr limit $\alpha\to0$ to land on the confluent Heun equation Eq.~\eqref{eq:ConfluentHeun}, so the transformation should send $r=r_3$ to $z=z_0$.
These two desiderata leave only two choices:
\begin{align}
    r\to z_+
    =-\pa{\frac{r_2-r_4}{r_1-r_2}}\pa{\frac{r-r_1}{r-r_4}},\qquad
    r\to z_-
    =\pa{\frac{r_1-r_4}{r_1-r_2}}\pa{\frac{r-r_2}{r-r_4}}.
\end{align}
The map to $z_+$ sends the (outer) event horizon $r=r_1$ to $z_+=0$ and the (inner) Cauchy horizon $r=r_2$ to $z_+=1$, while the map to $z_-$ does the reverse.
We define $z_0^\pm$ and $z_\infty^\pm$ as the images under these maps of $r=r_3$ and $r=\infty$, respectively:
\begin{gather}
    z_0^\pm = \mp\pa{\frac{r_{2,1}-r_4}{r_1-r_2}}\pa{\frac{r_3-r_{1,2}}{r_3-r_4}},\qquad
    z_\infty^\pm = \mp\pa{\frac{r_{2,1}-r_4}{r_1-r_2}}.
\end{gather}
Both M\"obius transformations transform Eq.~\eqref{eq:RadialIntermediate} into Eq.~\eqref{eq:HeunEq},
\begin{align}
    \label{eq:zRadialHeun}
    \frac{d^2H}{dz_\pm^2}+\pa{\frac{\gamma_\pm}{z_\pm}+\frac{\delta_\pm}{z_\pm-1}+\frac{\epsilon}{z-z_0^\pm}}\frac{dH}{dz_\pm}+\frac{\alpha_{\rm H}\beta z_\pm-q_\pm}{z_\pm\pa{z_\pm-1}\pa{z-z_0^\pm}}H(z_\pm)=0,
\end{align}
but with different parameters:
\begin{subequations}
\label{eq:RadialHeunParameters}
\begin{gather}
    \gamma_+=\delta_-
    =2\xi_1+s+1,\qquad
    \delta_+=\gamma_-
    =2\xi_2+s+1,\qquad
    \epsilon=2\xi_3+s+1,\\
    \alpha_{\rm H}=\xi_1+\xi_2+\xi_3+2s+1+\frac{i\Xi K_4}{\Delta'(r_4)},\qquad
    \beta=\xi_1+\xi_2+\xi_3+s+1-\frac{i\Xi K_4}{\Delta'(r_4)},\\
    q_\pm=\pm\frac{18\Xi^2}{\Lambda^2\triangle}\frac{(r_{2,1}-r_3)^2(r_{2,1}-r_4)^2(r_{1,2}-r_4)(r_3-r_4)}{r_1-r_2}\tilde{q}_\pm\pm\frac{6is\Xi}{\Lambda}\frac{\omega(r_{1,2}r_4+a^2)-am}{(r_1-r_2)(r_{1,2}-r_4)(r_3-r_4)}\notag\\
    \,\pm\frac{3}{\Lambda(r_1-r_2)(r_3-r_4)}\pa{\lambda_{\omega\ell m}^{(s)}+\frac{s\Lambda a^2}{3}}+(s+1)(2s+1)\br{z_\infty^\pm\pm\frac{2r_4^2}{(r_1-r_2)(r_3-r_4)}}\notag\\
    +2\xi_{1,2}\pa{z_0^\pm\xi_{2,1}+\xi_3}+(s+1)\br{\pa{z_0^\pm+1}\xi_{1,2}+z_0^\pm\xi_{2,1}+\xi_3},\\
    \tilde{q}_\pm=-\omega^2r_{1,2}^3\pa{r_1r_2+r_{1,2}r_3-2r_{2,1}r_3}-2a\omega(m-a\omega)r_{1,2}\pa{r_{2,1}r_3-r_{1,2}^2}\notag\\
    -a^2(m-a\omega)^2\pa{2r_{1,2}-r_{2,1}-r_3},
\end{gather}
\end{subequations}
where $\triangle$ is the discriminant of $-\frac{3}{\Lambda}\Delta(r)$,
\begin{align}
    \triangle=\prod_{1\leq i<j\leq4}(r_i-r_j)^2.
\end{align}
In the absence of additional boundary conditions, Eq.~\eqref{eq:zRadialHeun} admits two mode solutions, which take the simple forms \eqref{eq:Frobenius0} when expanded around $z_\pm=0$.
If we map $r\to z_+$, then we obtain the ``in'' and ``out'' radial modes \eqref{eq:RadialModes} that have a definite behavior at the event horizon $z_+=0$.

\subsection{Derivation of the radial Teukolsky--Starobinsky constants}
\label{app:RTSDerivation}

Finally, we derive the constants appearing in the radial Teukolsky--Starobinsky identities \eqref{eq:RTSI1},%
\begin{subequations}
\begin{align}
    \msc{D}_0^4\hat{R}_{\omega\ell m}^{(-2)\,{\rm in/out}}&=\hat{\msc{C}}_{\omega\ell m}^{\rm in/out}\hat{R}_{\omega\ell m}^{(+2)\,{\rm in/out}},\\
    \Delta^2\pa{\msc{D}_0^\dag}^4\Delta^2\hat{R}_{\omega\ell m}^{(+2)\,{\rm in/out}}&=\hat{\msc{C}}_{\omega\ell m}^{{\rm in/out}\,\prime}
    \hat{R}_{\omega\ell m}^{(-2)\,{\rm in/out}}.
\end{align}
\end{subequations}
As discussed in Sec.~\ref{subsec:TeukolskyStarobinsky}, their product $\mc{C}_{\omega\ell m}=\hat{\msc{C}}_{\omega\ell m}\hat{\msc{C}}_{\omega\ell m}^\prime$ can be determined from the second form \eqref{eq:RTSI2} of the radial Teukolsky--Starobinsky identities and is independent of the choice of modes.
Here, we compute $\hat{\msc{C}}_{\omega\ell m}$ and $\hat{\msc{C}}_{\omega\ell m}'$ for the particular modes we have chosen in Eq.~\eqref{eq:RadialModes}. 

Ori \cite{Ori2003} developed an elegant method for carrying out this computation.
The idea is to exploit the known behavior \eqref{eq:HorizonBehavior} of the ``in'' and ``out'' modes near the outer horizon:\footnote{Caveat lector: What we call the ``in'' and ``out'' modes, Ori calls ``down'' and ``up'' modes, and vice versa.}
\begin{subequations}
\label{eq:HorizonBehaviorBis}
\begin{align}
    \hat{R}_{\omega\ell m}^{(s)\,{\rm in}}(r)&\stackrel{r\to r_+}{\approx}c^{\rm in}\Delta^{-s}e^{-ikr_*},
    &&c^{\rm in}=\frac{\pa{r_1-r_2}^{iC_2k}}{(2M)^{i(C_1+C_2)k}}\pa{1-\frac{r_1}{r_3}}^{iC_3k}\pa{1-\frac{r_1}{r_4}}^{iC_4k},\\
    \hat{R}_{\omega\ell m}^{(s)\,{\rm out}}(r)&\stackrel{r\to r_+}{\approx}c^{\rm out}e^{ik r_*},
    &&c^{\rm out}=\frac{\pa{r_1-r_2}^{-iC_2k}}{(2M)^{-i(C_1+C_2)k}}\pa{1-\frac{r_1}{r_3}}^{-iC_3k}\pa{1-\frac{r_1}{r_4}}^{-iC_4k},
\end{align}
\end{subequations}
where $r_*$ is the tortoise coordinate \eqref{eq:TortoiseCoordinate}.
Near the horizon, the operators \eqref{eq:Dn} simplify to 
\begin{align}
    \msc{D}_0\approx\frac{\Xi\pa{r_1^2+a^2}}{\Delta}\pa{\pd_{r_*}-ik},\qquad
    \msc{D}_0^\dag\approx\frac{\Xi\pa{r_1^2+a^2}}{\Delta}\pa{\pd_{r_*}+ik}.
\end{align}
It follows that at leading order near the outer horizon, we have in terms of $w=2k\Xi(r_1^2+a^2)$:
\begin{subequations}
\begin{align}
    \msc{D}_0\pa{f(r)e^{ikr_*}}&\approx f'(r)e^{ikr_*},\\
    \msc{D}_0\pa{f(r)e^{-ikr_*}}&\approx\pa{f'(r)-\frac{iw}{\Delta}f(r)}e^{-ikr_*},\\
    \msc{D}_0^\dag\pa{f(r)e^{ikr_*}}&\approx\pa{f'(r)+\frac{iw}{\Delta}f(r)}e^{ikr_*},\\
    \msc{D}_0^\dag\pa{f(r)e^{-ikr_*}}&\approx f'(r)e^{-ikr_*}.
\end{align}
\end{subequations}
When $\msc{D}_0$ acts on $\hat{R}_{\omega\ell m}^{(-2)\,{\rm out}}$ or when $\msc{D}_0^\dag$ acts on $\Delta^2\hat{R}_{\omega\ell m}^{(+2)\,{\rm in}}$, the leading-order terms vanish by Eq.~\eqref{eq:HorizonBehaviorBis}.
As such, we focus on the following two expressions:
\begin{subequations}
\begin{align}
    \msc{D}_0^4\hat{R}_{\omega\ell m}^{(-2)\,{\rm in}}&\approx c^{\rm in}\msc{D}_0^4\pa{\Delta^2e^{-ikr_*}}
    \approx c^{\rm in}\frac{\Gamma}{\Delta^2}e^{-ikr_*},\\
    \Delta^2\pa{\msc{D}_0^\dag}^4\Delta^2\hat{R}_{\omega\ell m}^{(+2)\,{\rm out}}&\approx c^{\rm out}\Delta^2\pa{\msc{D}_0^\dag}^4\pa{\Delta^2e^{ikr_*}}
    \approx c^{\rm out}\wt{\Gamma}e^{ikr_*},
\end{align}
\end{subequations}
where $\Gamma$ and $\wt{\Gamma}$ are defined in Eq.~\eqref{eq:GammaSigmaW}.
We can now read off $\hat{\msc{C}}_{\omega\ell m}^{\rm in}$ and $\hat{\msc{C}}_{\omega\ell m}^{{\rm out}\,\prime}$.
Since the product of $\hat{\msc{C}}_{\omega\ell m}^{\rm in/out}$ and $\hat{\msc{C}}_{\omega\ell m}^{{\rm in/out}\,\prime}$ is fixed to be the known quantity $\mc{C}_{\omega\ell m}$ given in Eq.~\eqref{eq:RadialConstant}, this then also determines $\hat{\msc{C}}_{\omega\ell m}^{\rm out}$ and $\hat{\msc{C}}_{\omega\ell m}^{{\rm in}\,\prime}$, reproducing all the constants given in Eq.~\eqref{eq:ConstantsC}.

\section{Kerr limit}
\label{app:KerrLimit}

In this appendix, we take the limit of vanishing cosmological constant $\Lambda\to0$ (or equivalently, infinite (anti-)de Sitter radius $L\to\infty$) in which the Kerr--(anti-)de Sitter metric \eqref{eq:KerrdS} reduces to the usual Kerr geometry.
In particular, we focus on the limiting behavior of the angular and radial modes \eqref{eq:AngularModes} and \eqref{eq:RadialModes}, and check that they reproduce the expressions in Paper I.

\subsection{Kerr limit of angular modes}
\label{app:AngularKerrLimit}

Here, we check that the $L\to\infty$ limit of the angular modes \eqref{eq:AngularModes} recovers our Kerr modes from Paper I.
First, we expand the argument of the $\HeunG$ function in Eq.~\eqref{eq:AngularModes}:
\begin{align}
    \frac{1}{2}\pa{1-\frac{i}{\sqrt{\alpha}}}\frac{u+1}{u-\frac{i}{\sqrt{\alpha}}}&=\frac{u+1}{2}+\mc{O}\pa{\frac{1}{L}}.
\end{align}
Since $\mu_1$ and $\mu_2$ do not depend on $L$, we only need to compute [using Eq.~\eqref{eq:AngularModeParameters}]
\begin{subequations}
\begin{align}
    \mu_3&=\frac{i\omega L}{2\sigma}+\frac{s}{2}+\mc{O}\pa{\frac{1}{L}},\\
    \mu_4&=-\frac{i\omega L}{2\sigma}-\mu_1-\mu_2-\frac{s}{2}-1+\mc{O}\pa{\frac{1}{L}}.
\end{align}
\end{subequations}
We then focus on the pieces multiplying the $\HeunG$ function that depend on $L$.
We find that
\begin{align}
    \pa{\frac{u+\frac{i}{\sqrt{\alpha}}}{-1+\frac{i}{\sqrt{\alpha}}}}^{\mu_3}\pa{\frac{u-\frac{i}{\sqrt{\alpha}}}{-1-\frac{i}{\sqrt{\alpha}}}}^{\mu_4}=e^{a\omega(u+1)}+\mc{O}\pa{\frac{1}{L}},
\end{align}
while the remaining parameters in Eqs.~\eqref{eq:AngularModeParameters} obey
\begin{subequations}
\begin{align}
    \alpha_{\rm H}&=\mu_1+\mu_2+s+1
    =\frac{\alpha_{\rm Kerr}}{\epsilon_{\rm Kerr}},\qquad
    \gamma=2\mu_2+1
    =\gamma_{\rm Kerr},\qquad
    \delta=2\mu_1+1
    =\delta_{\rm Kerr},\\
    z_0&=-\frac{iL}{4a\sigma}+\frac{1}{2}+\mc{O}\pa{\frac{1}{L}},\qquad
    -\frac{\beta}{z_0}=4a\omega+\mc{O}\pa{\frac{1}{L}}
    =\epsilon_{\rm Kerr}+\mc{O}\pa{\frac{1}{L}},\\
    -\frac{q}{z_0}&=\lambda_{\omega\ell m}^{(s)}+2a\omega(2\mu_2+s+m+1)-2\mu_1\mu_2-\mu_1-\mu_2-\frac{m^2-s^2}{2}+s+\mc{O}\pa{\frac{1}{L}}\notag\\
    &=-p_{\rm Kerr}+\beta_{\rm Kerr}+\mc{O}\pa{\frac{1}{L}}.
\end{align}
\end{subequations}
By using all these expressions together with the identity \eqref{eq:HeunGToHeunC}, one can immediately see that the $L\to\infty$ limit of the angular modes \eqref{eq:AngularModes} reproduces Eqs.~(2.6) to (2.8) in Paper I.

\subsection{Kerr limit of radial modes}

Here, we check that the $L\to\infty$ limit of the radial modes \eqref{eq:RadialModes} recovers our Kerr modes from Paper I.
First, we use Eq.~\eqref{eq:RootExpansion} to expand the argument of the $\HeunG$ functions in Eq.~\eqref{eq:RadialModes}:
\begin{align}
    -\pa{\frac{r_2-r_4}{r_1-r_2}}\pa{\frac{r-r_1}{r-r_4}}=-\frac{r-r_1^{\rm Kerr}}{r_1^{\rm Kerr}-r_2^{\rm Kerr}}+\mc{O}\pa{\frac{1}{L}},\qquad
    r^{\rm Kerr}_{1,2}\equiv M\pm\sqrt{M^2-a^2}.
\end{align}
Next, we expand
\begin{subequations}
\begin{align}
    \xi_{1,2}&=\pm\frac{2iMr_{1,2}^{\rm Kerr}}{r_1^{\rm Kerr}-r_2^{\rm Kerr}}\pa{\omega-\frac{ma}{2Mr_{1,2}^{\rm Kerr}}}+\mc{O}\pa{\frac{1}{L^2}}
    =\xi_{1,2}^{\rm Kerr}+\mc{O}\pa{\frac{1}{L^2}},\\
    \xi_3 &= -\frac{i\sigma \omega L}{2}-i\omega M+\mc{O}\pa{\frac{1}{L}},\\
    \xi_4 &= \frac{i\sigma \omega L}{2}-i\omega M-2s-1+\mc{O}\pa{\frac{1}{L}}.
\end{align}
\end{subequations}
We then handle each of the pieces multiplying the $\HeunG$ function separately:
\begin{subequations}
\begin{align}
    \Delta'(r_1)&=r_1^{\rm Kerr}-r_2^{\rm Kerr}+\mc{O}\pa{\frac{1}{L^2}},\\
    \pa{r-r_1}^{\xi_1}\pa{\frac{r-r_2}{r_1-r_2}}^{\xi_2}&=\pa{r-r_1^{\rm Kerr}}^{\xi_1^{\rm Kerr}}\pa{\frac{r-r_2^{\rm Kerr}}{r_1^{\rm Kerr}-r_2^{\rm Kerr}}}^{\xi_2^{\rm Kerr}}+\mc{O}\pa{\frac{1}{L^2}},\\
    \pa{\frac{r-r_3}{r_1-r_3}}^{\xi_3}\pa{\frac{r-r_4}{r_1-r_4}}^{\xi_4}&=e^{i\omega\pa{r-r_1^{\rm Kerr}}}+\mc{O}\pa{\frac{1}{L}}.
\end{align}
\end{subequations}
Lastly, we find that the parameters in \eqref{eq:RadialModeParameters} obey
\begin{subequations}
\begin{align}
    \alpha_{\rm H}&=2s+1
    =\frac{\alpha_{\rm Kerr}}{\epsilon_{\rm Kerr}},\\
    \gamma&=2\xi_1^{\rm Kerr}+s+1+\mc{O}\pa{\frac{1}{L^2}}
    =\gamma_{\rm Kerr}+\mc{O}\pa{\frac{1}{L^2}},\\
    \delta&=2\xi_2^{\rm Kerr}+s+1+\mc{O}\pa{\frac{1}{L^2}}
    =\delta_{\rm Kerr}+\mc{O}\pa{\frac{1}{L^2}},\\
    z_0&=-\frac{\sigma L}{2\pa{r_1^{\rm Kerr}-r_2^{\rm Kerr}}}+\frac{1}{2}+\mc{O}\pa{\frac{1}{L}},\\
    -\frac{\beta}{z_0}&=-2i\omega\pa{r_1^{\rm Kerr}-r_2^{\rm Kerr}}+\mc{O}\pa{\frac{1}{L}}
    =\epsilon_{\rm Kerr}+\mc{O}\pa{\frac{1}{L}},\\
    -\frac{q}{z_0}&=\lambda_{\omega\ell m}^{(s)}-2i\omega r_1^{\rm Kerr}(2s+1)+\mc{O}\pa{\frac{1}{L}}
    =q_{\rm Kerr}+\mc{O}\pa{\frac{1}{L}}.
\end{align}
\end{subequations}
It is also useful to note that
\begin{subequations}
\begin{align}
    -\frac{q+(\epsilon-\delta)(1-\gamma)}{z_0}&=\pa{2\xi_1^{\rm Kerr}+s}\br{2i\omega\pa{r_1^{\rm Kerr}-r_2^{\rm Kerr}}+2\xi_2^{\rm Kerr}+s+1}+\mc{O}\pa{\frac{1}{L}}\notag\\
    &\phantom{=}+\lambda_{\omega\ell m}^{(s)}-2i\omega r_1^{\rm Kerr}(2s+1)\\
    &=q_{\rm Kerr}+\pa{\epsilon_{\rm Kerr}-\delta_{\rm Kerr}}\pa{1-\gamma_{\rm Kerr}}+\mc{O}\pa{\frac{1}{L}},\notag\\
    1+\alpha-\gamma&=\xi_2^{\rm Kerr}-\xi_1^{\rm Kerr}-2i\omega M+s+1+\mc{O}\pa{\frac{1}{L^2}}\\
    &=\frac{\alpha_{\rm Kerr}}{\epsilon_{\rm Kerr}}+\pa{1-\gamma_{\rm Kerr}}+\mc{O}\pa{\frac{1}{L^2}},\\
    -\frac{1+\beta-\gamma}{z_0}&=-2i\omega\pa{r_1^{\rm Kerr}-r_2^{\rm Kerr}}+\mc{O}\pa{\frac{1}{L}}
    =\epsilon_{\rm Kerr}+\mc{O}\pa{\frac{1}{L}}.
\end{align}
\end{subequations}
By using all these expressions together with the identity \eqref{eq:HeunGToHeunC}, one can immediately see that the $L\to\infty$ limit of the radial modes \eqref{eq:RadialModes} reproduces Eqs.~(2.17) to (2.19) in Paper I.

\bibliographystyle{utphys}
\bibliography{BGL2.bib}

\end{document}